%% file: ANIT_journal.tex
\documentclass[12pt,draftcls,onecolumn]{IEEEtran}

\usepackage{cite}
\usepackage{color}
\usepackage{standalone}
\usepackage{pstricks}
\usepackage{graphicx}
\usepackage{amsmath}
\usepackage{amssymb}
\usepackage{mathtools}
\usepackage{psfrag}
\usepackage{amsthm}
\usepackage{tikz,pgfplots}
\input{defns.tex}

\newrgbcolor{lineblue}{.2 .6 .9}
\newrgbcolor{linered}{.9 .3 .3}
\newrgbcolor{linegreen}{.3 .8 .3}
\newrgbcolor{linepurple}{.75 .5 .75}

\newrgbcolor{LineGray}{.6 .6 .6}
\newrgbcolor{AxisGray}{.3 .3 .3}
\newrgbcolor{LineBlue}{.2 .6 .9}
\newrgbcolor{LineRed}{.9 .3 .3}
\newrgbcolor{LineGreen}{.3 .8 .3}
\newrgbcolor{LinePurple}{.8 .3 .8}

\newtheorem{theorem}{Theorem}
\newtheorem{lemma}{Lemma}
\newtheorem{corollary}{Corollary}

\theoremstyle{definition}
\newtheorem{example}{Example}
\newtheorem{remark}{Remark}
\newtheorem{definition}{Definition}

\newcommand{\Rmax}{R_{\text{max}}}

\newcommand{\Rtmax}{\Rt_{\text{max}}}

\newcommand{\etab}{\boldsymbol{\eta}}

\newcommand{\nub}{\boldsymbol{\nu}}
\newcommand{\Gb}{\mathsf{G}}
\newcommand{\Ab}{\mathsf{A}}

\newcommand{\ab}{\boldsymbol{a}}
\newcommand{\bb}{\boldsymbol{b}}
\newcommand{\cb}{\boldsymbol{c}}
\newcommand*{\Scale}[2][4]{\scalebox{#1}{$#2$}}%
\newcommand{\intd}{{\,\operatorname{d}}}

\begin{document}

\title{A Joint Typicality Approach to Algebraic Network Information Theory}

\author{Sung Hoon Lim, Chen Feng, Adriano Pastore,\\ Bobak Nazer, and Michael Gastpar\\
\thanks{This paper was presented in part at the 2014 IEEE Information Theory Workshop, Hobart, Australia and the 2015 Allerton Conference on Communication, Control, and Computing, Monticello, IL.}
\thanks{Sung Hoon Lim, Adriano Pastore, and Michael Gastpar are with the School of Computer and Communication Sciences, Ecole Polytechnique F\'ed\'erale, 1015 Lausanne, Switzerland
 (e-mail: sung.lim@epfl.ch, adriano.pastore@epfl.ch, michael.gastpar@epfl.ch).}
\thanks{Chen Feng is with the School of Engineering, The University of British Columbia, Kelowna, BC, Canada (e-mail: chen.feng@ubc.ca).}
\thanks{Bobak Nazer is with the Department of Electrical and Computer Engineering, Boston University, Boston, MA (e-mail: bobak@bu.edu).}

%\bigskip
%
%\today
}

\maketitle
%\allowdisplaybreaks

\begin{abstract}
This paper presents a joint typicality framework for encoding and decoding nested linear codes for multi-user networks. This framework provides a new perspective on compute--forward within the context of discrete memoryless networks. In particular, it establishes an achievable rate region for computing the weighted sum of nested linear codewords over a discrete memoryless multiple-access channel (MAC). When specialized to the Gaussian MAC, this rate region recovers and improves upon the lattice-based compute--forward rate region of Nazer and Gastpar, thus providing a unified approach for discrete memoryless and Gaussian networks. Furthermore, this framework can be used to shed light on the joint decoding rate region for compute--forward, which is considered an open problem. Specifically, this work establishes an achievable rate region for simultaneously decoding two linear combinations of nested linear codewords from $K$ senders. 
\end{abstract}
\begin{IEEEkeywords}
Linear codes, joint decoding, compute--forward, multiple-access channel, relay networks
\end{IEEEkeywords}
 
%%%%%%%%%%%%%%%%%%%%%%%%%%%%%%%%%%%%%%%%%%
%%%%
%%%%  Introduction
%%%%
%%%%%%%%%%%%%%%%%%%%%%%%%%%%%%%%%%%%%%%%%%

\section{Introduction} \label{sec:intro}

In network information theory, random i.i.d.~ensembles serve as the foundation for the vast majority of coding theorems and analytical tools.  As elegantly demonstrated by the textbook of El Gamal and Kim~\cite{El-Gamal--Kim2011}, the core results of this theory can be unified via a few powerful packing and covering lemmas. However, starting from the many--help--one source coding example of K\"orner and Marton~\cite{Korner--Marton1979}, it has been well-known that there are coding theorems that seem to require random linear ensembles, as opposed to random i.i.d.~ensembles. Recent efforts have demonstrated that linear and lattice codes can yield new achievable rates for relay networks~\cite{Wilson--Narayanan--Pfister--Sprintson2010,Nam--Chung--Lee2010,Nazer--Gastpar2011,Niesen--Whiting2012,Song--Devroye2013,Hong--Caire2013,Ren--Goseling--Weber--Gastpar2014}, interference channels~\cite{Bresler--Parekh--Tse2010,Motahari--Gharan--Maddah-Ali--Khandani2014,Niesen--Maddah-Ali2013,Ordentlich--Erez--Nazer2014,Shomorony--Avestimehr2014,Ntranos--Cadambe--Nazer--Caire2013b,Padakandla--Sahebi--Pradhan2016}, distributed source coding~\cite{Krithivasan--Pradhan2009,Krithivasan--Pradhan2011,Wagner2011,Maddah-Ali--Tse2010,Yang--Xiong2014}, dirty-paper multiple-access channels~\cite{Philosof--Zamir2009,Philosof--Zamir--Erez--Khisti2011,Wang2012,Padakandla--Pradhan2013b}, and physical-layer secrecy~\cite{He--Yener2014,Vatedka--Kashyap--Thangaraj2015,Xie--Ulukus2014}. See~\cite{Nazer--Zamir2014} for a survey of lattice-based techniques for Gaussian networks.
 
Although there is now a wealth of examples that showcase the potential gains of random linear ensembles, it remains unclear if these examples can be captured as part of a general framework, i.e., an \textit{algebraic network information theory}, that is on par with the well-established framework for random i.i.d.~ensembles. The recent work of Padakandla and Pradhan~\cite{Padakandla--Pradhan2012,Padakandla--Pradhan2013b,Padakandla--Sahebi--Pradhan2016} has taken important steps towards such a theory, by developing joint typicality encoding and decoding techniques for nested linear code ensembles. In this paper, we take further steps in this direction by developing coding techniques and error bounds for nested linear code ensembles. For instance, we provide a packing lemma for analyzing the performance of linear codes under \textit{simultaneous} joint typicality decoding (in Sections~\ref{sec:comp-finite} and~\ref{sec:k-mac-proof}) and a Markov Lemma for linear codes (in Appendix~\ref{app:ML-LC}).

We will use the compute--forward problem as a case study for our approach. As originally stated in~\cite{Nazer--Gastpar2011}, the objective in this problem is to reliably decode one or more linear combinations of the messages over a Gaussian multiple-access channel (MAC). Within the context of a relay network, compute--forward allows relays to recover linear combinations of interfering codewords and send them towards a destination, which can then solve the resulting linear equations for the desired messages. Recent work has also shown that compute--forward is useful in the context of interference alignment. For instance, Ordentlich \textit{et al.}~\cite{Ordentlich--Erez--Nazer2014} approximated the sum capacity of the symmetric Gaussian interference channel via compute--forward. The achievable scheme from~\cite{Nazer--Gastpar2011} relies on nested lattice encoding combined with ``single-user'' lattice decoding, i.e., each desired linear combination is recovered independently of the others. Subsequent efforts~\cite{Ordentlich--Erez--Nazer2013,Ordentlich--Erez--Nazer2014,Nazer--Cadambe--Ntranos--Caire2015} developed a variation of successive cancellation for decoding multiple linear combinations.

In this paper, we generalize compute--forward beyond the Gaussian setting and develop single-letter achievable rate regions using joint typicality decoding. Within our framework, each encoder maps its message into a vector space over a field and the decoder attempts to recover a linear combination of these vectors. In particular, Theorem~\ref{thm:comp-Fq} establishes a rate region for recovering a finite-field linear combination over a MAC. This includes, as special cases, the problem of recovering a finite-field linear combination over a discrete memoryless (DM) MAC and a Gaussian MAC. In Theorem~\ref{thm:comp-real-discrete}, we develop a rate region for recovering an integer-linear combination of bounded, integer-valued vectors. Finally, in Theorem~\ref{thm:comp-real}, we use a quantization argument to obtain a rate region for recovering an integer-linear combination of real-valued vectors.

As mentioned above, the best-known rate regions for lattice-based compute--forward rely on successive cancellation decoding. One might expect that simultaneous decoding yields a larger rate region for recovering two or more linear combinations. However, for a random lattice codebook, a direct analysis of simultaneous decoding is challenging, due to the statistical dependencies induced by the shared linear structure~\cite{Ordentlich--Erez2013}. We are able to surmount this difficulty by carefully partitioning error events directly over the finite field from which the codebook is drawn. Overall, we obtain a rate region for simultaneously recovering two linear combinations in Theorem~\ref{thm:K-2-comp}.

Our results recover and improve upon the rate regions of~\cite{Nazer--Gastpar2011, Zhu--Gastpar2014a, Nazer--Cadambe--Ntranos--Caire2015}, thus providing a unified approach to compute--forward over both DM and Gaussian networks. Additionally, the single-letter rate region implicitly captures recent work~\cite[Example 3]{Zhu--Gastpar2015} that has shown that Gaussian input distributions are not necessarily optimal for Gaussian networks. One appealing feature of our approach is that the first-order performance analysis uses steps that closely resemble those used for random i.i.d.~ensembles. However, there are several technical subtleties that arise due to linearity, which require careful treatment in our error probability bounds.

\begin{figure}[t]
\begin{center}
\includegraphics[width=0.75\textwidth]{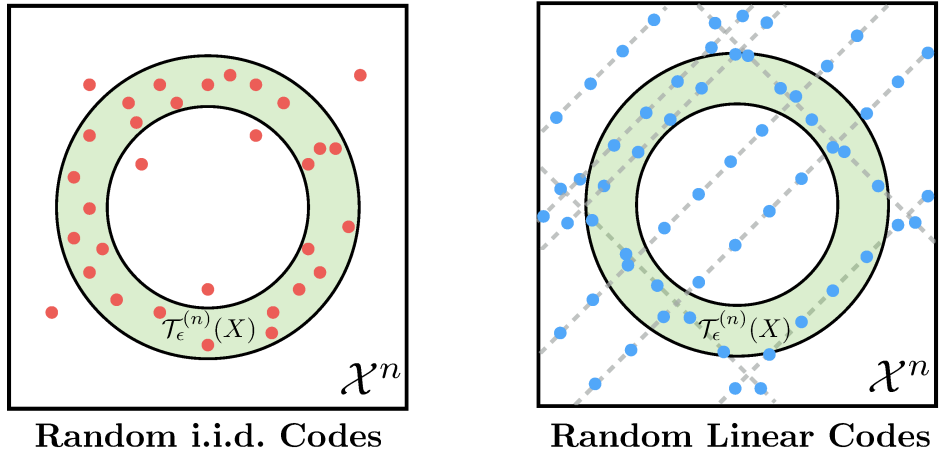} 
\end{center}
\caption{An illustration of the typicality of random i.i.d.~(red) and random linear (blue) codewords. Due to the weak law of large numbers, most random i.i.d.~codewords are typical for large $n$. In contrast, since random linear codewords are uniformly distributed, exponentially many codewords will be atypical with respect to non-uniform distributions. We resolve this issue via \textit{multicoding}, i.e., we generate exponentially more linear codewords than needed and use an auxiliary index to select the typical ones.}\label{fig:typicalsets}
\end{figure}

For a random linear codebook, each codeword is i.i.d.~uniformly distributed over the underlying finite field. This poses a challenge for generating non-uniform channel input distributions, and it is well-known that a direct application of a linear codebook cannot attain the point-to-point capacity in general~\cite{Ahlswede1971Group}. See Figure~\ref{fig:typicalsets} for an illustration. To get around this issue, we will use the nested linear coding architecture which first appeared in~\cite{Miyake2010, Padakandla--Pradhan2013}. This encoding architecture consists of the following components:
\begin{enumerate}
\item an auxiliary {\em linear} code (shared by all encoders)
\item a joint typicality encoder for multicoding
\item a symbol-by-symbol function of the auxiliary linear codeword.
\end{enumerate} Roughly speaking, the auxiliary linear code is designed at a higher rate than the target achievable rate, the joint typicality encoding is used to select codewords of the desired type, and the function is used to map the codeword symbols from the finite field to the channel input alphabet. The idea of using a joint typicality encoder for channel coding appears in the celebrated coding scheme by Gelfand and Pinsker~\cite{Gelfand--Pinsker1980a} for channels with state, Marton's coding scheme for the broadcast channel~\cite{Marton1979} and the hybrid coding scheme~\cite{Minero--Lim--Kim2015} for joint--source channel coding. In contrast to these applications, our joint typicality encoding step is used to find an auxiliary codeword that is {\em itself} typical with respect to a desired distribution, instead of with respect to a state or source sequence. The use of a symbol-by-symbol function is reminiscent of the Shannon strategy~\cite{Shannon1958a} for channels with states. 

The shared linear codebook creates subtle issues for the analysis of joint typicality encoding and decoding. Specifically, the users' choices of typical codewords depend upon the codebook, and thus the codewords are not independent across users. For this scenario, the standard Markov lemma (see, for instance,~\cite[Lemma 12.1]{El-Gamal--Kim2011}) does not directly apply. To overcome this issue, prior work by Padakandla and Pradhan proposed a Markov lemma for nested linear codes that required both a lower and an upper bound on the auxiliary rates~\cite{Padakandla--Pradhan2013b}. In Appendix~\ref{app:ML-LC}, we follow a different proof strategy, which enables us to remove the upper bound.

Furthermore, for a random linear codebook, the codewords are only pairwise independent. While this suffices to apply a standard packing lemma~\cite[Section~3.2]{El-Gamal--Kim2011} for decoding a single codeword, it creates obstacles for decoding multiple codewords. In particular, one has to contend with the fact that competing codewords may be linearly dependent on the true codewords. To cope with these linear dependencies, we develop a packing lemma for nested linear codes, which serves as a foundation for the achievable rate regions described above.

We closely follow the notation in~\cite{El-Gamal--Kim2011}. Let $\Xc$ denote the alphabet and $x^n$ a length-$n$ sequence whose elements belong to $\Xc$ (which can be either discrete or a subset of $\Real$). We use uppercase letters to denote random variables. For instance, $X$ is a random variable that takes values in $\Xc$. We follow standard notation for probability measures. Specifically, we denote the probability of an event $\Ac$ by $\P\{\Ac\}$ and use $P_X(x)$, $p_X(x)$, $f_X(x)$, and $F_X(x)$ to denote a probability distribution (i.e., measure), probability mass function (pmf), probability density function (pdf), and cumulative distribution function (cdf), respectively.

For finite and discrete $\Xc$, the type of $x^n$ is defined to be $\pi(x | x^n) := \big| \{ i: x_i = x \} \big| /n$ for $x \in \Xc$. Let $X$ be a discrete random variable over $\Xc$ with probability mass function $p_X(x)$. For any parameter $\e \in (0,1)$,  we define the set of $\e$-typical $n$-sequences
 $x^n$ (or the typical set in short)~\cite{Orlitsky--Roche2001} as
$\aep(X) = \{ x^n : | \pi(x|x^n) - p_X(x) | \le \e p_X(x)
\text{ for all } x \in \Xc \}$.  We use $\delta(\e) > 0$ to denote a generic function
 of $\e > 0$ that tends to zero as $\e \to 0$. One notable departure is that we define sets of message indices starting at zero rather than one, $[n] := \{0,\ldots, n-1\}$.

We use the notation $\Field$, $\mathbb{R}$, and $\Fq$ to denote a field, the real numbers, and the finite field of order $\q$, respectively. We denote deterministic row vectors either with lowercase, boldface font (e.g., $\ab \in \Fq^K$). Note that a deterministic row vector can also be written as a sequence (e.g., $u^n \in \Fq^n$). We will denote random sequences using uppercase font (e.g., $U^n \in \Fq^n$) and will not require explicit notation for random vectors. Random matrices will be denoted with uppercase, boldface font (e.g., $\mathbf{G} \in \Fq^{n \times \kappa}$)  and we will use uppercase, sans-serif font to denote realizations of random matrices (e.g., $\mathsf{G} \in \Fq^{n \times \kappa}$) or deterministic matrices.

%%%%%%%%%%%%%%%%%%%%%%%%%%%%%%%%%%%%%%%%%%%%%%%
%
%			PROBLEM STATEMENT
%
%%%%%%%%%%%%%%%%%%%%%%%%%%%%%%%%%%%%%%%%%%%%%%%%

\section{Problem Statement} \label{sec:problemstatement}

We now give a formal problem statement for compute--forward. Although the primary results of this paper focus on recovering one or two linear combinations, we state the general case of recovering $K$ linear combinations so that we can clearly state open questions. 

Consider the $K$-user memoryless multiple-access channel (MAC) 
\begin{align*}
(\Xc_1\times\cdots\times \Xc_K, P_{Y|X_1,\ldots,X_K}(y|x_1, \ldots, x_K), \Yc)
\end{align*}
which consists of $K$ sender alphabets $\Xc_k$, $k\in[1:K]$, one receiver alphabet
$\Yc$, and a collection of conditional probability distributions $P_{Y|X_1,\ldots,X_K}(y| x_1,\ldots,x_K)$. Since the channel is memoryless, we have that $$P_{Y^n | X_1^n,\ldots,X_K^n}(y^n | x_1^n,\ldots,x_K^n) = \prod_{i=1}^n P_{Y|X_1,\ldots,X_K}(y_i | x_{1i},\ldots,x_{Ki}).$$ In our considerations, the input alphabets $\Xc_k$ and receiver alphabet $\Yc$ are either finite or the real line. Note that discrete memoryless (DM) MACs and Gaussian MACs are special cases of this class of channels.

\input{probstatementfig}

Consider a field $\Field$ (not necessarily finite) and let $\mathbb{A}\subset\Field$ be a {\em discrete subset of $\Field$}. Let $\ab_1,\ldots,\ab_K \in \mathbb{A}^K$ denote the coefficient vectors, and let 
\begin{align}
\Ab = \begin{bmatrix} \ab_1 \\ \vdots \\ \ab_K \end{bmatrix} \in \mathbb{A}^{K\times K}
\end{align} 
denote the coefficient matrix. 

A $(2^{nR_1},\ldots,2^{nR_K},n,(\Field, \mathbb{A}),\Ab)$ code for compute--forward consists of 
\begin{itemize}
\item $K$ message sets $[2^{nR_k}]$, $k\in[1:K]$
\item $K$ encoders, where encoder $k$ maps each message $m_k \in [2^{nR_k}]$ to a pair of sequences $(u^n_k, x^n_k)(m_k)\in\Field^n\times\Xc_k^n$ such that $u^n_k(m_k)$ is {\em bijective}, 
\item $K$ linear combinations for each message tuple $(m_1,\ldots, m_K)$
\begin{align*}
\begin{bmatrix}
w^n_{\ab_1}(m_1,\ldots, m_K)\\
\vdots\\
w^n_{\ab_K}(m_1,\ldots, m_K)
\end{bmatrix} = \Ab \begin{bmatrix}
u^n_{1}(m_1)\\
\vdots \\
u^n_{K}(m_K)
\end{bmatrix},
\end{align*}
where the linear combinations are defined over the vector space $\Field^n$, and

\item a decoder that assigns estimates $(\hat{w}^n_{\ab_1},\ldots,\hat{w}^n_{\ab_K}) \in \Field^n \times \cdots \times \Field^n$ to each received sequence $y^n \in \Yc^n$.

\end{itemize}

Each message $M_k$ is independently and uniformly drawn from $[2^{nR_k}]$. The average probability of error is defined as $\pen = \P\big\{(\hat{W}^n_{\ab_1},\ldots,\hat{W}^n_{\ab_K}) \neq (W^n_{\ab_1},\ldots,W^n_{\ab_K})\big\}$. We say that a rate tuple $(R_1,\ldots,R_K)$ is achievable for recovering the linear combinations with coefficient matrix $\Ab$ if there exists 
a sequence of $(2^{nR_1},\ldots,2^{nR_K},n,(\Field, \mathbb{A}),\Ab)$ codes such that $\lim_{n\rightarrow \infty} \pen = 0$.

The role of the mappings $u^n_k(m_k)$ is to embed the messages into the vector space $\Field^n$, so that it is possible to take linear combinations. The restriction to bijective mappings ensures that it is possible to solve the linear combinations and recover the original messages (subject to appropriate rank conditions).

The goal is for the receiver to recover the linear combinations
\begin{align}
	w_{\ab_\ell}^n(m_1,\ldots, m_K)
	= \sum_{k=1}^K a_{\ell,k} u_k^n(m_k), \qquad \ell\in[1:K]. \label{eq:linear-combination}
\end{align} 
where $a_{\ell,k}$ is the $(\ell,k)^{\text{th}}$ entry of $\Ab$ and the multiplication and summation operations are over $\Field$. The matrix $\Ab$ can be of any rank, for example, setting $\ab_2=\cdots=\ab_K=\mathbf{0}$ and $\ab_1 = \ab$ corresponds to the case where the receiver only wants a single linear combination $w^n_{\ab}(m_1, \ldots, m_K)$.

One natural example is to take the field as the reals, $\Field = \Real$, and the set of possible coefficients as the integers, $\mathbb{A} = \Integer$. This corresponds to the Gaussian compute--forward problem statement from~\cite{Nazer--Gastpar2011} where the receiver's goal is to recover integer-linear combinations of the real-valued codewords. Another example is to set $\mathbb{A} = \Field = \Fq$, i.e., linear combinations are taken over the finite field of order $\q$. This will be the starting point for our coding schemes.

\begin{remark} We could also attempt to define compute--forward formally for \textit{any} choice of deterministic functions of the messages. See~\cite{Niesen--Whiting2012} for an example. However, all known compute--forward schemes, have focused on the special case of linear functions. Moreover, certain applications, such as interference alignment, take explicit advantage of the connection to linear algebra. Therefore, we find it more intuitive to directly frame the problem in terms of linear combinations.
\end{remark}

%%%%%%%%%%%%%%%%%%%%%%%%%%%%%%%%%%%%%%%%%%%%%%%%%%%%%
%
%				MAIN RESULTS
%
%%%%%%%%%%%%%%%%%%%%%%%%%%%%%%%%%%%%%%%%%%%%%%%%%%%%%%

\section{Main Results}

We now state our achievability theorems and work out several examples. For the sake of clarity and simplicity, we begin with the special case of $K = 2$ transmitters and a receiver that only wants a single linear combination. Theorem~\ref{thm:comp-Fq} describes an achievable rate region for finite-field linear combinations, Theorem~\ref{thm:comp-real-discrete} provides a rate region for recovering integer-linear combinations of integer-valued random variables, and Theorem~\ref{thm:comp-real} establishes a rate region for recovering integer-linear combinations of real-valued random variables. Afterwards, in Theorem~\ref{thm:K-2-comp}, we provide a rate region for recovering two finite-field linear combinations of $K$ codewords, and Theorem~\ref{thm:mac} argues that, if $K = 2$, this corresponds to a multiple-access strategy.

\subsection{Computing One Linear Combination Over a Two-User MAC}

\input{cfrateregion}
In this subsection, we consider the special case of a receiver that wants a single linear combination of $K = 2$ transmitters' codewords. Specifically, we set $\ab_2=\mathbf{0}$ and, for notational simplicity, denote $\ab_1$ by $\ab=[a_1,\, a_2]$.
 
In order to state our main result, we need to define two rate regions. See Figure~\ref{fig:cfrateregion} for an illustration. The first region can be interpreted as the rates available for directly recovering the linear combination $w^n_{\ab}(m_1, m_2)$ from the received sequence $Y^n$ via ``single-user'' decoding,
\begin{align}   \label{R_CF}
\Rc_{\mathsf{CF}}(\ab) :=\{(R_1, R_2): R_1 < I_{\mathsf{CF},1}(\ab),\, R_2 < I_{\mathsf{CF},2}(\ab)\},
\end{align}
where $I_{\mathsf{CF},1}(\ab)$ and $I_{\mathsf{CF},2}(\ab)$ will be specified in the following theorems.

The second rate region can be interpreted as the rates available for recovering both messages individually via multiple-access with a shared nested linear codebook:
\begin{subequations} \label{eq:lmac}
\begin{align}
\Rc_{\mathsf{LMAC}} &:= \Rc_{\mathsf{LMAC},1} \cup  \Rc_{\mathsf{LMAC},2} \\
\Rc_{\mathsf{LMAC},1} &:= \Big\{ (R_1,R_2):  R_1 < \max_{\bb \in \mathbb{A}^2 \setminus \{\mathbf{0}\}}\min\{I_{\mathsf{CF},1}(\bb),  I(X_1, X_2; Y)-I_{\mathsf{CF},2}(\bb)\}, \\
&\qquad \qquad \qquad ~R_2 < I(X_2; Y|X_1),\nn \\
&\qquad ~~~~~~R_1+R_2 < I(X_1, X_2; Y) \Big\}, \nn \\
\Rc_{\mathsf{LMAC},2} &:= \Big\{ (R_1,R_2):  R_1 < I(X_1; Y|X_2),\\
&\qquad \qquad \qquad ~R_2 < \max_{\bb \in \mathbb{A}^2 \setminus \{\mathbf{0}\}}\min\{I_{\mathsf{CF},2}(\bb),  I(X_1, X_2; Y)-I_{\mathsf{CF},1}(\bb)\}, \nn \\
&\qquad ~~~~~~R_1+R_2 < I(X_1, X_2; Y) \Big\}. \nn
\end{align} 
\end{subequations}
Notice that $\Rc_{\mathsf{LMAC}}$ does not correspond, in general, to the classical multiple-access rate region.

We are ready to state our main theorems. Note that all of our theorems apply to both discrete and continuous input and output alphabets $\Xc_k$ and $\Yc$, and are distinguished from another by the alphabet of the auxiliary random variables $U_k$.

The theorem below gives an achievable rate region for recovering a single linear combination over $\Fq$. 

\begin{theorem}[Finite-Field Compute--Forward]\label{thm:comp-Fq}
Set $(\Field,\mathbb{A}) = (\Fq,\Fq)$ and let $\ab \in \Fq^2$ be the desired coefficient vector. A rate pair $(R_1, R_2)$ is achievable if it is included in $\Rc_{\mathsf{CF}}\cup\Rc_{\mathsf{LMAC}}$ for some input pmf $p_{U_1}(u_1)p_{U_2}(u_2)$ and symbol mappings $x_1(u_1)$ and $x_2(u_2)$, where $\Uc_k\subseteq\Fq$, 
\begin{subequations}
\begin{align}
I_{\mathsf{CF},1}(\ab)&=H(U_1)-H(W_{\ab}|Y),\label{eq:I1} \\
I_{\mathsf{CF},2}(\ab)&=H(U_2)-H(W_{\ab}|Y), \label{eq:I2} 
\end{align}
\end{subequations}
and 
\begin{align}
W_{\ab}=a_1U_1 \oplus a_2 U_2, \label{eq:linear-combination-Fq}
\end{align}
where the addition and multiplication operations in \eqref{eq:linear-combination-Fq} are over $\Fq$.
\end{theorem} 
\begin{remark}
We have omitted the use of time-sharing random variables for the sake of simplicity. We note that the achievability results in this paper can be extended to include a time-sharing random variable following the standard coded time-sharing method~\cite[Sec. 4.5.3]{El-Gamal--Kim2011}.
\end{remark}

\begin{remark}
Prior work by Padakandla and Pradhan proposed a finite-field compute--forward scheme for communicating the sum of codewords over a two-user MAC~\cite{Padakandla--Pradhan2013}, resulting in the achievable rate region $\Rc_{\mathsf{PP}} = \{(R_1,R_2): R_k \leq \min(H(U_1),\ H(U_2)) - H(U_1 \oplus U_2 |Y),~k=1,2\}$. Note that this region is included in $\Rc_{\mathsf{CF}}([1~1])$ from Theorem~\ref{thm:comp-Fq}, and corresponds to the special case where the rates are set to be equal $R_1 = R_2$. 
\end{remark}

We prove Theorem~\ref{thm:comp-real-discrete} in two steps in Section~\ref{sec:comp-finite}. First, we develop an achievable scheme for a \textit{DM-MAC}, which will serve as a foundation for the remainder of our achievability arguments. Afterwards, we use a quantization argument to extend this scheme to real-valued receiver alphabets.

\begin{example}
Consider the binary multiplying MAC with channel output $Y =  X_1 \cdot X_2$ and binary sender and receiver alphabets, $\Xc_1 = \Xc_2 = \Yc = \{0,1\}$. The receiver would like to recover the sum $W = U_1 \oplus U_2$ over the binary field  $\q = 2$ where $U_k \sim \Bern(p_k)$ and $x_k(u_k) = u_k$, $k = 1,2$. The highest symmetric rate $R_1 = R_2 = R_{\text{sym}}$ achievable via Theorem~\ref{thm:comp-Fq} is $R_{\text{sym}} = 0.6656$, which is attained with $p_1 = p_2 = 0.7331$. Note that, if we send both $U_1$ and $U_2$ to the receiver via classical multiple-access, the highest symmetric rate possible is $R_{\text{sym}} = 0.5$.
\end{example}

In many settings, it will be useful to recover a real-valued sum of the codewords, rather than the finite-field sum. Below, we provide two theorems for recovering integer-linear combinations of codewords over the real field. The first restricts the $U_k$ random variables to (bounded) integer values, which in turn allows us to express the rate region in terms of discrete entropies. The second allows the $U_k$ to be continuous--valued random variables (subject to mild technical constraints), and the rate region is written in terms of differential entropies.

\begin{theorem}\label{thm:comp-real-discrete}
%[Integer Compute--Forward]
Set $(\Field,\mathbb{A}) = (\Real,\Integer)$ and let $\ab \in \Integer^2$ be the desired coefficient vector. Assume that $\Uc_k \subset \Integer$ and $| \Uc_k | < \infty$. A rate pair $(R_1, R_2)$ is achievable if it is included in $\Rc_{\mathsf{CF}}\cup\Rc_{\mathsf{LMAC}}$ for some input pmf $p_{U_1}(u_1)p_{U_2}(u_2)$ and symbol mappings $x_1(u_1)$ and $x_2(u_2)$, where
\begin{subequations}
\begin{align*}
I_{\mathsf{CF},1}(\ab)&=H(U_1)-H(W_{\ab}|Y),\\
I_{\mathsf{CF},2}(\ab)&=H(U_2)-H(W_{\ab}|Y),
\end{align*}
\end{subequations}
and 
\begin{align}
W_{\ab}=a_1U_1 + a_2 U_2, \label{eq:linear-combination-Zq}
\end{align}
where the addition and multiplication in \eqref{eq:linear-combination-Zq} are over $\Real$.
\end{theorem} The proof of Theorem~\ref{thm:comp-real-discrete} is given in Section~\ref{sec:comp-finite}.  Notice that, while the $U_k$ are restricted to integer values, the $x_k(u_k)$ are free to map to any real values.

\begin{definition}[Weak continuity of random variables]
Consider a family of cdfs $\{ F_{\tv} \}$ that are parametrized by $\tv \in \mathbb{R}^K$ and denote random variables $X_{\tv} \sim F_{\tv}$. The family $\{ F_{\tv} \}$ is said to be {\em weakly continuous} at $\tv_0$ if $X_{\tv}$ converges in distribution to $X_{\tv_0}$ as $\tv \to \tv_0$.
\end{definition}

\begin{theorem}[Continuous Compute--Forward]\label{thm:comp-real}
Set $(\Field,\mathbb{A}) = (\Real,\Integer)$ and let $\ab \in \Integer^2$ be the desired coefficient vector. Let $U_1$ and $U_2$ be two independent real-valued random variables with absolutely continuous distributions described by pdfs $f_{U_1}$ and $f_{U_2}$, respectively. Also, assume that the family of cdfs $\{ F_{Y|\Uv}(\cdot|\uv) \}$ is weakly continuous in $\uv$ almost everywhere. Finally, assume that the following finiteness conditions on entropies and differential entropies hold:
\begin{enumerate}
	\item	$h(U_1) < \infty$ and $h(U_2) < \infty$
	\item	$H( \lceil U_1 \rfloor ) < \infty$ and $H(\lceil U_2 \rfloor) < \infty$
\end{enumerate} 
where $\lceil u \rfloor$ rounds $u$ to the nearest integer. A rate pair $(R_1, R_2)$ is achievable if it is included in $\Rc_{\mathsf{CF}}(\ab) \cup\Rc_{\mathsf{LMAC}}$ for some input pdf $f_{U_1}(u_1)f_{U_2}(u_2)$ and symbol mappings $x_1(u_1), x_2(u_2)$, where
\begin{subequations}
\begin{IEEEeqnarray}{rCl}
	I_{\mathsf{CF},1}(\ab,\beta)
	&:=& h(U_1) - h(W_{\ab}|Y) + \log\gcd(\ab) \\
	I_{\mathsf{CF},2}(\ab,\beta)
	&:=& h(U_2) - h(W_{\ab}|Y) + \log\gcd(\ab),
\end{IEEEeqnarray} 
\end{subequations} and \begin{align}
W_{\ab}=a_1 U_1 + a_2 U_2, \label{eq:linear-combination-real}
\end{align}
where the addition and multiplication in \eqref{eq:linear-combination-real} are over $\Real$ and $\gcd(\ab)$ denotes the greatest common divisor of $|a_1|$ and $|a_2|$.
\end{theorem}
The proof of this theorem is deferred to Section~\ref{sec:comp-real}.

\begin{remark}
The $\log\gcd(\ab)$ term neutralizes the penalty for choosing a coefficient vector $\ab$ with $\gcd(\ab) > 1$. For example, set $\ab = [ 1~1]$ and $\boldsymbol{\tilde{a}} = [2 ~ 2]$ and note that $\gcd(\ab) = 1$ and $\gcd(\boldsymbol{\tilde{a}}) = 2$. Since $h(W_{\boldsymbol{\tilde{a}}} | Y) = h(W_{\ab} | Y) + \log(2)$, we find that the $\log\gcd(\boldsymbol{\tilde{a}})$ term compensates exactly for the penalty in the conditional entropy. Previous work on compute--forward either ignored the possibility of a penalty~\cite{Nazer--Gastpar2011} or compensated by taking an explicit union over all integer coefficient matrices with the same row span~\cite{Nazer--Cadambe--Ntranos--Caire2015}. 
\end{remark}

Consider the Gaussian MAC
\begin{align}\label{eq:gaussian-mac}
	Y=h_{1}X_1+h_{2}X_2+Z,
\end{align}
with channel gains $h_k \in \Real$, average power constraints $\sum_{i=1}^n x^2_{k,i}(m_k) \leq nP_k$, $k = 1,2$,  and zero-mean additive Gaussian noise with unit variance. Specializing Theorem~\ref{thm:comp-real} by setting $f_{U_k}$ to be $\Nc(0, \frac{P_k}{\beta_k^2})$ and $x_k(u_k) = \beta_k \, u_k$ for some $\beta_k\in \Real$, we establish the following corollary, which includes the Gaussian compute--forward rate regions in~\cite{Nazer--Gastpar2011, Zhu--Gastpar2014b, Nazer--Cadambe--Ntranos--Caire2015}.

\begin{corollary}[Gaussian Compute--Forward] \label{cor:comp-gaussian}
Consider a Gaussian MAC and set $(\Field,\mathbb{A}) = (\Real,\Integer)$ and let $\ab \in \Integer^2$ be the desired coefficient vector. A rate pair $(R_1, R_2)$ is achievable if it is included in $\Rc_{\mathsf{CF}}(\ab) \cup\Rc_{\mathsf{LMAC}}$ for some $\beta_k\in\Real$, $k=1,2,$ where
\begin{align*}
I_{\mathsf{CF},1}(\ab,\beta_1)&:=\frac{1}{2}\log\left(\frac{\beta_1^2(1+h_1^2P+h_2^2P)}{(a_1\beta_1h_2-a_2\beta_2 h_1)^2P+(a_1\beta_1)^2+(a_2\beta_2)^2}\right)+\log\gcd(\ab),\\
	I_{\mathsf{CF},2}(\ab,\beta_2)&:=\frac{1}{2}\log\left(\frac{\beta_2^2(1+h_1^2P+h_2^2P)}{(a_1\beta_1h_2-a_2\beta_2h_1)^2P+(a_1\beta_1)^2+(a_2\beta_2)^2}\right)+\log\gcd(\ab),\\
	I(X_1,X_2; Y) &=  \C(h_1^2 P_1+h_2^2 P_2), \\
	I(X_1 ; Y | X_2) &= \C(h_1^2 P_1), \\
	I(X_2 ; Y | X_1) &= \C(h_2^2 P_2),
\end{align*}
and $\C(x):=\frac{1}{2}\log(1+x)$.
\end{corollary}

\begin{figure}
\begin{center}
\input{gaussianmacplot_sumrate}
\caption{Performance comparison for sending the sum of codewords over a symmetric Gaussian MAC $Y= X_1 + X_2 + Z$.}
\label{fig:gaussianmacplot}
\end{center}
\end{figure}

\begin{figure}
\begin{center}
\input{magnified}
\caption{Example showing that the compute--forward scheme in Theorem~\ref{thm:comp-real-discrete} with $|\Uc_1|=|\Uc_2|=3$ can outperform both compute--forward with Gaussian inputs and i.i.d.~Gaussian coding.}
\label{fig:example2b}
\end{center}
\end{figure}

\begin{example} \label{ex:example2}
We now apply each of the theorems above to the problem of sending the sum of two codewords over a symmetric Gaussian MAC with channel output $Y = X_1 + X_2 + Z$ where $Z \sim \Nc(0,1)$ is independent, additive Gaussian noise and we have the usual power constraints $\sum_{i=1}^n x^2_{k,i}(m_k)  \leq nP$, $k=1,2$. Specifically, we would like to send the linear combination with coefficient vector $\ab = [1~1]$ at the highest possible sum rate $R_{\text{sum}}=R_1 + R_2$. In Figure~\ref{fig:gaussianmacplot}, we have plotted the sum rate for several strategies with respect to $\mathsf{SNR} = 10 \log_{10}(P)$. 

The upper bound $R_{\text{sum}} \leq \log(1 + P)$ follows from a simple cut-set bound.
Corollary~\ref{cor:comp-gaussian} with $\beta_1 = \beta_2 = 1$ yields the sum rate $R_{\text{sum}} = \max(\log(\frac{1}{2} + P),\ \frac{1}{2} \log(1 + 2P))$. Note that this is the best-known\footnote{The performance can be slightly improved if the transmitters remain silent part of the time, and increase their power during the remainder of the time. Specifically, this approach would achieve $R_1=R_2 = \max(\sup_{\alpha \in [0,1)} \frac{\alpha}{2}\log(\frac{1}{2} + \frac{P}{1 - \alpha}),\ \frac{1}{4} \log(1 + 2P))$. Note that this requires the use of a time--sharing auxiliary random variable.} performance for the Gaussian two-way relay channel~\cite{Wilson--Narayanan--Pfister--Sprintson2010,Nam--Chung--Lee2010,Nazer--Gastpar2011}. The best-known performance for i.i.d.~Gaussian codebooks is $R_{\text{sum}} = \frac{1}{2}\log(1 + 2P)$. 

We have also plotted two examples of Theorem~\ref{thm:comp-Fq} with $\q =2$ and $\q = 4$. For the binary field $\q = 2$, we take $U_k \sim \mathrm{Unif}(\Field_2)$, $x_k(0) = -\sqrt{P}$, and $x_k = \sqrt{P}$, $k=1,2$. For $\q = 4$, we take $U_k \sim \mathrm{Unif}(\Field_4)$, $x_k(0) = -3\sqrt{\frac{P}{5}}$, $x_k(1) = -\sqrt{\frac{P}{5}}$, $x_k(2) = \sqrt{\frac{P}{5}}$, and $x_k(3) = 3\sqrt{\frac{P}{5}}$, $k = 1,2$.

Finally, we have plotted an example of Theorem~\ref{thm:comp-real-discrete} with $\Uc_k = \{-3,\,-1,\,1,\,3\}$, $U_k \sim \mathrm{Unif}(\Uc_k)$, and $x_k(u_k) = \sqrt{\frac{P}{5}}\, u_k$, $k = 1,2$. Note that this outperforms the $\q = 4$ strategy in Theorem~\ref{thm:comp-Fq}, which effectively uses the same input distributions. If we were to set $\Uc_k = \{-1,\ 1\}$, $U_k \sim \mathrm{Unif}(\Uc_k)$, and $x_k(u_k) = \sqrt{P}\, u_k$, $k=1,2$, we would match the achievable rate of Theorem~\ref{thm:comp-Fq} with $\q=2$ exactly (not shown on the plot).
\end{example}

\begin{example}
Consider the Gaussian MAC channel in Example~\ref{ex:example2}. In Figure~\ref{fig:example2b}, we have plotted an example of Theorem~\ref{thm:comp-real-discrete} with $\Uc_k = \{-1,0,1\}$, pmfs $p_{U_k}=\{\frac{1-p_k}{2}, p_k, \frac{1-p_k}{2}\}$, and $X_k=U_k\sqrt{\frac{P}{1-p_k}}$, which we optimize over $p_k\in[0,1)$. For SNR near $1.8$ dB, we can see that the strategy in Theorem~\ref{thm:comp-real-discrete} strictly outperforms both the Gaussian-input compute--forward (and thus the lattice-based compute--forward in~\cite{Nazer--Gastpar2011}) and i.i.d.~Gaussian coding. The suboptimality of Gaussian inputs for compute--forward was first observed by Zhu and Gastpar~\cite{Zhu--Gastpar2015}.
\end{example}

%%%%%%%%%%%%%%%%%%%%%%%%%%%%%%%%%%%%%%%%%%%%%%%%%%%%%%%%%%%%%%%%%%%%%%

\subsection{Computing Two Linear Combinations Over a $K$-User MAC}

In this subsection, we extend the results of the previous section to compute two linear combinations over a $K$-user MAC. The problem of recovering multiple linear combinations at a single receiver was previously studied in~\cite{Feng--Silva--Kschischang2013,Zhan--Nazer--Erez--Gastpar2014,Ordentlich--Erez--Nazer2014,Ordentlich--Erez--Nazer2013,Nazer--Cadambe--Ntranos--Caire2015, Zhu--Gastpar2015}. Applications include lattice interference alignment~\cite{Ordentlich--Erez--Nazer2014}, multiple-access~\cite{Ordentlich--Erez--Nazer2014,Ordentlich--Erez--Nazer2013,Nazer--Cadambe--Ntranos--Caire2015, Zhu--Gastpar2015}, and low--complexity MIMO receiver architectures~\cite{Zhan--Nazer--Erez--Gastpar2014,Ordentlich--Erez--Nazer2013}. Prior to this paper, the largest available rate region relied on successive cancellation decoding~\cite{Ordentlich--Erez--Nazer2013,Nazer--Cadambe--Ntranos--Caire2015} and was limited to the Gaussian setting. Here, we derive an achievable rate region for the discrete memoryless setting using \textit{simultaneous joint typicality decoding}. 

There are $K$ transmitters and a single receiver that wants to recover two linear combinations with coefficient vectors $\ab_1, \ab_2 \in \mathbb{A}^K$. Without loss of generality, we assume that $\ab_1$ and $\ab_2$ are linearly independent.  (Otherwise, we can use the results for recovering a single linear combination described above.) 

\begin{theorem}[Two Linear Combinations]\label{thm:K-2-comp}
Let $(\Field,\mathbb{A}) = (\Fq,\Fq)$ and $\ab_1, \ab_2 \in \Fq^K$ be the desired coefficient vectors. Assume that $\ab_1$ and $\ab_2$ are linearly independent and define $\Kc_\ell=\{k\in[1:K]: a_{\ell k}\neq 0\}$, $\ell=1,2$ as well as 
\begin{align}
W_{\ab_1}&=\sum_{k=1}^K a_{1k}U_k, \label{eq:thm4eq1}\\
W_{\ab_2}&=\sum_{k=1}^K a_{2k}U_k,\label{eq:thm4eq2}\\ 
V_{\bb}&=b_1W_{\ab_1} + b_2W_{\ab_2}, \label{eq:thm4eq3}
\end{align}
where $\bb \in \Fq^2 \setminus \{ \mathbf{0}\}$ and the multiplications and summations are over $\Fq$. A rate tuple $(R_1, \ldots, R_K)$ is achievable if
\begin{align*}
R_{k} &< \max_{\bb\in\Fq^2\setminus\{\mathbf{0}\}}\min\{H(U_k)-H(V_{\bb}|Y), H(U_k)-H(W_{\ab_1}, W_{\ab_2}|Y, V_{\bb})\}, ~~k\in\Kc_1\\
R_{j} &< I(W_{\ab_2}; Y,W_{\ab_1})-H(W_{\ab_2})+H(U_j), ~~j\in \Kc_2,\\
R_{k}+R_{j} &< I(W_{\ab_1}, W_{\ab_2}; Y)-H(W_{\ab_1}, W_{\ab_2})+H(U_k)+H(U_j), ~~k\in\Kc_1, j\in\Kc_2
\end{align*}
or 
\begin{align*}
R_{k} &< I(W_{\ab_1}; Y,W_{\ab_2})-H(W_{\ab_1})+H(U_k), ~~k\in \Kc_1,\\
R_{j} &<  \max_{\bb\in\Fq^2\setminus\{\mathbf{0}\}}\min\{H(U_j)-H(V_{\bb}|Y), H(U_j)-H(W_{\ab_1}, W_{\ab_2}|Y, V_{\bb})\}, ~~j\in\Kc_2,\\
R_{k}+R_{j} &< I(W_{\ab_1}, W_{\ab_2}; Y)-H(W_{\ab_1}, W_{\ab_2})+H(U_k)+H(U_j), ~~k\in\Kc_1, j\in\Kc_2
\end{align*}
for some input pmf $\prod_{k=1}^K p_{U_k}(u_k)$, symbol mappings $x_k(u_k)$, $k\in[1:K]$, where $\Uc_k\subseteq \Fq$. 
\end{theorem}

\begin{remark}\label{rmk:extension}
Theorem~\ref{thm:K-2-comp} can be easily extended to the case $(\Field, \mathbb{A})=(\Real, \Integer)$ with $\Uc_k\subset \Integer$, $|\Uc_k|<\infty$ (similar to Theorem~\ref{thm:comp-real-discrete}). For this case, we would replace $(\Field, \mathbb{A})=(\Fq, \Fq)$ with $(\Field, \mathbb{A})=(\Real, \Integer)$, set $\Uc_k\subset \Integer$, $|\Uc_k|<\infty$, and take the summations in $\eqref{eq:thm4eq1}$ to $\eqref{eq:thm4eq3}$ are over $\Real$.
\end{remark}
We defer to Section~\ref{sec:k-2-comp-proof} for a detailed description of the decoder, the proof of Theorem~\ref{thm:K-2-comp}, and the proof of Remark~\ref{rmk:extension}.

\begin{remark} \label{rem:tupledecoding}
The rate region from Theorems~\ref{thm:comp-Fq} and~\ref{thm:comp-real-discrete} demonstrate that, even if we are interested in recovering a single linear combination, a joint typicality decoder will sometimes implicitly recover both messages. (This occurs for rates that fall in $\Rc_{\mathsf{LMAC}}$.) It seems likely that, for recovering two linear combinations with coefficient vectors $\ab_1$ and $\ab_2$, a complete analysis of a joint typicality decoder should also include the rate regions for decoding linear combinations with all coefficient matrices $\Ab$ of rank $2$ or greater whose rowspan includes $\ab_1$ and $\ab_2$. This is not the case for Theorem~\ref{thm:K-2-comp}, due to the fact that our error analysis can only handle pairs of indices. The analysis of the simultaneous joint typicality decoder for more than two indices is left as an open problem.
\end{remark}

We now consider the special case of $K=2$ users and a coefficient matrix $\Ab$ with rank $2$, which, by the bijective mapping assumption on $U^n_k(M_k)$, is equivalent to recovering both messages $(M_1, M_2)$. 

\begin{theorem}[Multiple-Access via Compute--Forward]\label{thm:mac}
Consider the sequences of code pairs that achieves the rate region in Theorems~\ref{thm:comp-Fq},~\ref{thm:comp-real-discrete}, and~\ref{thm:comp-real}  for some input distribution $p_{U_1}(u_1)p_{U_2}(u_2)$ and symbol mappings $x_1(u_1)$ and $x_2(u_2)$. Then, the rate pair $(R_1, R_2)$ is also achievable for recovering the individual messages with the {\em same} sequence of codes, if it is included in $\Rc_{\mathsf{LMAC}}$.
\end{theorem}
The proof is deferred to Section~\ref{sec:mac-proof}.

The following corollary is a Gaussian specialization of Theorem~\ref{thm:mac}.
\begin{corollary}[Gaussian Multiple-Access via Compute--Forward]\label{cor:mac-gaussian}
Consider the sequences of code pairs that achieves the rate region in Corollary~\ref{cor:comp-gaussian} for some Gaussian MAC. 
Then, the rate pair $(R_1, R_2)$ is also achievable for recovering the messages with the same sequence of codes if it is included in $\Rc_{\mathsf{LMAC}}$ for some $\beta_k\in\Real$.
\end{corollary}

The following example considers a compound MAC where one receiver only wants the sum of the codewords. It demonstrates that simultaneous joint typicality decoding can outperform successive cancellation decoding for compute--forward, even after time-sharing. It also shows that our strategy outperforms the best known random i.i.d.~coding scheme.

\begin{figure}[h!]
\begin{center}
\footnotesize
\psfrag{m1}[c]{$M_1$}
\psfrag{m2}[c]{$M_2$}
\psfrag{x1}[c]{$X^n_1$}
\psfrag{x2}[c]{$X^n_2$}
\psfrag{y1}[c]{$Y^n_1$}
\psfrag{y2}[c]{$Y^n_2$}
\psfrag{z1}[c]{$Z^n_1$}
\psfrag{z2}[c]{$Z^n_2$}
\psfrag{g11}[c]{}
\psfrag{g21}[c]{}
\psfrag{g12}[c]{$h$}
\psfrag{g22}[c]{}
\psfrag{e1}[c]{Encoder 1}
\psfrag{e2}[c]{Encoder 2}
\psfrag{d1}[c]{Decoder 1}
\psfrag{d2}[c]{Decoder 2}
\psfrag{mh1}[c]{$~(\Mh_1, \Mh_2)$}
\psfrag{mh2}[c]{$~~~\Wh_{\ab}^n$}
\includegraphics[width=0.6\textwidth]{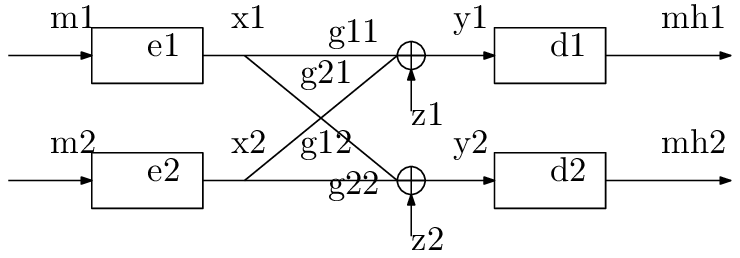}
\end{center}
\caption{A two-sender two-receiver network. Decoder 1 wishes to recover both messages and Decoder 2 wishes to compute the sum of the channel inputs, $W_{\ab}^n = X_1^n(M_1) + X_2^n(M_2)$.}\label{fig:22net}
\end{figure}

\input{gaussianregion}

\begin{example}\label{ex:binary2mac2}
Consider the two-sender, two-receiver Gaussian network depicted in Figure~\ref{fig:22net}.
The channel outputs are given by
\begin{align*}
Y_1&= X_1+ hX_2+Z_1\\
Y_2&= X_1+ X_2+Z_2,
\end{align*}
where $Z_1$ and $Z_2$ are independent Gaussian noise components with zero mean and unit variance, $h=\sqrt{2}$, and $P_1=25$ and $P_2=18$ where $P_1$ and $P_2$ are the power constraints on $X_1$ and $X_2$, respectively. Here, we assume that Receiver 1 wishes to recover both messages separately while Receiver 2 wishes to recover the sum of the codewords, 
\begin{align}
W_{\ab}^n = X^n_1(M_1)+X_2^n(M_2), \label{eq:cwsum}
\end{align}
where $\ab = [ 1 ~ 1]$.

To explicitly compute the linear combinations of the transmitted codewords~\eqref{eq:cwsum}, we fix $x_1(u_1)$ and $x_2(u_2)$ to be identity mappings in Corollary~\ref{cor:mac-gaussian}. By Corollary~\ref{cor:mac-gaussian}, decoding is possible at Receiver $1$ if the rates are included in $\Rc_{\mathsf{LMAC}}$ (with $\beta_1 = \beta_2 = 1$) for the induced MAC. By Corollary~\ref{cor:comp-gaussian}, decoding is possible at Receiver $2$ if the rates are included in $\Rc_{\mathsf{CF}}([1~1]) \cup \Rc_{\mathsf{LMAC}}$ (with $\beta_1 = \beta_2 = 1$) for the induced MAC. In Figure~\ref{fig:gaussianregion}, we have plotted these rate constraints, followed by their intersection, and the convexification of this region allowed by time--sharing. We have also plotted the performance available to nested lattice codes combined with successive cancellation decoding as derived in~\cite[Theorem 7]{Nazer--Cadambe--Ntranos--Caire2015}. Finally, we have plotted the performance of random i.i.d.~codes coupled with simultaneous joint typicality decoding, which corresponds to the rates available for a compound Gaussian MAC. While our strategy strictly outperforms the other two strategies in this scenario, it is not known to be optimal in general.
\end{example}

In the following two sections, we introduce the {\em nested linear coding architecture} which will form the foundation of our achievability strategies.

\section{Point-to-Point Channels Revisited} \label{sec:ptop}
\begin{figure*}[t]
\footnotesize
\begin{center}
\psfrag{m}[c]{$M$}
\psfrag{v1}[c]{$\nub(M)$}
\psfrag{u1}[c]{$U^n$}
\psfrag{x1}[c]{$X^n$}
\psfrag{y}[c]{$Y^n$}
\psfrag{yx}[c]{$p(y|x)$}
\psfrag{e1}[c]{Encoder}
\psfrag{d}[c]{Decoder}
\psfrag{e2}[c]{Decoder}
\psfrag{mh}[c]{$\Mh$}
\includegraphics[width=0.8\textwidth]{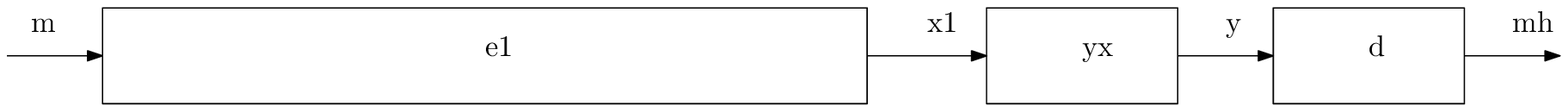}
\caption{A point-to-point communication system.}\label{fig:p2p}
\end{center}
\end{figure*}

To better explain the intuition and structure of our coding strategies, we will first revisit and explain the {\em nested linear code architecture} for point-to-point communication. Consider the point-to-point communication system depicted in Figure~\ref{fig:p2p}, where a sender wishes to reliably communicate a message $M$ at a rate $R$ bits per transmission to a receiver over the discrete memoryless channel (DMC) $p(y|x)$.

\begin{figure*}[t]
\footnotesize
\begin{center}
\psfrag{m}[c]{$M$}
\psfrag{v1}[c]{}
\psfrag{u1}[c]{$U^n$}
\psfrag{x1}[c]{$X^n$}
\psfrag{y}[c]{$Y^n$}
\psfrag{yx}[c]{$p(y|x)$}
\psfrag{e1}[c]{Encoder}
\psfrag{le1}[c]{Linear}
\psfrag{le2}[c]{code}
\psfrag{jte}[c]{Multicoding}
\psfrag{ex}[c]{$x(u)$}
\psfrag{d}[c]{Decoder}
\psfrag{c}[c]{Channel}
\psfrag{e2}[c]{Decoder}
\psfrag{mh}[c]{$\Mh$}
\includegraphics[width=0.8\textwidth]{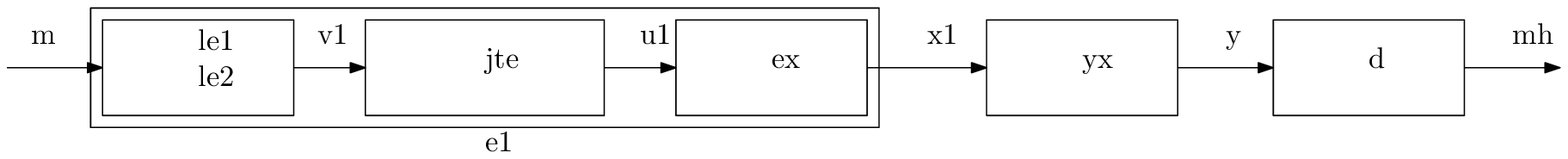}
\end{center}
\caption{A joint typicality encoding architecture for point-to-point communication based on nested linear codes.}\label{fig:arc2}
\end{figure*}

The celebrated channel coding theorem of Shannon~\cite{Shannon1948} states that 
the capacity $C$ of the discrete memoryless channel $p(y|x)$ is given by the capacity formula
\begin{equation} \label{eq:shannon}
C=\max_{p(x)}I(X; Y).
\end{equation}
The classic achievability proof for the channel coding theorem relies on a random coding argument. Specifically, the codeword symbols are randomly and independently generated from the capacity achieving distribution $p(x)$ and the receiver employs joint typicality decoding.

As an alternative strategy, consider the linear coding architecture in Figure~\ref{fig:arc2}. 
This architecture is based on three components, an auxiliary linear code, a joint typicality encoder for multicoding, and a symbol-by-symbol mapping function $x(u)$. Multicoding is often used in the context of Gelfand-Pinsker (i.e., dirty-paper) coding~\cite{Gelfand--Pinsker1980a} to find codewords that are jointly typical with respect to the observed state sequence. In constrast, the proposed architecture uses multicoding to select linear codewords that are typical with respect to the desired input distribution (as opposed to the uniform distribution). This linear coding architecture was studied by Miyake~\cite{Miyake2010} in the context of sparse codes for point-to-point channels and by Padakandla and Pradhan for three-user broadcast channels~\cite{Padakandla--Pradhan2012}, recovering the sum of discrete memoryless sources over a discrete memoryless MAC with distributed state information~\cite{Padakandla--Pradhan2013}, and three-user interference channels~\cite{Padakandla--Sahebi--Pradhan2016}. Below, we provide an overview of the codebook construction, encoding and decoding operations, and error analysis for this linear coding architecture in the context of a memoryless point-to-point channel. This will help build useful intuition for our main theorems.
  
\smallskip

\noindent{\bf Codebook generation.} 
Fix a finite field $\Fq$ and a parameter $\epsilon' \in (0,1)$. In addition to the messages $m\in[2^{nR}]$, we use auxiliary indices $l\in[2^{n\Rh}]$, with rates $R$ and $\Rh$, respectively. Randomly generate a $\kappa\times n$ matrix, $\Gb\in\Fq^{\kappa\times n}$, and a vector $d^n\in\Fq^n$ where each element of $\Gb$ and $d^n$ are independently and randomly generated according to $\U(\Fq)$, and  $\kappa=\ceil{nR/\log(\q)}+\ceil{n\Rh/\log(\q)}$.

Generate a linear code $\Cc$ with parameters $(R, \Rh, n, \q)$ by
\begin{align}
u^n(m,l) &= [\nub(m), \nub(l)] \Gb \oplus d^n, 
\end{align}
for $m\in[2^{nR}]$, $l\in[2^{n\Rh}]$, where $\nub(m)$ is the $\q$-ary expansion of the index $m\in[2^{nR}]$ with length $\tilde{\kappa}=\ceil{nR/\log(\q)]}$ and $\nub(l)$ is the $\q$-ary expansion of the index $l\in[2^{n\Rh}]$ with length $\ceil{n\Rh/\log(\q)}$, and
\begin{align*}
\Gb =
\left[
\begin{array}{cccc}
g_{11} & g_{12} & \cdots & g_{1n}\\
g_{21} & g_{22} & \cdots & g_{2n}\\
\vdots & \vdots & \ddots & \vdots\\
g_{\tilde{\kappa},1} & g_{\tilde{\kappa},2} & \cdots & g_{\tilde{\kappa},n}\\
\hline
g_{\tilde{\kappa}+1,1} &g_{\tilde{\kappa}+1,2} &\cdots& g_{\tilde{\kappa}+1, n} \\
g_{\tilde{\kappa}+2,1} & g_{\tilde{\kappa}+2,2} &\cdots& g_{\tilde{\kappa}+2, n} \\
\vdots &\vdots&\ddots & \vdots \\
g_{\kappa,1} &g_{\kappa,2} &\cdots& g_{\kappa,n} 
\end{array}
\right] .
\end{align*}
\smallskip
Note that from this construction, the codewords are pairwise independent
\begin{align}
\P\{U^n(m, l)=u^n, U^n(\mt, \lt)=\ut^n\}=\prod_{i=1}^n p_\q(u_i)p_\q(\ut_i),\quad (m, l)\neq (\mt, \lt), \label{eq:code-dist}
\end{align}
where $p_\q=\U(\Fq)$. The general joint distribution of the codewords resulting from this construction can be found in~\cite[Theorem~1]{Domb--Zamir--Feder2013}.

\noindent{\bf Encoding.} Fix a pmf $p(u)$ and a function $x: \Fq \rightarrow \mathcal{X}$.
For each $m\in[2^{nR}]$, find an index $l\in[2^{n\Rh}]$ such that $ u^n(m,l)\in\aepvar(U)$. If there is more than one, choose one randomly from such indices. 
If there is none, randomly choose an index from $[2^{n\Rh}]$. 

To send message $m\in[2^{nR}]$, transmit $x_i(u_i(m,l))$ for $i=1,\ldots,n$, where $l$ is the chosen index from the above encoding step.
\smallskip
From Lemma~\ref{lem:mm-covering} in Appendix~\ref{app:mm-packing-covering}, the probability of encoding error tends to zero as $n\to\infty$ if
\begin{align}\label{eq:p2p-enc-cond}
\Rh &> D(p_{U}\|p_\q)+\d(\e'). 
\end{align}

\noindent{\bf Decoding.} Select a parameter $\epsilon > \e'$. Upon observing $y^n$, the receiver searches for a unique message $m\in[2^{nR}]$ such that
\begin{align*}
 (u^n(m,l), y^n)\in\aep,
\end{align*}
for some $l\in[2^{n\Rh}]$.
If there is none or more than one such message, it declares an error.
From Lemma~\ref{lem:mm-packing} in Appendix~\ref{app:mm-packing-covering}, the probability of encoding error tends to zero as $n\to\infty$ if
\begin{align}\label{eq:p2p-dec-cond}
R+\Rh &< I(U; Y)+D(p_{U}\|p_\q)-\d(\e).
\end{align}

By eliminating $\Rh$ from~\eqref{eq:p2p-enc-cond} and~\eqref{eq:p2p-dec-cond}, and sending $\e\to 0$, any rate $R$ that satisfies
\begin{align*}
R < \max_{p(u), x(u)} I(U; Y)
\end{align*}
is achievable. Finally, for $\q\ge |\Xc|$, we can simply select an injective function $x: \Fq \rightarrow \mathcal{X}$ and a pmf $p(u)$ so that $X = x(U)$ has the capacity-achieving input distribution. Thus, we can achieve the point-to-point capacity \eqref{eq:shannon} using nested linear codes. 

As mentioned earlier, the above argument can be viewed as a special case of~\cite[Theorem 5.1]{Miyake2010} or~\cite[Theorem 1]{Padakandla--Pradhan2013}. In the following sections, we generalize this technique and use it to develop a discrete memoryless version of compute--forward.

\section{Compute--Forward with Multicoding} \label{sec:highlevel}

Consider a relay in a Gaussian network that observes a noisy linear combination of several codewords. Classical relaying strategies for this scenario can be viewed as variations on three fundamental strategies: decode--forward~\cite[Th.~1]{Cover--El-Gamal1979}, compress--forward~\cite[Th.~6]{Cover--El-Gamal1979}, and amplify--forward~\cite{Schein--Gallager2000}. Recent work~\cite{Nazer--Gastpar2011} has introduced a novel strategy, compute--forward, which enables a relay to decode a linear combination of \textit{$\q$-ary expansions of the messages}. Recall that, in our problem formulation from Section~\ref{sec:problemstatement}, the messages $m_k$ are mapped to representative sequences $u_k^n(m_k) \in \Field^n$, and the goal of the decoder is to recover linear combinations of the $u_k^n(m_k)$. Below, we provide intuition for why this generalization is useful to move beyond the Gaussian, equal power setting. Afterwards, we provide a formal description of our codebook generation and encoding procedure.

\subsection{High-Level Overview}

To begin, consider a scenario with $K$ transmitters and a single receiver that operate with blocklength $n$. The $k^{\text{th}}$ transmitter has a message $m_k \in [2^{nR_k}]$ where $R_k \geq 0$ denotes its rate. An appealing approach is to view the messages as vectors in a vector space over the finite field $\Fq$. Specifically, let $\nub(m_k)$ denote the $\q$-ary expansion of $m_k$ into a vector of length $nR_k / \log(\q)$.\footnote{For the remainder of the paper, we will assume that $n R_k / \log(\q)$ is integer-valued in order to simplify our notation.} For the special case of symmetric rates, $R_1 = \cdots = R_K$, we can define the class of desired linear functions as those of the form 
\begin{align*}
\bigoplus_{k=1}^K a_k \nub(m_k) %\label{eqn:oldlinearcomb}
\end{align*} for some $a_k \in \Fq$. This is the approach taken in~\cite{Nazer--Gastpar2011} for transmitters with equal power constraints.

Unfortunately, it seems that this framework is not rich enough to handle the setting where each transmitter has a different input distribution. Specifically,  this is due to the use of multicoding to select linear codewords with the desired types. A similar issue arises in the Gaussian setting with unequal powers across transmitters~\cite{Nazer--Cadambe--Ntranos--Caire2015}. Our solution is to broaden the notion of recovering a linear combination. 
\input{linearcombination}

As part of our coding scheme, the $k^{\text{th}}$ transmitter will have an auxiliary index $l_k \in [2^{n\hat{R}_k}]$ for some auxiliary rate $\hat{R}_k$ that represents its selection during the multicoding step. Define $\tilde{R}_k = R_k + \hat{R}_k$ and $\tilde{R}_{\text{max}} = \max_k \tilde{R}_k$. We will map each transmitter's message and auxiliary indices into $\Fq^{\kappa}$ where $\kappa = n \tilde{R}_{\text{max}} / \log(\q)$. This is accomplished by concatenating the $\q$-ary expansions, followed by zero-padding (if necessary), resulting in 
\begin{align*}
\etab(m_k,l_k) := [\nub(m_k) ~ \nub(l_k) ~ \mathbf{0}], 
\end{align*} which is then mapped to the the linear codeword
\begin{align*}
u_k^n(m_k,l_k) &= \etab(m_k, l_k) \Gb \oplus d_k^n, 
\end{align*} 
where $\Gb \in \Fq^{\kappa \times n}$ is the generator matrix and $d_k^n \in \Fq^n$ is the dither vector.

The goal of the receiver is to recover up to $K$ linear combinations, each of which can be expressed as a linear codeword,
\begin{align*}
w^n_{\ab_\ell}(m_1,\ldots,m_K) &= \bigoplus_{k=1}^K a_{\ell, k} u^n_k(m_k,l_k) \\
&= \bigoplus_{k=1}^K a_{\ell, k} \big(\etab(m_k, l_k) \Gb \oplus d_k^n\big) \\
&=  \bigg(\bigoplus_{k=1}^K a_{\ell, k} \etab(m_k,l_k)\bigg) \Gb \oplus \bigoplus_{k=1}^K a_{\ell,k} d_k^n.  %\label{eqn:linearcomb}  
\end{align*} 

It will be convenient to associate each linear combination with a unique index. First, notice that the effective rate for a linear combination is determined by the maximum rate of all participating messages,
\begin{align}
\Rt(\ab_\ell):=\max\{\Rt_k: a_{\ell, k}\neq 0, k\in[1:K]\} \ . \label{eq:A-rates}
\end{align} Let $s_{\ab_\ell} \in [2^{n\Rt(\ab_\ell)}]$ be the unique index whose $\q$-ary expansion satisfies 
\begin{align}
[\nub(s_{\ab_\ell}) ~ \mathbf{0}] = \bigoplus_{k=1}^K a_{\ell, k} \etab(m_k,l_k).  \label{eqn:linearcombindex}  
\end{align} 
Now, with a slight abuse of notation, we can refer to each possible linear combination as follows
\begin{align}
w_{\ab_\ell}^n(s_{\ab_{\ell}}) = \nub(s_{\ab_\ell}) \Gb\, \oplus\, \bigoplus_{k=1}^K a_{\ell,k} d_k^n. \label{eqn:linearcomb}
\end{align}

\begin{remark} From an algebraic perspective, the set $\big\{\etab(m_k,l_k) : l_k \in [2^{nR_k}] \big\}$ corresponds to a coset for the message $m_k$. Similarly, we can view the linear combinations from~\eqref{eqn:linearcomb} as linear combinations of cosets. 
\end{remark}

%%%%%%%%%%%%%%%%%%%%%%%%%%%%%%%%%%%%%%%%%%%%%%%%%%%%%
%
%		NESTED LINEAR CODING ARCHITECTURE
%
%%%%%%%%%%%%%%%%%%%%%%%%%%%%%%%%%%%%%%%%%%%%%%%%%%%%%

\subsection{Nested Linear Code Architecture} \label{sec:arc}

We now specify the nested linear codes that will be used as our encoding functions throughout the paper. 
In addition to the messages $m\in[2^{nR_k}]$, $k=1,\ldots, K$, we use auxiliary indices $l\in[2^{n\Rh_k}]$, $k=1,\ldots, K$, with rates $R_k$ and $\Rh_k$, respectively. We define 
$\Rt_k:=R_k+\Rh_k$, $\Rmax:=\max\{R_1, R_2,\ldots, R_K\}$, and $\Rtmax:=\max\{\Rt_1, \Rt_2, \ldots, \Rt_K\}$. 
Let $\nub(m_k)$ denote the length $\ceil{nR_k/\log(\q)}$ $\q$-ary expansion of $m_k\in[2^{nR_k}]$. 
Similarly, let $\nub(l_k)$ denote the length $\ceil{n\Rh_k/\log(\q)}$ $\q$-ary expansion of $l_k\in[2^{n\Rh_k}]$. For simplicity, we assume that $nR_k/\log(\q)$ and $n\Rh_k/\log(\q)$ are integers for all rates in the sequel.
Further define  
\begin{align*}
\etab(m_k, l_k) &=[\nub(m_k), \nub(l_k), \mathbf{0}], \quad k\in[1:K],
\end{align*}
where $\etab(m_k, l_k)\in\Fq^\kappa$, $\kappa=n\Rtmax/\log(\q)$, and $\mathbf{0}$ is a vector of zeros with length $n(\Rtmax-\Rt_k)/\log(\q)$.
Note that all $\etab(m_k, l_k)$ have the same length due to zero padding.
\smallskip

We define a $(2^{nR_1},\ldots, 2^{nR_K},2^{n\Rh_1},\ldots, 2^{n\Rh_K},\Fq, n)$ {\em nested linear code} as the collection of $K$ codebooks generated by the following procedure.
\smallskip

Fix a pmf $\prod_{k=1}^Kp(u_k)$ and functions $x_k(u_k)$, $k\in[1:K]$.

\noindent{\bf Codebook generation.} 
Fix a finite field $\Fq$ and a parameter $\epsilon' \in (0,1)$.  
Randomly generate a $\kappa\times n$ matrix, $\Gb\in\Fq^{\kappa\times n}$, and sequences $d^n_k\in\Fq^n$, $k=1,\ldots, K$ where each element of $\Gb$ and $d_k^n$ are independently and randomly generated according to $\U(\Fq)$, and  $\kappa=n\Rtmax/\log(\q)$.

For each $k\in[1:K]$, generate a linear code $\Cc_k$ with parameters $(R_k, \Rh_k, n, \q)$ by
\begin{align}
u_k^n(m_k,l_k) &= \etab(m_k, l_k) \Gb \oplus d_k^n, 
\end{align}
for $m_k\in[2^{nR_k}]$, $l_k\in[2^{n\Rh_k}]$.
Note that from this construction, the codewords are pairwise independent and i.i.d. distributed, i.e., 
\begin{align}
\P\{U^n_k(m_k, l_k)=u_k^n, U_k^n(\mt_k, \lt_k)=\ut_k^n\}=\prod_{i=1}^n p_\q(u_i)p_\q(\ut_i),\quad (m, l)\neq (\mt, \lt), \label{eq:code-dist2}
\end{align}
where $p_\q=\U(\Fq)$. The general joint distribution of the codewords resulting from this construction can be found in~\cite[Theorem~1]{Domb--Zamir--Feder2013}.

\noindent{\bf Encoding.} For $k\in[1:K]$, given $m_k\in[2^{nR_k}]$, find an index $l_k\in[2^{n\Rh_k}]$ such that $ u_k^n(m_k,l_k)\in\aepvar(U_k)$. If there is more than one, select one randomly and uniformly. If there is none, randomly choose an index from $[2^{n\Rh_k}]$. Node $k$ transmits $x_{ki}(u_{ki})$, $i=1,\ldots, n$.

In the following section, we propose a decoding strategy that establishes Theorem~\ref{thm:comp-Fq}.

%%%%%%%%%%%%%%%%%%%%%%%%%%%%%%%%%%%%%%%%%%%%%%%%%%%%%%%%
%
%			DISCRETE MEMORYLESS COMPUTE--FORWARD
%
%%%%%%%%%%%%%%%%%%%%%%%%%%%%%%%%%%%%%%%%%%%%%%%%%%%%%%%%

\section{Proof of Theorems~\ref{thm:comp-Fq} and~\ref{thm:comp-real-discrete}} \label{sec:comp-finite}

\subsection{Proof of Theorem~\ref{thm:comp-Fq}}\label{sec:comp-Fq}

In the following, we provide achievable rate regions for the important special case of two transmitters and a receiver that wants a single linear combination over a finite field $\Fq$. As we will demonstrate, the rate region can be viewed as a union of the rates available to a ``single-user'' decoder that attempts to directly recover the desired linear combination and the rates available to a ``multiple-access'' decoder that recovers the messages individually and then takes the linear combination. Moreover, the achievability argument follows naturally via simultaneous joint typicality decoding, rather than a deliberate combination of two specialized decoders. 

We will break up the proof into two steps. First, we will establish Theorem~\ref{thm:comp-Fq} for the special case when the channel is a discrete memoryless MAC. 
Afterwards, we will use a standard quantization argument to extend this result to a the case, $\Yc = \Real$.  

\noindent{\bf{Step 1: Discrete memoryless MAC}}

\begin{figure*}[t!]
\centering
\small
\psfrag{v1}[c]{}
\psfrag{v2}[c]{}
\psfrag{m1}[c]{$M_1$}
\psfrag{m2}[c]{$M_2$}
\psfrag{u1}[c]{$U^n_1$}
\psfrag{u2}[c]{$U^n_2$}
\psfrag{x1}[c]{$X^n_1$}
\psfrag{x2}[c]{$X^n_2$}
\psfrag{y}[c]{$Y^n$}
\psfrag{yx}[c]{$p(y|x_1, x_2)$}
\psfrag{jte1}[c]{Multicoding}
\psfrag{jte2}[c]{Multicoding}
\psfrag{e1}[c]{Encoder 1}
\psfrag{e2}[c]{Encoder 2}
\psfrag{l1e1}[c]{Nested}
\psfrag{l1e2}[c]{linear code}
\psfrag{l2e1}[c]{Nested}
\psfrag{l2e2}[c]{linear code}
\psfrag{ex1}[c]{$x_1(u_1)$}
\psfrag{ex2}[c]{$x_2(u_2)$}
\psfrag{d}[c]{Decoder}
\psfrag{c}[c]{Channel}
\psfrag{mh}[c]{$\Sh_{\ab}$}
\includegraphics[scale=0.75]{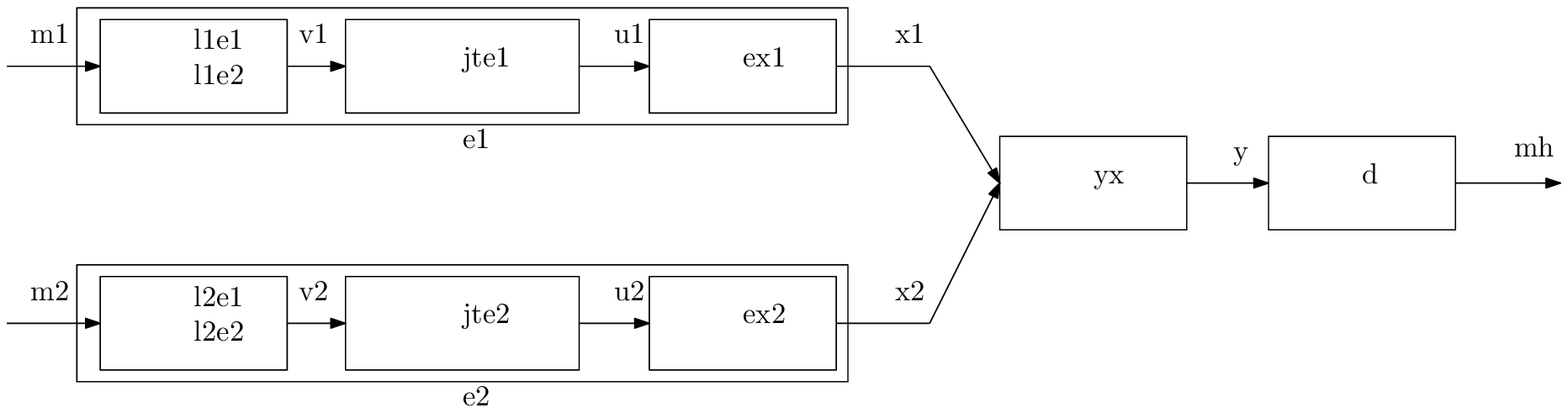}
\caption{Nested linear coding architecture for computing a linear combination with coefficient vector $\ab \in \Fq^2$ over a two-user DM-MAC. Each user selects, via multicoding, a linear codeword $U_k^n$ of the desired type, maps it into the channel input alphabet via the function $x_k(u_k)$, and transmits it as $X_k^n$. The receiver observes $Y^n$ over the DM-MAC specified by $p(y|x_1,x_2)$ and outputs an estimate $\Sh_{\ab}$. Decoding is successful if $\Sh_{\ab} = S_{\ab}$ where $S_{\ab}$ is the index whose $\q$-ary expansion corresponds to the linear combination with coefficient vector $\ab$ in the sense of~\eqref{eqn:linearcomb}.}
\label{fig:cf-mac}
\end{figure*}

Fix $\Fq$, pmf $p(u_1)p(u_2)$, and functions $x_1(u_1)$, $x_2(u_2)$.
The codebook construction and encoding steps follow the nested linear coding architecture in Section~\ref{sec:arc}. Without loss of generality, we assume that $a_1 \neq 0$ and $a_2 \neq 0$. (If one coefficient is equal to zero, the problem degenerates to the point-to-point communication case.)

\noindent{\bf Decoding.}
Let $\e'<\e$. Upon receiving $y^n$, the decoder searches for a unique index $s_{\ab} \in[2^{n\Rtmax}]$ such that
\begin{align}
	(u_1^n(m_1, l_1), u_2^n(m_2, l_2), y^n)\in\aep(U_1, U_2, Y),
\end{align}
for some $(m_1, l_1, m_2, l_2)\in[2^{nR_1}]\times [2^{n\Rh_1}]\times[2^{nR_2}]\times [2^{n\Rh_2}]$ such that
\begin{align*}
\nub(s_{\ab})=a_1 \etab(m_1, l_1)\oplus a_2 \etab(m_2, l_2).
\end{align*} 
If there is no such index, or more than one, the decoder declares an error.

\noindent{\bf Analysis of the probability of error.} 
Let $M_1, M_2$ be the messages, $L_1, L_2$ be the indices chosen by the encoders, and $S_{\ab}$ be the (unique) index of the linear combination $W_{\ab}^n(S_{\ab})$ such that
\begin{align}
\nub(S_{\ab})&=a_{1}\etab(M_1, L_1)\oplus a_{2}\etab(M_2, L_2). 
\end{align}
Then, the decoder makes an error only if one or more of the following events occur,
\begin{align*}
\Ec_1&=\{U_k^n(m_k, l_k) \not\in\aepvar \text{ for all } l_k, \text{ for some } m_k, k=1,2\},\\
\Ec_2&=\{(U_1^n(M_1, L_1), U_2^n(M_2, L_2), Y^n)\not\in\aep\},\\
\Ec_3&=\{(U_1^n(m_1, l_1), U_2^n(m_2, l_2), Y^n) \in \aep \text{ for some } (m_1, l_1,m_2,l_2)\\
&\qquad\qquad \text{ such that } \nub(S_{\ab})\neq a_{1}\etab(m_1,l_1)\oplus a_{2}\etab(m_2, l_2)\}.
\end{align*}
Then, by the union of events bound,
\begin{align}
\P(\Ec) &\le \P(\Ec_1)+\P(\Ec_2\cap\Ec_1^c)+\P(\Ec_3\cap\Ec_1^c).
\end{align}
By Lemma~\ref{lem:mm-covering} in Appendix~\ref{app:mm-covering-packing}, the probability $\P(\Ec_1)$ tends to zero as $n\to\infty$ if
\begin{align}
\Rh_k>  D(p_{U_k}\|p_\q)+\d(\e'),\quad k=1,\ldots,K. \label{eq:cf-cover}
\end{align}

Define $\Mc:=\{M_1=0, M_2=0, L_1=0, L_2=0\}$ as the event where both messages are zero and the chosen auxiliary indices are zero as well. By symmetry of the codebook construction and encoding steps, we have that $P(\Ec_2\cap\Ec_1^c)=P(\Ec_2\cap\Ec_1^c|\Mc)$ and $P(\Ec_3\cap\Ec_1^c)=P(\Ec_3\cap\Ec_1^c|\Mc)$.

\begin{remark}
To bound the second probability term, we need a non-trivial proof to establish that the pair of selected codewords are jointly typical with the channel output. If each encoder employed an independent random codebook, this could be shown via a standard application of the Markov lemma~\cite[Lemma 12.1]{El-Gamal--Kim2011}. However, due to the shared generator matrix, the codebooks are dependent across the users. Prior work by Padakandla and Pradhan~\cite{Padakandla--Pradhan2013b} established that the channel inputs and output are jointly typical for $K = 2$ users under the additional constraint that $\Rh_k < D(p_{U_k}\|p_\q)+3\d(\e')$. In Appendix~\ref{app:ML-LC}, we provide an alternative proof that removes this constraint and generalizes to $K > 2$ users.
\end{remark}

By Lemma~\ref{lem:ML-LC} in Appendix~\ref{app:ML-LC}, the second term $\P(\Ec_2\cap\Ec_1^c|\Mc)$ tends to zero as $n\to\infty$ if~\eqref{eq:cf-cover} is satisfied. 

We bound the probability $P(\Ec_3\cap\Ec_1^c|\Mc)$ in two ways. The first bounds the event that an incorrect linear combination is jointly typical with the channel output. The second bounds the event that incorrect codewords are jointly typical with the channel output, regardless of the resulting linear combination. Note that the event $\Mc$ implies that $S_{\ab}=0$. Let $\Sc =\{(m_1, l_1, m_2, l_2): a_1\etab(m_1, l_1)\oplus a_2\etab(m_2, l_2)=\mathbf{0}\}$ denote the set of indices that yield the correct linear combination.
For the first bound,
\begin{align}
&\P\left( \Ec_3\cap\Ec_1^c|\Mc \right)\nn\\
&= \P\{(U_1^n(m_1,l_1), U_2^n(m_2,l_2), Y^n)\in\aep, \Ec_1^c, \nn\\
&\qquad~~\text{ for some } (m_1, l_1, m_2, l_2)\notin \Sc |\Mc\}\nn\\
&\stackrel{(a)}{=} \P\{(W_{\ab}^n(s_{\ab}), U_1^n(m_1,l_1), U_2^n(m_2,l_2), Y^n)\in\aep, \nn\\
&\qquad~~\Ec_1^c, \text{ for some } (m_1, l_1, m_2, l_2)\notin \Sc |\Mc\}\nn\\ 
&\stackrel{(b)}{\le} \P\{(W^n_{\ab}(s_{\ab}), Y^n)\in\aep, \Ec_1^c, \text{ for some } s_{\ab} \neq 0 |\Mc\} \label{eq:upper1}
\end{align}
where 
\begin{align*}
W^n_{\ab}(s_{\ab}) &= a_1 U^n_1(m_1, l_1)\oplus a_2 U^n_2(m_2, l_2),
\end{align*}
step $(a)$ follows from the fact that $W^n_{\ab}(s_{\ab})$ is a deterministic function of $(U_1^n(m_1,l_1), U_2^n(m_2,l_2))$, and step $(b)$ follows from the fact that $(W^n_{\ab}(s_{\ab}), U_1^n(m_1,l_1), U_2^n(m_2,l_2), Y^n)\in\aep$ implies $(W^n_{\ab}(s_{\ab}), Y^n)\in\aep$.
Define 
\begin{align*}
\tilde{\Ec}(s_{\ab}) &=\{ (W^n_{\ab}(s_{\ab}),  Y^n) \in \aep, U^n_1(0,0)\in\aepvar, U^n_2(0,0)\in\aepvar \}.
\end{align*}
 
Then, by the union of events bound,
\begin{align}
P(\Ec_3\cap\Ec_1^c|\Mc) \le \sum_{s_{\ab}\neq 0} \P(\tilde{\Ec}(s_{\ab})|\Mc). \label{eq:e3unionbound}
\end{align}
\begin{lemma}\label{lem:cfbound}
Let $\tilde{D}_U=D(p_{U_1}\|p_\q)+D(p_{U_2}\|p_\q)$. Then,
\begin{align*}
&\P(\tilde{\Ec}(s_{\ab})|\Mc)\le 2^{n(\Rh_1+\Rh_2)} 2^{-n(I(W_{\ab}; Y)+D(p_{W_{\ab}}\|p_\q)-\d(\e))}2^{-n(\tilde{D}_U-\d(\e))}.
\end{align*}
\end{lemma}
\begin{IEEEproof}
\begin{align*}
&\P(\tilde{\Ec}(s_{\ab})|\Mc) \\
&\le \P\{ (W^n_{\ab}(s_{\ab}),  Y^n) \in \aep, U^n_1(0,0)\in\aep, U^n_2(0,0)\in\aep |\Mc\}\\
&= \sum_{u^n_1\in\aep, \, u^n_2\in\aep}\sum_{(w^n, y^n)\in\aep} \P\{ W^n_{\ab}(s_{\ab})=w^n,  Y^n=y^n, U^n_1(0,0)=u^n_1, U^n_2(0,0)=u^n_2 |\Mc\}\\
&\stackrel{(a)}{=} \sum_{u^n_1\in\aep, \, u^n_2\in\aep}\sum_{y^n\in\aep}\sum_{w^n\in\aep(W_{\ab}|y^n)} \P\{ Y^n=y^n| U^n_1(0,0)=u^n_1, U^n_2(0,0)=u^n_2, \Mc\}\\
&\quad\times \P\{ W^n_{\ab}(s_{\ab})=w^n,  U^n_1(0,0)=u^n_1, U^n_2(0,0)=u^n_2 |\Mc\}\\
&\stackrel{(b)}{\leq} 2^{n(\Rh_1+\Rh_2)} \sum_{u^n_1\in\aep, \, u^n_2\in\aep}\sum_{y^n\in\aep} p(y^n|u_1^n, u_2^n)\\
&\quad\times \sum_{w^n\in\aep(W_{\ab}|y^n)}\P\{ W^n_{\ab}(s_{\ab})=w^n,  U^n_1(0,0)=u^n_1, U^n_2(0,0)=u^n_2 \}\\
&\stackrel{(c)}{=} 2^{n(\Rh_1+\Rh_2)} \sum_{u^n_1\in\aep, \, u^n_2\in\aep}\sum_{y^n\in\aep} p(y^n|u_1^n, u_2^n)\\
&\quad\times \sum_{w^n\in\aep(W_{\ab}|y^n)}\P\{ W^n_{\ab}(s_{\ab})=w^n\}  \P\{ U^n_1(0,0)=u^n_1 \} \P\{U^n_2(0,0)=u^n_2 \}\\
&\stackrel{(d)}{=} 2^{n(\Rh_1+\Rh_2)} \sum_{u^n_1\in\aep,\, u^n_2\in\aep}\sum_{y^n\in\aep}p(y^n|u_1^n, u_2^n)\\
&\quad\times \sum_{w^n\in\aep(W_{\ab}|y^n)}  2^{-n(H(W_{\ab})+D(p_{W_{\ab}}\|p_\q))}2^{-n(H(U_1)+D(p_{U_1}\|p_\q))}2^{-n(H(U_2)+D(p_{U_2}\|p_\q))}\\
&\le 2^{n(\Rh_1+\Rh_2)} \sum_{u^n_1\in\aep, \, u^n_2\in\aep}\sum_{y^n\in\aep} p(y^n|u_1^n, u_2^n) \\
&\quad\times 2^{-n(I(W_{\ab}; Y)+D(p_{W_{\ab}}\|p_\q)-\d(\e))}2^{-n(H(U_1)+D(p_{U_1}\|p_\q))}2^{-n(H(U_2)+D(p_{U_2}\|p_\q))}\\
&\le 2^{n(\Rh_1+\Rh_2)} \sum_{u^n_1\in\aep,\, u^n_2\in\aep}2^{-n(I(W_{\ab}; Y)+D(p_{W_{\ab}}\|p_\q)-\d(\e))}2^{-n(H(U_1)+D(p_{U_1}\|p_\q))}2^{-n(H(U_2)+D(p_{U_2}\|p_\q))}\\
&\le 2^{n(\Rh_1+\Rh_2)} 2^{-n(I(W_{\ab}; Y)+D(p_{W_{\ab}}\|p_\q)-\d(\e))}2^{-n(D(p_{U_1}\|p_\q)-\d(\e))}2^{-n(D(p_{U_2}\|p_\q)-\d(\e))},
\end{align*}
where step $(a)$ follows from the fact that conditioned on $\Mc$, we have the Markov relation
\begin{align*}
Y^n \to (U^n_1(0,0), U^n_2(0,0)) \to W^n_{\ab}(s_{\ab}),
\end{align*}
step $(b)$ follows from Lemma~\ref{lem:message} in Appendix~\ref{app:message}, step $(c)$ follows from the fact that $W^n_{\ab}(s_{\ab})$, $U^n_1(0,0)$, and $U^n_2(0,0)$ are independent due to the dithers and that $s_{\ab} \neq 0$, and step $(d)$ uses the fact that $W^n_{\ab}(s_{\ab})$, $U^n_1(0,0)$, and $U^n_2(0,0)$ are each uniformly distributed over $\Fq^n$ and that, for any pmf $p_V(v)$, we can use the relation
\begin{align}
\log{\q} = H(V) + D(p_V \| p_{\q}), \label{eqn:relation}
\end{align} 
where $p_\q = \mathrm{Unif}(\Fq)$ to write
\begin{align*}
\frac{1}{\q^n} = 2^{-n(H(V)+D(p_{V} \|p_\q ))}.
\end{align*} 
\end{IEEEproof}

Plugging the bound from Lemma~\ref{lem:cfbound} back into~\eqref{eq:e3unionbound}, we find that 
\begin{align*}
P(\Ec_3\cap\Ec_1^c|\Mc) \le 2^{n(\Rtmax + \Rh_1+\Rh_2)} 2^{-n(I(W_{\ab}; Y)+D(p_{W_{\ab}}\|p_\q)-\d(\e))}2^{-n(\tilde{D}_U-\d(\e))}.
\end{align*} Thus, the probability of $P(\Ec_3\cap\Ec_1^c|\Mc)$ tends to zero if as $n\to\infty$ if
\begin{align*}
R_1+2\Rh_1+\Rh_2 &< I(W_{\ab}; Y)+D(p_{W_{\ab}}\|p_\q)+\tilde{D}_U- 2\d(\e),\\
R_2+\Rh_1+2\Rh_2 &< I(W_{\ab}; Y)+D(p_{W_{\ab}}\|p_\q)+\tilde{D}_U- 2\d(\e).
\end{align*}
By eliminating $\Rh_1$ and $\Rh_2$, setting $\Rh_1=D(p_{U_1}\|p_\q)+2\d(\e')$ and $\Rh_2=D(p_{U_2}\|p_\q)+2\d(\e')$ in order to satisfy~\eqref{eq:cf-cover}, and sending $\e\to 0$, we have shown that a rate pair $(R_1, R_2)$ is achievable if
\begin{align*}
R_1 &< H(U_1)-H(W_{\ab}|Y),\\
R_2 &< H(U_2)-H(W_{\ab}|Y),
\end{align*} 
where we have used the relation~\eqref{eqn:relation} to simplify the expression.

Next, we show the second bound on $P(\Ec_3\cap\Ec_1^c|\Mc)$ by the following steps:
\begin{align}
&\P\left(\Ec_3\cap\Ec_1^c|\Mc\right)\nn\\
&= \P\{(U_1^n(m_1,l_1), U_2^n(m_2,l_2), Y^n)\in\aep, \Ec_1^c, \nn\\
&\qquad~~\text{ for some } (m_1, l_1, m_2, l_2)\notin \Sc |\Mc\}\nn\\
&\le \P\{(U_1^n(m_1,l_1), U_2^n(m_2,l_2), Y^n)\in\aep, \Ec_1^c, \nn\\
&\qquad~~\text{ for some } (m_1, l_1, m_2, l_2)\neq (0,0,0,0) |\Mc\}. \label{eq:upper2}
\end{align}
Define 
\begin{align*}
\Scale[0.94]{\tilde{\Ec}(m_1, l_1, m_2, l_2)} &=\Scale[0.94]{\{(U^n_1(m_1, l_1), U^n_2(m_2, l_2),  Y^n) \in \aep,}\\
&\qquad \Scale[0.94]{U^n_1(0,0)\in\aepvar, U^n_2(0,0)\in\aepvar} \},
\end{align*}
and subsets of $[2^{nR_1}]\times[2^{n\Rh_1}]\times[2^{nR_2}]\times[2^{n\Rh_2}]$ as
\begin{align*}
\Scale[0.95]{\Ac} &=\Scale[0.95]{\{(m_1, l_1, m_2, l_2): (m_1, l_1, m_2, l_2)\neq (0,0,0,0)\},}\\
\Scale[0.95]{\Ac_{1}} &=\Scale[0.95]{\{(m_1, l_1, m_2, l_2): (m_1, l_1)\neq (0,0), (m_2, l_2)= (0,0) \},}\\
\Scale[0.95]{\Ac_{2}} &=\Scale[0.95]{\{(m_1, l_1, m_2, l_2):  (m_1, l_1)= (0,0), (m_2, l_2) \neq (0,0) \},}\\
\Scale[0.95]{\Ac_{12}} &=\Scale[0.95]{\{(m_1, l_1, m_2, l_2): (m_1, l_1) \neq (0,0), (m_2, l_2) \neq (0,0)\},}\\
\Scale[0.95]{\Lc} &=\Scale[0.95]{\{(m_1, l_1, m_2, l_2)\in\Ac_{12}: \etab(m_1, l_1), \etab(m_2, l_2)}\\
&\qquad\Scale[0.95]{ \text{ are linearly dependent}\},}\\
\Scale[0.95]{\Lc^c} &= \Scale[0.95]{\{(m_1, l_1, m_2, l_2)\in\Ac_{12}: \etab(m_1, l_1), \etab(m_2, l_2)}\\
&\qquad\Scale[0.95]{\text{ are linearly independent}\}}.
\end{align*}
Further, for some $\bb\in\Fq^2$ such that $\bb\neq \mathbf{0}$, define
\begin{align*}
\Lc_1(\bb) &=\{(m_1, l_1, m_2, l_2)\in\Lc: \\
&\qquad b_1\etab(m_1, l_1)\oplus b_2\etab(m_2, l_2) \neq \mathbf{0} \},\\
\Lc_2(\bb) &=\{(m_1, l_1, m_2, l_2)\in\Lc: \\
&\qquad b_1\etab(m_1, l_1)\oplus b_2\etab(m_2, l_2) = \mathbf{0} \}.
\end{align*}
Note that, for any $\bb\in\Fq^2$ that is not the all-zero vector, we have
\begin{align*}
&\Ac \subseteq (\Ac_{1} \cup \Ac_{2} \cup \Ac_{12}),\\
&\Ac_{12}=\Lc \cup \Lc^c,\\
&\Lc=\Lc_1(\bb) \cup \Lc_2(\bb),
\end{align*}
and thus, $\Ac \subseteq (\Ac_{1} \cup \Ac_{2} \cup \Lc^c \cup \Lc_1(\bb) \cup \Lc_2(\bb))$.
Furthermore, the cardinality of these sets can be upper bounded by
\begin{align}
&|\Ac_1|\le 2^{n(R_1+\Rh_1)}, \nn\\
&|\Ac_2|\le 2^{n(R_2+\Rh_2)}, \nn\\
&|\Ac_{12}|\le 2^{n(R_1+\Rh_1+R_2+\Rh_2)}, \nn\\
&|\Lc|\le 2^{n(\min\{ R_1 + \Rh_1, R_2 + \Rh_2 \} )}(\mathsf{q}-1). \label{eq:cardinality1}
\end{align}

Then, 
\begin{align}
&\P(\Ec_3\cap\Ec_1^c|\Mc) =\nn\\
& \P(\tilde{\Ec}(m_1, l_1, m_2, l_2)\text{ for some } (m_1, l_1, m_2, l_2)\in\Ac | \Mc )\nn\\
&\le \sum_{(m_1, l_1, m_2, l_2)\in\Ac}\P(\tilde{\Ec}(m_1,l_1, m_2, l_2)\big|\Mc)\nn\\
&\le \sum_{(m_1, l_1, m_2, l_2)\in\Ac_1}\P(\tilde{\Ec}(m_1,l_1, m_2, l_2)\big|\Mc)\nn\\
&~~+\sum_{(m_1, l_1, m_2, l_2)\in\Ac_2}\P(\tilde{\Ec}(m_1,l_1, m_2, l_2)\big|\Mc)\nn\\
&~~+\sum_{(m_1, l_1, m_2, l_2)\in\Lc^c}\P(\tilde{\Ec}(m_1,l_1, m_2, l_2)\big|\Mc)\nn\\
&~~+\sum_{(m_1, l_1, m_2, l_2)\in\Lc_1(\bb)}\P(\tilde{\Ec}(m_1,l_1, m_2, l_2)\big|\Mc)\nn\\
&~~+\sum_{(m_1, l_1, m_2, l_2)\in\Lc_2(\bb)}\P(\tilde{\Ec}(m_1,l_1, m_2, l_2)\big|\Mc). \label{eq:error1}
\end{align}
We establish upper bounds on $\P(\tilde{\Ec}(m_1,l_1, m_2, l_2)\big|\Mc)$ in the following lemma.
\begin{lemma}\label{lem:jtl-linear}
Let $\tilde{D}_U=D(p_{U_1}\|p_\q)+D(p_{U_2}\|p_\q)$. The probability $\P(\tilde{\Ec}(m_1,l_1, m_2, l_2)\big|\Mc)$ can be upper bounded by considering the following cases:
\begin{enumerate}
\item For $(m_1, l_1, m_2, l_2)\in\Ac_1$, 
\begin{align*}
&\P(\tilde{\Ec}(m_1,l_1, m_2, l_2)\big|\Mc) \le 2^{n(\Rh_1+\Rh_2)}\\
&\quad \times2^{-n(I(U_1; Y | U_2)+D(p_{U_1}\|p_\q)+\tilde{D}_U-\d(\e))}.
\end{align*}
\item For $(m_1, l_1, m_2, l_2)\in\Ac_2$, 
\begin{align*}
&\P(\tilde{\Ec}(m_1,l_1, m_2, l_2)\big|\Mc) \le 2^{n(\Rh_1+\Rh_2)}\\
&\quad \times2^{-n(I(U_2; Y |  U_1)+D(p_{U_2}\|p_\q)+\tilde{D}_U-\d(\e))}.
\end{align*}
\item For $(m_1, l_1, m_2, l_2)\in\Lc^c$, 
\begin{align*}
&\P(\tilde{\Ec}(m_1,l_1, m_2, l_2)\big|\Mc)\le 2^{n(\Rh_1+\Rh_2)}\\
&\quad \times2^{-n(I(U_1, U_2; Y)+2\tilde{D}_U-\d(\e))}.
\end{align*}
\item For $(m_1, l_1, m_2, l_2)\in\Lc_1(\bb)$, 
\begin{align*}
&\P(\tilde{\Ec}(m_1,l_1, m_2, l_2)\big|\Mc)\le 2^{n(\Rh_1+\Rh_2)}\\
&\quad \times2^{-n(I(W_{\bb}; Y)+D(p_{W_{\bb}}\|p_\q)+\tilde{D}_U-\d(\e))},
\end{align*}
where $W_{\bb}=b_1U_1\oplus b_2U_2$.
\item For $(m_1, l_1, m_2, l_2)\in\Lc_2(\bb)$, 
\begin{align*}
&\P(\tilde{\Ec}(m_1,l_1, m_2, l_2)\big|\Mc) \le 2^{n(\Rh_1+\Rh_2)}\\
&\quad \times 2^{-n(I(W_{\cb}; Y, W_{\bb})+D(p_{W_{\cb}}\|p_\q)+\tilde{D}_U-\d(\e))},
\end{align*}
for some non-zero vector $\cb=[c_1,\, c_2]\in\Fq^2$ that is linearly independent of $\bb$ where 
\begin{align*}
W_{\cb}=c_1U_1\oplus c_2U_2.
\end{align*}
\end{enumerate}
\end{lemma}

The proof is given in Appendix~\ref{app:proof-jtl}.

From the cardinality bounds given in \eqref{eq:cardinality1} and Lemma~\ref{lem:jtl-linear}, the probability terms in~\eqref{eq:error1} tends to zero as $n\to\infty$ if
\begin{align}
R_1 + 2\Rh_1+ \Rh_2 &< I(U_1; Y|U_2)+D(p_{U_1}\|p_\q)+\tilde{D}_U-\d(\e),\label{eq:cond2}\\ 
R_2 +\Rh_1+2\Rh_2 &< I(U_2; Y|U_1)+D(p_{U_2}\|p_\q)+\tilde{D}_U-\d(\e),\label{eq:cond3}\\
R_1+R_2+2\Rh_1+2\Rh_2  &< I(U_1, U_2; Y)+2\tilde{D}_U-\d(\e),\label{eq:cond4}\\
\min\{R_1+\Rh_1, R_2+\Rh_2\}+\Rh_1+\Rh_2  &< I(W_{\bb}; Y)+D(p_{W_{\bb}}\|p_\q)+\tilde{D}_U-\d(\e),\label{eq:cond5}\\
\min\{R_1+\Rh_1, R_2+\Rh_2\}+\Rh_1+\Rh_2&< I(W_{\cb}; Y, W_{\bb})+D(p_{W_{\cb}}\|p_\q)+\tilde{D}_U-\d(\e). \label{eq:cond6}
\end{align}

By choosing the auxiliary rates $\Rh_k=D(p_{U_k}\|p_\q)+2\d(\e')$, $k=1,2$, in order to satisfy~\eqref{eq:cf-cover}, 
using the relation~\eqref{eqn:relation}, and taking $\e\to 0$, we can conclude that any rate pair $(R_1, R_2)$ satisfying
\begin{align}
R_1 &< I(U_1; Y| U_2), \nn\\ 
R_2 &< I(U_2; Y| U_1), \nn\\
R_1+R_2 &<  I(U_1, U_2; Y)\nn\\
\min\bigl(R_1-H(U_1), R_2-H(U_2)\bigr) &< I(W_{\bb}; Y)-H(W_{\bb}),\nn\\
\min\bigl(R_1-H(U_1), R_2-H(U_2)\bigr) &< I(W_{\cb}; Y, W_{\bb})-H(W_{\cb}), \label{eq:equiv-rr}
\end{align}
for some pmf $p(u_1)p(u_2)$, functions $x_1(u_1)$,  $x_2(u_2)$, and non-zero linearly independent vectors $\bb,\cb\in\Fq^2$ is achievable.

Finally, in Appendix~\ref{app:equiv-rr}, we show that the above rate region is equivalent to the rate region $\Rc_{\mathsf{LMAC}}$ which concludes the proof for the DM-MAC. We now generalize this result to the case where the channel output is real-valued, $\Yc = \Real$.

\noindent{\bf Step 2: Real-valued channel outputs}

Assume that $\Yc = \Real$. Let $[y]_j$ denote the output of a uniform quantizer that maps $y \in \Real$ to the closest point in
\begin{align*}
\left\{-j\Delta, -(j-1)\Delta, \ldots, -\Delta, 0, \Delta,\ldots, (j-1)\Delta, j\Delta \right\},
\end{align*}
where the step size is $\Delta = 1/\sqrt{j}$.

From the proof in Step 1 above, the rate region in Theorem~\ref{thm:comp-Fq} is achievable with $Y$ replaced by $[Y]_j$. Since the real line $\mathbb{R}$ is a {\em standard space} according to the nomenclature of \cite[Section~1.4]{Gray2011}, and since as $\Delta \to 0$, the quantization partitions generated by $\Delta \mathbb{Z}$ asymptotically recover the Borel field of the real line, by \cite[Lem.~7.18]{Gray2011} we have the limits
\begin{align*}
	\lim_{j \to \infty} H\bigl(W\big|[Y]_{j}\bigr)
	&= H(W|Y),\\ 
	\lim_{j \to \infty} I( X_1 ; [Y]_{j} | X_2 \bigr)
	&= I(X_1; Y | X_2), \\
	\lim_{j \to \infty} I(X_2 ; [Y]_{j} | X_1 \bigr)
	&= I(X_2 ; Y | X_1), \\
	\lim_{j \to \infty}  I(X_1, X_2 ; [Y]_{j}\bigr)
	&= I(X_1, X_2; Y).
\end{align*} 
This completes the proof of Theorem~\ref{thm:comp-Fq}.

\subsection{Proof of Theorem~\ref{thm:comp-real-discrete}}
Our approach is to show that integer-linear combinations of bounded integers can be viewed as linear combinations over a sufficiently large, prime-sized finite field. This will enable us to apply Theorem~\ref{thm:comp-Fq}.

Let $\q$ be a prime number. Consider the finite field $\Fq=\Integer/ \q \mathbb{Z}$,
\begin{align*}
\Fq =\left\{-\frac{\q-1}{2},\ldots,-1, 0, 1, \ldots, \frac{\q-1}{2}\right\},
\end{align*}
where, for $a,b\in \Fq$, the addition and multiplication operations are defined as  
\begin{align*}
a\oplus b = [a+b]\bmod\q\\ 
a b = [a\cdot b]\bmod\q,
\end{align*} 
respectively, with the modulo operation taken over the residue system $\mathbb{Z} / \q \mathbb{Z}$. That is, $[a]\bmod\q=r$ where $r\in \mathbb{Z} / \q \mathbb{Z}$ is the unique element satisfying $a=i\q+r$ (over the reals) for some integer $i$.

Notice that, for any $a,b \in \Field_{\q}$ the addition and multiplication operations over $\Field_{\q}$ can be expressed as 
\begin{align*}
a\oplus b &=  [ a + b ] \bmod\q,\\ 
ab &=  [ a \cdot b ] \bmod\q,
\end{align*} 
respectively.

The next lemma will allow us to translate our integer-linear combinations over $\Real$ into linear combinations over $\Field_{\q}$.
\begin{lemma}[Translation Lemma] \label{lem:translation}
Select $a_{\ell, k} \in \Integer, \ell,k \in [1:K]$ and assume that $U_k$ take values on a bounded subset of $\Integer$. Then, for prime $\q$ large enough and $\ell = 1,\ldots,K$, we have that 
\begin{align*}
\sum_{k=1}^K a_{\ell, k} U_k = \bigoplus_{k=1}^K \tilde{a}_{\ell,k} U_k 
\end{align*}
where $\tilde{a}_{\ell,k} = [a_{\ell,k}] \bmod{\q}$ and the multiplication and summation operations are taken over $\Real$ on the left-hand side and over $\Field_{\q}$ on the right-hand side.
\end{lemma}
\begin{IEEEproof}
Since the $U_k$'s are bounded, there exists a $\Gamma > 0$ such that $| U_k | \leq \Gamma$, $k = 1,\ldots,K$. Select a prime $\q$ large enough to satisfy the following relation 
\begin{align*}
\max \bigg\{ \max_{\ell, k}  |a_{\ell, k}| , \Gamma \bigg\}  \leq \bigg\lfloor \sqrt{\frac{\q - 1}{2K}} \bigg\rfloor.
\end{align*} 
It follows that $| a_{\ell, k} U_k | \leq ( \q -1)/(2K)$ over $\Real$ and that $\tilde{a}_{\ell,k} = a_{\ell,k}$. Therefore, $\big| \sum_{k=1}^K a_{\ell, k} U_k \big| \leq ( \q -1)/2$ over $\Real$, and the $\bmod~\q$ operation will not be used in any of the addition or multiplication operations over $\Field_{\q,\Delta}$, i.e., 
\begin{align*}
\bigoplus_{k=1}^K  a_{\ell,k} U_k = \bigg[ \sum_{k=1}^K a_{\ell,k} U_k \bigg] \bmod\q = \sum_{k=1}^K a_{\ell,k} U_k .
\end{align*} 
\end{IEEEproof}

Now, using the Translation Lemma, select $\q$ large enough so that $a_1 U_1 + a_2 U_2 = \tilde{a}_1 U_1 \oplus \tilde{a}_2 U_2$ where the operations on the left-hand side are over $\Real$ while those on the right-hand side are over $\Field_{\q}$ and $\tilde{a}_k = [a_k] \bmod{\q}$. Now, invoking Theorem~\ref{thm:comp-Fq} with finite field $\Field_{\q}$, input pmf $p_{U_1}(u_1)p_{U_2}(u_2)$, and symbol mappings $x_1(u_1)$ and $x_2(u_2)$, we obtain the desired achievable rate region.

%%%%%%%%%%%%%%%%%%%%%%%%%%%%%%%%%%%%%%%%%%%%%%%%%%%%%%%%%
%
%		Proof of Theorem 2
%
%%%%%%%%%%%%%%%%%%%%%%%%%%%%%%%%%%%%%%%%%%%%%%%%%%%%%%%%%%%

\section{Proof of Theorem~\ref{thm:comp-real}} \label{sec:comp-real}

We will use a quantization argument to establish Theorem~\ref{thm:comp-real} starting from Theorem~\ref{thm:comp-real-discrete}. In particular, we will use a variant on the approach in~\cite[Chapter 3.4.1]{El-Gamal--Kim2011} that will enable us to view addition and multiplication on the quantized variables as operations over the reals. Let us first assume that $U_1$ and $U_2$ are compactly supported (an assumption we will relax at the end of the proof by means of a truncation argument).

For a given resolution $\Delta > 0$, define
\begin{align*}
	\lceil u \rfloor_\Delta = \argmin_{\tilde{u} \in \Delta \Integer} |u - \tilde{u}|
\end{align*}
to be the quantization of $u$ to the closest point in $\Delta \Integer$, ties being broken in any arbitrary way.

Now, define the variables
\begin{IEEEeqnarray*}{rCl}
	W
	&\triangleq& a_1 U_1 + a_2 U_2 \\
	W'
	&\triangleq& a'_1 U_1 + a'_2 U_2 \\
	W_\Delta
	&\triangleq& a_1 \lceil U_1 \rfloor_\Delta + a_2 \lceil U_2 \rfloor_\Delta \\
	W'_\Delta
	&\triangleq& a'_1 \lceil U_1 \rfloor_\Delta + a'_2 \lceil U_2 \rfloor_\Delta
\end{IEEEeqnarray*}
where $a'_1 \triangleq a_1/\gcd(\ab)$ and $a'_2 \triangleq a_2/\gcd(\ab)$ denote the gcd-reduced  coefficients.
Let $Y_\Delta$ denote the channel output variable induced by the quantized input variables $\lceil U_1 \rfloor_\Delta$ and $\lceil U_2 \rfloor_\Delta$. That is, conditional on $(\lceil U_1 \rfloor_\Delta,\lceil U_2 \rfloor_\Delta) = (u_1,u_2)$, the variable $Y_\Delta$ is distributed with a cdf $Y_\Delta \sim F_{Y|U_1,U_2}(\cdot|u_1,u_2)$.

Note that in Theorem~\ref{thm:comp-real-discrete}, the assumption $\mathcal{U}_k \subset \mathbb{Z}$ can be equivalently replaced by $\mathcal{U}_k \subset \Delta\mathbb{Z}$ with some positive scaling factor $\Delta > 0$ without affecting the achievable rate region (which is invariant under this scaling).
Owing to the compact support assumption on $U_1$ and $U_2$, the quantized auxiliaries $\lceil U_1 \rfloor_\Delta$ and $\lceil U_2 \rfloor_\Delta$ are finitely supported for any $\Delta > 0$. Hence the following compute--forward rate region is achievable by Theorem~\ref{thm:comp-real-discrete}:
\begin{IEEEeqnarray*}{rCl}
	R_1
	&<& H\bigl(\lceil U_1 \rfloor_\Delta\bigr) - H\bigl(W_\Delta\big|Y_\Delta\bigr) \\
	R_2
	&<& H\bigl(\lceil U_2 \rfloor_\Delta\bigr) - H\bigl(W_\Delta\big|Y_\Delta\bigr).
\end{IEEEeqnarray*}
We will calculate the limit of this achievable rate region as we take the quantization step $\Delta$ to zero. It suffices to prove the following three statements in order to conclude the proof of Theorem~\ref{thm:comp-real}:
\begin{subequations}
\begin{IEEEeqnarray}{rCl}
	\lim_{\Delta \to 0} \bigl\{ H([U_1]_{\Delta}) + \log(\Delta) \bigr\} &=& h(U_1)   \label{quod_est_1} \\
	\lim_{\Delta \to 0} \bigl\{ H([U_2]_{\Delta}) + \log(\Delta) \bigr\} &=& h(U_2)   \label{quod_est_1_bis} \\
	\limsup_{\Delta \to 0} \bigl\{ H(W_{\Delta}|Y_{\Delta}) + \log(\Delta) \bigr\} &\leq& h(W|Y) - \log\gcd(\ab).   \label{quod_est_2}
\end{IEEEeqnarray}
\end{subequations}

Let us first state a classical result by R\'enyi.

\begin{lemma}[{\cite[Theorem~1]{Renyi1959}}]   \label{lem:Renyi}
Let $\Xv$ be an $\mathbb{R}^K$-valued random vector with an absolutely continuous distribution such that $H(\lceil\Xv\rfloor)$ and $h(\Xv)$ are finite. Then
\begin{equation*}
	\lim_{\Delta \to 0} \bigl\{ H(\lceil \Xv \rfloor_\Delta) + K \log(\Delta) \bigr\}
	= h(\Xv).
\end{equation*}
\end{lemma}
Note that \eqref{quod_est_1} and \eqref{quod_est_1_bis} follow directly from Lemma~\ref{lem:Renyi}. Next, we will need a recent result of Makkuva and Wu~\cite{Makkuva--Wu2016}.

\begin{lemma}[{\cite[Lemma~1]{Makkuva--Wu2016}}]   \label{lem:Makkuva_Wu}
Let $X_1,\dots,X_K$ be mutually independent, continuous random variables with compact support such that $H(\lceil X_i \rfloor)$ and $h(X_i)$ are finite for all $i=1,\dotsc,K$. Then for relatively prime integer coefficients $(a_1,\dotsc,a_K) \in \mathbb{Z}^K$,
\begin{equation*}
	\lim_{\Delta \to 0} \left\{ H\left( \left\lceil \sum_{i=1}^K a_i X_i\right \rfloor_\Delta \right) - H\left( \sum_{i=1}^K a_i \left\lceil X_i\right \rfloor_\Delta \right) \right\}
	= 0.
\end{equation*}
\end{lemma}

To prove the remaining statement~\eqref{quod_est_2}, note that
\begin{IEEEeqnarray*}{rCl}
	\limsup_{\Delta \to 0} \bigl\{ H(W_\Delta|Y_\Delta) + \log(\Delta) \bigr\}
	&=& \limsup_{\Delta \to 0} \bigl\{ H(W_\Delta) - I(W_\Delta;Y_\Delta) + \log(\Delta) \bigr\} \\
	&\leq& \limsup_{\Delta \to 0} \bigl\{ H(W_\Delta) + \log(\Delta) \bigr\} - \liminf_{\Delta \to 0} I(W_\Delta;Y_\Delta).   \IEEEeqnarraynumspace\IEEEyesnumber\label{entropy_limit}
\end{IEEEeqnarray*}
For the first limit, we have
\begin{IEEEeqnarray*}{rCl}
	\lim_{\Delta \to 0} \bigl\{ H(W_\Delta) + \log(\Delta) \bigr\}
	&\stackrel{(a)}{=}& \lim_{\Delta \to 0} \bigl\{ H(W'_\Delta) + \log(\Delta) \bigr\} \\
	&\stackrel{(b)}{=}& \lim_{\Delta \to 0} \bigl\{ H(\lceil W' \rfloor_\Delta) + \log(\Delta) \bigr\} \\
	&\stackrel{(c)}{=}& h(W') \\
	&=& h(W) - \log\gcd(\ab)   \IEEEyesnumber\label{first_limit}
\end{IEEEeqnarray*} where step $(a)$ follows from scale invariance of discrete entropy, step $(b)$ is due to Lemma~\ref{lem:Makkuva_Wu}, and step $(c)$ is due to Lemma~\ref{lem:Renyi}.

For the second limit, we will prove that $(W_\Delta,Y_\Delta)$ converges in distribution to $(W,Y)$ as $\Delta \to 0$. This convergence will imply, by the lower semi-continuity of relative entropy \cite[Thm.~1]{Posner1975}, \cite[Thm.~19]{Erven--Harremoes2014} that
\begin{equation}
	\liminf_{\Delta \to 0} I(W_\Delta;Y_\Delta) \geq I(W;Y)   \IEEEyesnumber\label{second_limit}
\end{equation}
which, combined with \eqref{entropy_limit}--\eqref{second_limit}, will conclude the proof of \eqref{quod_est_2}. To prove this weak convergence property, first observe that the pair of quantized variables $\lceil \Uv \rfloor_\Delta = (\lceil U_1 \rfloor_\Delta,\lceil U_2 \rfloor_\Delta)$ converges in probability (and hence in distribution) to the unquantized pair $\Uv = (U_1,U_2)$.

Since by assumption, we have that for almost all $\uv$ belonging to the support of $\Uv$, the family of cdfs $F_{Y|\Uv}(\cdot|\uv)$ is continuous in $\uv$ (in the sense of weak convergence of random variables), it follows by the Portmanteau Theorem~\cite[Theorem~2.8.1]{AsDo00} that for any continuous and bounded $\varphi \colon \mathbb{R}^3 \to \mathbb{R}$, the associated function
\begin{IEEEeqnarray*}{rCl}
	\tilde{\varphi}(\uv)
	&\triangleq& \E\bigl[ \varphi(Y,\uv) \big| \Uv = \uv \bigr] \\
	&=& \int \varphi(\uv,y) F_{Y|\Uv}(\mathrm{d}y|\uv)
\end{IEEEeqnarray*}
is continuous almost everywhere and bounded.
It further follows that the pair $(\lceil \Uv \rfloor_\Delta,Y_\Delta)$ converges in distribution to $(\Uv,Y)$ as $\Delta \to 0$, because for any continuous bounded function $\varphi \colon \mathbb{R}^3 \to \mathbb{R}$, we have
\begin{IEEEeqnarray*}{rCl}
	\lim_{\Delta \to 0} \E[ \varphi(\lceil \Uv \rfloor_\Delta,Y_\Delta)]
	&=& \lim_{\Delta \to 0} \int \varphi(\uv,y) F_{\lceil \Uv \rfloor_\Delta,Y_\Delta}(\mathrm{d}\uv,\mathrm{d}y) \\
	&\stackrel{(a)}{=}& \lim_{\Delta \to 0} \int \left( \int \varphi(\uv,y) F_{Y|\Uv}(\mathrm{d}y|\uv) \right) F_{\lceil \Uv \rfloor_\Delta}(\mathrm{d}\uv) \\
	&=& \lim_{\Delta \to 0} \int \tilde{\varphi}(\uv) F_{\lceil \Uv \rfloor_\Delta}(\mathrm{d}\uv) \\
	&\stackrel{(b)}{=}&  \E[\varphi(\Uv,Y)].
\end{IEEEeqnarray*}
Here, equality $(a)$ holds by Fubini's Theorem, which is applicable since $\varphi$ is bounded and the integrals are taken with respect to probability measures; equality $(b)$ holds because $\tilde{\varphi}$ is continuous and bounded (as argued above), and $\lceil \Uv \rfloor_\Delta$ converges in distribution to $\Uv$, which by assumption is absolutely continuous. In particular, if we set $\varphi$ to be any function of the form
\begin{equation*}
	\varphi(\uv,y) = \psi(a'_1 u_1 + a'_2 u_2, y)
\end{equation*}
with an arbitrary continuous bounded function $\psi$, it will hold that
\begin{IEEEeqnarray*}{rCl}
	\lim_{\Delta \to 0} \E[\psi(W_\Delta,Y_\Delta)]
	&=& \lim_{\Delta \to 0} \E[\psi(a'_1\lceil U_1 \rfloor_\Delta + a'_2\lceil U_2 \rfloor_\Delta,Y_\Delta)] \\
	&=& \lim_{\Delta \to 0} \E[\varphi(\lceil \Uv \rfloor_\Delta,Y_\Delta)] \\
	&=& \E[\varphi(\Uv,Y)] \\
	&=& \E[\psi(W,Y)].
\end{IEEEeqnarray*}
Hence, $(W_\Delta,Y_\Delta)$ tends in distribution to $(W,Y)$, which concludes the proof of \eqref{quod_est_2}.

Thus far, we have proven Theorem~\ref{thm:comp-real} for the case where $U_1$ and $U_2$ are compactly supported. To relax this assumption, it suffices to show that for arbitrarily supported $(U_1,U_2)$, the differential entropies $h(U_1)$, $h(U_2)$ and $h(W|Y)$ can be represented as the limiting differential entropies of sequences of compactly supported variables. For this purpose, consider arbitrarily supported variables $U_1 \in \mathbb{R}$ and $U_2 \in \mathbb{R}$ complying with the assumptions set forth by Theorem~\ref{thm:comp-real}, and their respective truncated versions $\langle U_1 \rangle_{\tau}$ and $\langle U_2 \rangle_\tau$ with pdfs defined as follows:
\begin{IEEEeqnarray*}{rCl}
	f_{\langle U_1 \rangle_\tau}(u_1)
	&\triangleq& f_{U_1}(u_1) \frac{\ind\{|u_1| < \tau\}}{\P\{|U_1| < \tau\}} \\
	f_{\langle U_2 \rangle_\tau}(u_2)
	&\triangleq& f_{U_2}(u_2) \frac{\ind\{|u_2| < \tau\}}{\P\{|U_2| < \tau\}} 
\end{IEEEeqnarray*} 
where $\ind\{\cdot\}$ represents the indicator function.
Let us further define
\begin{equation*}
	W'_\tau
	\triangleq a'_1 \langle U_1 \rangle_\tau + a'_2 \langle U_2 \rangle_\tau
\end{equation*}
and let $Y_\tau$ denote the output variable induced by the truncated auxiliaries $\langle U_1 \rangle_\tau$ and $\langle U_2 \rangle_\tau$.\footnote{We commit a slight abuse of notation here, since $W'_\Delta$ and $Y_\Delta$, defined earlier, have a different meaning than $W'_\tau$ and $Y_\tau$.} That is, conditional on $(\langle U_1 \rangle_\tau, \langle U_2 \rangle_\tau) = (u_1,u_2)$, the variable $Y_\tau$ is distributed as $Y_\tau \sim P_{Y|U_1,U_2}(\cdot|u_1,u_2)$. Then the following holds:
\begin{lemma}[Truncation]   \label{lem:truncation}
In the limit as $\tau \to \infty$, the following holds:
\begin{subequations}
\begin{IEEEeqnarray}{rCl}
	h(U_1)
	&=& \lim_{\tau \to \infty} h(\langle U_1 \rangle_\tau)   \label{truncation_1} \\
	h(U_2)
	&=& \lim_{\tau \to \infty} h(\langle U_2 \rangle_\tau)   \label{truncation_2} \\
	h(W'|Y)
	&\geq& \limsup_{\tau \to \infty} h(W'_\tau|Y_\tau).   \label{truncation_3}
\end{IEEEeqnarray}
\end{subequations}
\end{lemma}
\begin{IEEEproof}
The first two equalities can be proven by standard arguments. In fact, they follow directly from \cite[Lem.~2]{Makkuva--Wu2016}. As to the inequality \eqref{truncation_3}, the joint cdf of $W'_\tau$ and $Y_\tau$ is expressible as
\begin{IEEEeqnarray*}{rCl}
	F_{W'_\tau,Y_\tau}(w,y)
	&=& \P\{ W'_\tau \leq w, Y_\tau \leq y \} \\
	&=& \iint\limits_{[-\tau,\tau]^2} \P\left\{ W'_\tau \leq w, Y_\tau \leq y \middle| \langle U_1 \rangle_\tau = u_1, \langle U_2 \rangle_\tau = u_2 \right\} f_{\langle U_1 \rangle_\tau}(u_1) f_{\langle U_2 \rangle_\tau}(u_2) \intd u_1 \intd u_2 \\
	&=& \iint\limits_{[-\tau,\tau]^2} \ind\{a'_1 u_1 + a'_2 u_2 \leq w\} F_{Y|\Uv}(y|\uv) \frac{f_{U_1}(u_1)}{\P\{ |U_1| < \tau \}} \frac{f_{U_2}(u_2)}{\P\{ |U_2| < \tau \}} \intd u_1 \intd u_2.
\end{IEEEeqnarray*}
Hence, the joint cdf converges pointwise on the continuity set, because
\begin{IEEEeqnarray*}{rCl}
	\lim_{\tau \to \infty} F_{W'_\tau,Y_\tau}(w,y)
	&=& \iint\limits_{\mathbb{R}^2} \ind\{a'_1 u_1 + a'_2 u_2 \leq w\} F_{Y|U_1,U_2}(y) f_{U_1}(u_1) f_{U_2}(u_2) \intd u_1 \intd u_2 \\
	&=& F_{W',Y}(w,y)
\end{IEEEeqnarray*}
for each point $(w,y)$ at which $F_{W',Y}$ is continuous.
It follows in particular that the marginals converge weakly, i.e.,
\begin{IEEEeqnarray*}{rCl}
	\lim_{\tau \to \infty} F_{W'_\tau}(w)
	&=& F_{W'}(w) \\
	\lim_{\tau \to \infty} F_{Y_\tau}(y)
	&=& F_Y(y)
\end{IEEEeqnarray*}
for all $w$ and $y$ being continuity points of $F_{W'}$ and $F_Y$, respectively.
Consequently, the joint distribution and the product distribution of marginals converge as
\begin{IEEEeqnarray*}{rCl}
	P_{W'_\tau,Y_\tau} &\xrightarrow[]{\tau \to \infty}& P_{W',Y} \\
	P_{W'_\tau} \times P_{Y_\tau} &\xrightarrow[]{\tau \to \infty}& P_W' \times P_Y
\end{IEEEeqnarray*}
in the sense of weak convergence. Since relative entropy is lower semi-continuous in the weak topology \cite[Theorem~1]{Posner1975}, \cite[Theorem~19]{Erven--Harremoes2014}, it follows that
\begin{equation}
	\liminf_{\tau \to \infty} I(W'_\tau;Y_\tau)
	\geq I(W';Y)
	= I(W;Y).
\end{equation}
By \cite[Lemma~2]{Makkuva--Wu2016}, we further know that $\lim_{\tau \to \infty} h(W'_\tau) = h(W')$. It thus follows that
\begin{IEEEeqnarray*}{rCl}
	\liminf_{\tau \to \infty} \bigl\{ h(\langle U_1 \rangle_\tau) - h(W'_\tau|Y_\tau) \bigr\}
	&=& \liminf_{\tau \to \infty} \bigl\{ h(\langle U_1 \rangle_\tau) - h(W'_\tau) + I(W'_\tau;Y_\tau) \bigr\} \\
	&\geq& h(U_1) - h(W'|Y).
\end{IEEEeqnarray*}
which concludes the proof of \eqref{truncation_3} and hence the proof of Lemma~\ref{lem:truncation}.
\end{IEEEproof}
It follows from Lemma~\ref{lem:truncation} that the compactness assumption on the support sets of $U_1$ and $U_2$ can be removed, which establishes Theorem~\ref{thm:comp-real}.%###}}}

\section{Proofs of Theorem~\ref{thm:K-2-comp} and Theorem~\ref{thm:mac}}\label{sec:k-mac-proof}

\subsection{Proof of Theorem~\ref{thm:K-2-comp}}\label{sec:k-2-comp-proof}
Fix $\Fq$, pmf $\prod_{k=1}^Kp(u_k)$, and functions $x_k(u_k)$, $k\in[1:K]$.
The codebook construction and encoding steps follow the nested linear coding architecture in Section~\ref{sec:arc}.

\noindent{\bf Decoder.} Let $\e'<\e$. Upon receiving $y^n$, the decoder finds a unique index pair $(s_{\ab_1}, s_{\ab_2})$, such that
\begin{align*}
	(w_{\ab_1}^n(s_{\ab_1}), w_{\ab_2}^n(s_{\ab_2}), y^n)\in\aep,
\end{align*}
for some $s_{\ab_1}\in[2^{n\Rt(\ab_1)}]$ and  $s_{\ab_2}\in[2^{n\Rt(\ab_2)}]$, where $w_{\ab_1}^n(s_{\ab_1})$ and $w_{\ab_2}^n(s_{\ab_2})$ are defined in~\eqref{eqn:linearcomb} and $\Rt(\ab_1)$ and $\Rt(\ab_2)$ are defined in~\eqref{eq:A-rates}. If there is no such index pair, or more than one, the decoder declares an error.
\medskip

\noindent{\bf Analysis of the probability of error.} 
In the following analysis, we will omit some steps which are simple extensions of the proof steps in the previous section.
Let $M_1,\ldots,M_K$ be the chosen messages, $L_1,\ldots,L_K$ be the indices chosen by the encoders, and $S_{\ab_1}$, $S_{\ab_2}$ be the indices of the desired linear combinations $W_{\ab_1}^n(S_{\ab_1})$, $W_{\ab_2}^n(S_{\ab_2})$.

Then, the decoder makes an error only if one or more of the following events occur,
\begin{align*}
\Ec_1&=\{U_k^n(m_k, l_k) \not\in\aepvar \text{ for all } l_k, \text{ for some } m_k, k\in[1:K]\},\\
\Ec_2&=\{(W_{\ab_1}^n(S_{\ab_1}), W_{\ab_2}^n(S_{\ab_2}), Y^n)\not\in\aep\},\\
\Ec_3&=\{(W_{\ab_1}^n(s_{\ab_1}), W_{\ab_2}^n(s_{\ab_2}),  Y^n) \in \aep \text{ for some } (s_{\ab_1}, s_{\ab_2})\neq (S_{\ab_1}, S_{\ab_2}) \}.
\end{align*}
Then, by the union of events bound,
\begin{align}
\P(\Ec) &\le \P(\Ec_1)+\P(\Ec_2\cap\Ec_1^c)+\P(\Ec_3\cap\Ec_1^c).
\end{align}
By Lemma~\ref{lem:mm-covering} in Appendix~\ref{app:mm-packing-covering}, the probability $\P(\Ec_1)$ tends to zero as $n\to\infty$ if
\begin{align}
\Rh_k>  D(p_{U_k}\|p_\q)+\d(\e'),\quad k=1,\ldots,K. \label{eq:kcomp-cover}
\end{align} 
Define $\Mc=\{M_1=\cdots=M_K=0, L_1=\cdots=L_K=0\}$ as the event where all messages are zero and the chosen auxiliary indices are zero as well. Note that, conditioned on the event $\Mc$, the correct indices are zero, $S_{\ab_1}=S_{\ab_2}=0$. By symmetry of the codebook construction and encoding steps, we have that $P(\Ec_2\cap\Ec_1^c)=P(\Ec_2\cap\Ec_1^c|\Mc)$ and $P(\Ec_3\cap\Ec_1^c)=P(\Ec_3\cap\Ec_1^c|\Mc)$.

By Lemma~\ref{lem:ML-LC} in Appendix~\ref{app:ML-LC} and the conditional typicality lemma~\cite[\S 2.5]{El-Gamal--Kim2011} the probability $\P(\Ec_2\cap\Ec_1^c| \Mc)$ tends to zero as $n\to\infty$ if~\eqref{eq:kcomp-cover} is satisfied.
Define 
\begin{align*}
{\tilde{\Ec}(s_{\ab_1}, s_{\ab_2})} &={\{(W_{\ab_1}^n(s_1), W_{\ab_2}^n(s_2),  Y^n) \in \aep,}\\
&\qquad {U^n_j(0,0)\in\aepvar, j\in[1:K]} \},
\end{align*} and partitions of the index pairs by
\begin{align*}
\Ac &=\{(s_{\ab_1}, s_{\ab_2}): (s_{\ab_1}, s_{\ab_2})\neq(0,0)\},\\
\Ac_{1} &=\{(s_{\ab_1}, s_{\ab_2}): s_{\ab_1}\neq0, s_{\ab_2}=0 \},\\
\Ac_{2} &=\{(s_{\ab_1}, s_{\ab_2}): s_{\ab_1}=0, s_{\ab_2}\neq0 \},\\
\Ac_{12} &=\{(s_{\ab_1}, s_{\ab_2}): s_{\ab_1}\neq 0, s_{\ab_2}\neq0 \},\\
\Lc &=\{(s_{\ab_1}, s_{\ab_2}) \in\Ac_{12}: \etab({s}_{a_1}), \etab({s}_{a_2}) \text{ are linearly dependent}\},\\
\Lc^c &=\{(s_{\ab_1}, s_{\ab_2}) \in\Ac_{12}: \etab({s}_{a_1}), \etab({s}_{a_2}) \text{ are linearly independent}\}.
\end{align*}
Furthermore, for $\bb\in\Fq^2$, $\bb\neq \mathbf{0}$, define the sets
\begin{align}
{\Lc}_1(\bb) &=\{(s_{\ab_1}, s_{\ab_2})\in\Lc:  b_{1}\etab({s}_{a_1})\oplus b_{2}\etab({s}_{a_2}) \neq \mathbf{0} \},\label{eq:def-lc1}\\
{\Lc}_2(\bb) &=\{(s_{\ab_1}, s_{\ab_2})\in\Lc:  b_{1}\etab({s}_{a_1})\oplus b_{2}\etab({s}_{a_2}) = \mathbf{0} \}. \label{eq:def-lc2}
\end{align}
Note that, for any $\bb\in\Fq^2$ that is not the all-zero vector, we have
\begin{align*}
&\Ac \subseteq (\Ac_{1} \cup \Ac_{2} \cup \Ac_{12}),\\
&\Ac_{12}=\Lc \cup \Lc^c,\\
&\Lc={\Lc}_1(\bb) \cup {\Lc}_2(\bb),
\end{align*}
and thus, $\Ac = (\Ac_{1} \cup \Ac_{2} \cup \Lc^c \cup {\Lc}_1(\bb) \cup {\Lc}_2(\bb))$.
Furthermore, the cardinality of these sets can be upper bounded by
\begin{align}
|\Ac_1| &\le 2^{n\Rt(\ab_1)}, \nn\\
|\Ac_2| &\le 2^{n\Rt(\ab_2)}, \nn\\
|\Ac_{12}| &\le 2^{n(\Rt(\ab_1)+\Rt(\ab_2))}, \nn\\
|\Lc| &\le \q2^{n\min(\Rt(\ab_1), \Rt(\ab_2))}. \label{eq:cardinality}
\end{align}

Then, 
\begin{align}
&\P(\Ec_3\cap\Ec_1^c|\Mc) =\nn\\
& \P\{\tilde{\Ec}(s_{\ab_1}, s_{\ab_2})\text{ for some } (s_{\ab_1}, s_{\ab_2})\in\Ac | \Mc \}\nn\\
&\le \sum_{(s_{\ab_1}, s_{\ab_2})\in\Ac}\P\{\tilde{\Ec}(s_{\ab_1}, s_{\ab_2})\big|\Mc\}\nn\\
&\le \sum_{(s_{\ab_1}, s_{\ab_2})\in\Ac_1}\P\{\tilde{\Ec}(s_{\ab_1}, s_{\ab_2})\big|\Mc\}+\sum_{(s_{\ab_1}, s_{\ab_2})\in\Ac_2}\P\{\tilde{\Ec}(s_{\ab_1}, s_{\ab_2})\big|\Mc\}\nn\\
&~~+\sum_{(s_{\ab_1}, s_{\ab_2})\in\Lc^c}\P\{\tilde{\Ec}(s_{\ab_1}, s_{\ab_2})\big|\Mc\}+\sum_{(s_{\ab_1}, s_{\ab_2})\in\Lc_1(\bb)}\P\{\tilde{\Ec}(s_{\ab_1}, s_{\ab_2})\big|\Mc\}\nn\\
&~~+\sum_{(s_{\ab_1}, s_{\ab_2})\in\Lc_2(\bb)}\P\{\tilde{\Ec}(s_{\ab_1}, s_{\ab_2})\big|\Mc\}. \label{eq:error}
\end{align}
Let $\tilde{D}_U=D(p_{U_1}\|p_\q)+\cdots+D(p_{U_K}\|p_\q)$ and define
\begin{align*}
V_{\bb}=b_{1}W_{\ab_1}\oplus b_{2}W_{\ab_2},\\
V_{\cb}=c_{1} W_{\ab_1}\oplus c_{2} W_{\ab_2},
\end{align*}
where $\cb=[c_{1} ,\, c_{2}]\in\Fq^2$ is a non-zero vector that is linearly independent of $\bb$.

By the cardinality bounds in \eqref{eq:cardinality} and by closely following the steps in Lemma~\ref{lem:jtl-linear} (by replacing $U_k$ with $W_{\ab_k}$, $k=1,2$, replacing $W_{\bb}$ with $V_{\bb}$, and replacing $W_{\cb}$ with $V_{\cb}$), the probability terms in~\eqref{eq:error} tend to zero as $n\to\infty$ if
\begin{align}
\Rt(\ab_1) + \Rh_\Sigma &< I(W_{\ab_1}; Y,W_{\ab_2})+D(p_{W_{\ab_1}}\|p_\q)+\tilde{D}_U-\d(\e),\\ 
\Rt(\ab_2) + \Rh_\Sigma &< I(W_{\ab_2}; Y,W_{\ab_1})+D(p_{W_{\ab_2}}\|p_\q)+\tilde{D}_U-\d(\e),\\
\Rt(\ab_1) +\Rt(\ab_2) + \Rh_\Sigma &< I(W_{\ab_1}, W_{\ab_2}; Y)+I(W_{\ab_1}; W_{\ab_2})\\
&\quad+D(p_{W_{\ab_1}}\|p_\q)+D(p_{W_{\ab_2}}\|p_\q)+\tilde{D}_U-\d(\e),\\
\min\bigl(\Rt(\ab_1), \Rt(\ab_2)\bigr)+ \Rh_\Sigma  &< I(V_{\bb}; Y)+D(p_{V_{\bb}}\|p_\q)+\tilde{D}_U-\d(\e),\\
\min\bigl(\Rt(\ab_1), \Rt(\ab_2)\bigr)+ \Rh_\Sigma &< I(V_{\cb}; Y, V_{\bb})+D(p_{V_{\cb}}\|p_\q)+\tilde{D}_U-\d(\e),
\end{align}
where $\Rh_\Sigma=\Rh_1+\cdots+\Rh_K$. Finally, the rate region in Theorem~\ref{thm:K-2-comp} is established by eliminating the auxiliary rates by choosing $\Rh_k=D(p_{U_k}\|p_\q)+2\d(\e')$, $k\in[1:K]$ to satisfy~\eqref{eq:kcomp-cover}, using the relation~\eqref{eqn:relation}, following the steps in Appendix~\ref{app:equiv-rr} to simplify the rate region expression into the form without $V_{\cb}$, and taking $\e\to 0$. This concludes the proof for $(\Field, \mathbb{A})=(\Fq, \Fq)$. 

Finally, by Lemma~\ref{lem:translation}, we can find a large enough $\q$ such that the linear combinations in \eqref{eq:thm4eq1}, \eqref{eq:thm4eq2}, and \eqref{eq:thm4eq3}, can be translated to linear combinations in $(\Real, \Integer)$, which concludes the proof of Remark~\ref{rmk:extension}.

\subsection{Proof of Theorem~\ref{thm:mac}}\label{sec:mac-proof}

First, note that by the achievability proof of Theorems~\ref{thm:comp-Fq} and~\ref{thm:comp-real-discrete}, there exists a sequence of nested linear coding architectures with rates $(R_1, R_2)$ that are achievable for computing $(\Fq, \Fq)$ and $(\Real, \Integer_\q)$ linear combinations. Moreover, note that the rate region in Theorem~\ref{thm:K-2-comp} simplifies to $\Rc_{\mathsf{LMAC}}$ when specialized to the case $K=2$ and $\Ab$ is the identity matrix. Thus, by the achievability proof of Theorem~\ref{thm:K-2-comp}, the same nested linear coding architecture recovers the message pair, if the sequence of codes have rate pairs $(R_1, R_2)\in \Rc_{\mathsf{LMAC}}$. To prove the theorem for $(\Real, \Integer)$ computation codes (Theorem~\ref{thm:comp-real}), the same quantization method in Section~\ref{sec:comp-real} applies to the rate region $\Rc_{\mathsf{LMAC}}$.

\section{Concluding Remarks}

Looking ahead, the framework of joint typicality is a promising approach for exploring the performance of random structured codes. Here, we have generalized prior work on Gaussian compute--forward and developed a compute--forward framework for memoryless MACs where the goal is either to recover a linear combination over $\Fq$ or an integer-linear combination of real-valued codewords. Furthermore, we have analyzed the performance of simultaneous joint typicality decoding for recovering two linear combinations. As discussed in Remark~\ref{rem:tupledecoding}, an open problem is to extend our analysis of simultaneous joint typicality decoding from recovering pairs of messages to recovering more than two messages.

\section*{Acknowledgments}
The authors would like to thank Aditya Gangrade, Young-Han Kim, Olivier L\'ev\^eque and Or Ordentlich for helpful discussions.

%%%%%%%%%%%%%%%%%%%%%%%%%%%%%%%%%%%%%%%%%%%%%%%%%
%
%			APPENDICES
%
%%%%%%%%%%%%%%%%%%%%%%%%%%%%%%%%%%%%%%%%%%%%%%%%%%%

\appendices

\section{Joint typicality lemma for\\ mismatched distributions} \label{app:mm-jtl}

\begin{lemma}\label{lem:mm-dist}
Let $X \sim p_{X}(x)$ and let $\tilde{p}_X(x)$ be another distribution on $\Xc$ such that $D_X=D(p_X\|\tilde{p}_X)<\infty$.
Then, for $x^n\in\aep(X)$,
\begin{align}
2^{-n(D_X+H(X)+\d(\e))}\le \prod_{i=1}^n \tilde{p}_X(x_i) &\le 2^{-n(D_X+H(X)-\d(\e))}.\label{eq:qx-bound}
\end{align}
\end{lemma}

\begin{IEEEproof}
To prove the first statement, observe that, $\prod_{i=1}^n \tilde{p}_X(x_i)=\prod_{x\in\Xc}\tilde{p}_X(x)^{n \pi(x|x^n)}$, where recall that $\pi(x|x^n)$ is the empirical pmf of $x^n$.
Then,
\begin{align*}
\log \tilde{p}_X(x^n) &= \sum_{x\in \Xc} n \pi(x|x^n)\log \tilde{p}_X(x) \nonumber\\
&= \sum_{x\in \Xc} n( \pi(x|x^n)-p_X(x)+p_X(x))\log \tilde{p}_X(x)\nonumber\\
&= n \sum_{x\in \Xc}p_X(x)\log \tilde{p}_X(x) - n\sum_{x\in \Xc}\left( \pi(x|x^n)-p_X(x)\right)(-\log \tilde{p}_X(x))\\
&= -n\left( D(p_X\|\tilde{p}_X) +H(X) \right) - n\sum_{x\in \Xc}\left( \pi(x|x^n)-p_X(x)\right)(-\log \tilde{p}_X(x)).
\end{align*}
Since $x^n\in \aep(X)$,
\begin{align*}
&\left|\sum_{x\in\Xc} \left(\pi(x|x^n)-p_X(x)\right)(-\log \tilde{p}_X(x))\right|\\
 &\le \sum_{x\in\Xc} \left| \pi(x|x^n)-p_X(x)\right|(-\log \tilde{p}_X(x)) \\
&\le -\e\sum_{x\in\Xc} p_X(x) \log \tilde{p}_X(x) \\
&= \e (D(p_X\|\tilde{p}_X)+H(X))
\end{align*}
\end{IEEEproof}
\smallskip

\begin{lemma}\label{lem:jtl}
Let $(X, Y)\sim p_{X, Y}(x, y)$ and $\tilde{p}_X(x)$ be another distribution on $\Xc$ such that $D(p_X\|\tilde{p}_X)<\infty$. Let $\e'<\e$. Then, there exists $\d(\e)>0$ that tends to zero as $\e\to 0$ such that the following statement holds:
\begin{enumerate}
\item If $\yt^n$ is an arbitrary sequence and $\Xt^n\sim \prod_{i=1}^n \tilde{p}_{X}(\xt_i)$, then
\begin{align*}
&\P\{( \Xt^n, \yt^n)\in\aep(X, Y)\}\\
& \le 2^{-n(I(X;Y)+D(p_X\|\tilde{p}_X)-\d(\e))}
\end{align*}
\item If $\yt^n\in\aepvar(Y)$ and $\Xt^n\sim \prod_{i=1}^n \tilde{p}_{X}(\xt_i)$, then for $n$ sufficiently large,
\begin{align*}
&\P\{( \Xt^n, \yt^n)\in\aep(X, Y)\} \\
&\ge 2^{-n(I(X;Y)+D(p_X\|\tilde{p}_X)+\d(\e))}
\end{align*}
\end{enumerate}
\end{lemma} 
The proof follows from Lemma~\ref{lem:mm-dist} and standard cardinality bounds on the conditional typical set $\aep(X|y^n)$.

\section{Packing and covering lemmas for\\ mismatched distributions}\label{app:mm-packing-covering}\label{app:mm-covering-packing}

\begin{lemma}[Mismatched Covering Lemma] \label{lem:mm-covering}
Let $(X,\Xh)\sim p_{X, \Xh}(x,\xh)$ and $\tilde{p}_\Xh(\xh)$ be a distribution on $\hat{\Xc}$ such that $D(p_{\Xh}\|\tilde{p}_{\Xh})<\infty$. Let $X^n$ be a random sequence with $\lim_{n\to\infty}\P\{X^n\in\aep(X)\}=1$ and let $\Xt^n(m),\, m \in \Cc$, where $|\Cc| \ge 2^{nR}$, be pairwise independent and independent of $X^n$, each distributed according to $\prod_{i=1}^n \tilde{p}_\Xh(\xt_i)$. Then, there exists a $\d(\e)$ that tends to zero as $\e\to 0$ such that
\begin{align*}
\lim_{n\to\infty} \P\{(X^n, \Xt^n(m))\not \in\aep(X, \Xh) \text{ for all } m\in\Cc\} =0,
\end{align*}
if $R>I(X; \Xh)+D(p_\Xh\|\tilde{p}_\Xh)+\d(\e)$.
\end{lemma}
\begin{IEEEproof}
Let $\Ac = \{m\in[1:2^{nR}]: (X^n, \Xt^n(m))\in\aep(X, \Xh)\}$. Then, by the Chebyshev lemma,
\begin{align*}
\P\{|\Ac|=0\} & \le \frac{\var(|\Ac|)}{(\E|\Ac|)^2}.
\end{align*}
For $m\in[1:2^{nR}]$, define the indicator random variables
\begin{align*}
E(m) = \begin{cases}
1 & \mbox{ \text{if } $(X^n, \Xt^n(m))\in\aep(X, \Xh)$,}\\
0 & \mbox{ \text{otherwise},}
\end{cases}
\end{align*}
and let $p_1 := \P\{E(1)=1\}$ and $p_2 := \P\{E(1)=1, E(2)=1\}= p_1^2$.
Then,
\begin{align*}
\E(|\Ac|) &= \sum_{m} \P\{ (X^n, \Xt(m)) \in \aep(X, \Xh)\} = 2^{nR} p_1,\\
\E(|\Ac|^2) &= \sum_{m} \P\{ (X^n, \Xt(m)) \in \aep(X, \Xh)\}\\
&\quad+\sum_{m}\sum_{m'\neq m} \P\{ (X^n, \Xt(m)) \in \aep(X, \Xh),\\[-1em]
&\qquad\qquad\qquad\quad~ (X^n, \Xt(m')) \in \aep(X, \Xh)\}\\
&\le 2^{nR} p_1+2^{n2R} p_2.
\end{align*}
Thus, $\var(|\Ac|) \leq 2^{nR} p_1$.
From Lemma~\ref{lem:jtl}, for sufficiently large $n$, we have
\begin{align*}
p_1 &\le 2^{-n(I(X; Y)+D(p_X\|\tilde{p}_X)-\d(\e))},\\
p_1 &\ge 2^{-n(I(X; Y)+D(p_X\|\tilde{p}_X)+\d(\e))},
\end{align*}
and hence,
\begin{align*}
\frac{\var(|\Ac|)}{(\E|\Ac|)^2} \le 2^{-n(R-I(X; Y)-D(p_X\|\tilde{p}_X)-\d(\e))},
\end{align*}
which tends to zero as $n\to\infty$ if
\begin{align*}
R> I(X; Y)+D(p_X\|\tilde{p}_X)+\d(\e).
\end{align*}
\end{IEEEproof}

\begin{lemma}[Mismatched Packing Lemma] \label{lem:mm-packing}
Let $(X, Y)\sim p_{X, Y}(x, y)$ and $\tilde{p}_X(x)$ be a distribution on $\Xc$ such that $D(p_{X}\|\tilde{p}_{X})<\infty$. Let $\Yt^n$ be an arbitrarily distributed random sequence, and $\Xt^n(m),\, m \in \Cc$, where $|\Cc| \le 2^{nR}$ and each sequence is distributed according to $\prod_{i=1}^n \tilde{p}_X(x_i)$. Further assume that $\Xt^n(m), m\in\Cc$ is pairwise independent of $\Yt^n$, but is arbitrarily dependent on other $\Xt^n$ sequences. Then, there exists $\d(\e)$ that tends to zero as $\e\to 0$ such that
\begin{align*}
\lim_{n\to\infty} \P\{(\Xt^n(m),\Yt^n)\in\aep(X, Y) \text{ for some } m\in\Cc\} =0,
\end{align*}
if $R<I(X; Y)+D(p_X\|\tilde{p}_X)-\d(\e)$.
\end{lemma}
The proof of this lemma follows directly from the union of events bound and Lemma~\ref{lem:jtl}.

\section{Lemma~\ref{lem:message}} \label{app:message}

\begin{lemma}\label{lem:message}
Let $\Mc=\{M_k=0, L_k=0, k\in[1:K]\}$ and $\Ac$ be an arbitrary event that is independent of the event $\{M_1=0, \ldots, M_K=0\}$. Then,
\begin{align*}
\P(\Ac|\Mc) \le 2^{n(\Rh_1\,+\,\cdots\,+\,\Rh_K)}\P(\Ac).
\end{align*}
\end{lemma}
\begin{IEEEproof}
From the relation
\begin{align*}
\P(\Ac|\Mc)=\frac{\P(\Mc| \Ac)}{\P(\Mc)}\P(\Ac),
\end{align*}
and
\begin{align*}
\P(\Mc |\Ac)&\le \P(M_1=0,\ldots, M_K=0 |\Ac)\\
&=\P(M_1=0,\ldots, M_K=0  ),
\end{align*}
it is sufficient to show that
\begin{align*}
\P(\Mc) &= \P(L_1=0,\ldots, L_K=0 |M_1=0,\ldots, M_K=0 )\P(M_1=0,\ldots, M_K=0 )\\
&=\frac{1}{2^{n(\Rh_1+\cdots+\Rh_K)}}\P(M_1=0,\ldots, M_K=0 ),\\
&= \frac{1}{2^{n(R_1+\cdots+R_K+\Rh_1+\cdots+\Rh_K)}},
\end{align*}
i.e., the tuple of messages and indices are uniformly distributed, which follows from the symmetry of the codebook construction. 
To be precise, in the following we will show that
\begin{align*}
\P(L_1=0,\ldots, L_K=0|M_1=0,\ldots, M_K=0) &= \P(L_1=l_1,\ldots, L_K=l_K|M_1=0,\ldots, M_K=0),
\end{align*}
for $(l_1,\ldots, l_K)\in [2^{n\Rh_1}]\times\cdots\times[2^{n\Rh_K}]$.

Let $\tilde\Mc=\{M_1=0,\ldots, M_K=0\}$. Then, we have
\begin{align}
\P&(L_k=l_k, k\in[1:K]| \tilde\Mc)\nn\\
&=\sum_{u^n_1,\ldots,u^n_K}\sum_{\Gb}\P(\mathbf{G}=\Gb, U^n_k(0, l_k)=u^n_k, L_k=l_k, k\in[1:K]|\tilde\Mc), \label{eq:st}
\end{align}
and
\begin{align}
&\P(\mathbf{G}=\Gb, U^n_k(0, l_k)=u^n_k, L_k=l_k, k\in[1:K]|\tilde\Mc)\nn\\
&=\P(\mathbf{G}=\Gb, \etab(0,l_k)\Gb\oplus D^n_k=u^n_k, L_k=l_k, k\in[1:K]|\tilde\Mc)\nn\\
&=\P(\mathbf{G}=\Gb, D^n_k=u^n_k\ominus \etab(0,l_k)\Gb, L_k=l_k, k\in[1:K]|\tilde\Mc)\nn\\
&\stackrel{(a)}{=}\P([U^n(0, l_k')=\etab(0, l_k')\Gb\oplus u^n_k\ominus \etab(0,l_k)\Gb: l_k'\neq l_k], \mathbf{G}=\Gb,\nn\\
&\qquad\qquad D^n_k=u^n_k\ominus \etab(0,l_k)\Gb, L_k=l_k, k\in[1:K]|\tilde\Mc)\nn\\
&=\P([U^n(0,l_k')=(\etab(0,l_k')\ominus \etab(0,l_k))\Gb\oplus u^n_k : l_k'\neq l_k], \mathbf{G}=\Gb,\nn\\
&\qquad\qquad  D^n_k=u^n_k\ominus \etab(0,l_k)\Gb, L_k=l_k, k\in[1:K]|\tilde\Mc)\nn\\
&=\P([\Uh^n(0,\lh_k)=\etab(0,\lh_k)\Gb\oplus u^n_k : \lh_k \neq 0], \mathbf{G}=\Gb, D^n_k=u^n_k, L_k=0, k\in[1:K]|\tilde\Mc)\nn\\
&=\P(\mathbf{G}=\Gb, D^n_k=u^n_k, L_k=0, k\in[1:K]|\tilde\Mc) \label{eq:L0} 
\end{align}
where $\Uh^n_k(0, \lh_k)$, $\lh_k\in[2^{n\Rh_k}]$ is a permuted codebook of $U^n(0, l_k')$, $\lh_k'\in[2^{n\Rh_k}]$ with respect to $l_k$ such that
$$\etab(0,\lh_k)=\etab(0,l_k')\ominus \etab(0,l_k),$$
and step $(a)$ follows from the fact that that $\Gb$ and $D^n_k=u^n_k\ominus \etab(0,l_k)\Gb$ determines the rest of the codewords.
Finally, plugging in \eqref{eq:L0} into \eqref{eq:st} completes the proof.
\end{IEEEproof}

%%%%%%%%%%%%%%%%%%%%%%%%%%%%%%%%%
%
%	SECTION: Proof of Lemma  lem:jtl-linear
%
%%%%%%%%%%%%%%%%%%%%%%%%%%%%%%%%%%

\section{Proof of Lemma~\ref{lem:jtl-linear}} \label{app:proof-jtl}

In this section, we prove Lemma~\ref{lem:jtl-linear}. 
We show upper bounds on $\P(\tilde{\Ec}(m_1, l_1, m_2, l_2)\big|\Mc)$ for the following cases.
\subsection{Case $(m_1, l_1, m_2, l_2)\in\Ac_1$:}
\begin{align*}
\P\{&(U^n_1(m_1, l_1), U^n_2(0,0), Y^n)\in\aep, U^n_1(0,0)\in\aepvar|\Mc\}\\
&\leq\sum_{\substack{(u^n_1,u^n_2):\\u^n_1\in\aep,u^n_2\in\aep}} \P\{(U^n_1(m_1, l_1), u^n_2, Y^n)\in\aep, U^n_1(0,0)=u^n_1, U^n_2(0,0)=u^n_2 |\Mc\}\\
&=\sum_{\substack{(u^n_1,u^n_2):\\u^n_1\in\aep,u^n_2\in\aep}} \sum_{\substack{(\ut^n_1, y^n):\\(\ut^n_1, u^n_2, y^n)\in\aep}}\P\{U^n_1(m_1, l_1) = \ut^n_1, Y^n=y^n, U^n_1(0,0)=u^n_1, U^n_2(0,0)=u^n_2 |\Mc\}\\
&\stackrel{(a)}{=}\sum_{\substack{(u^n_1,u^n_2):\\u^n_1\in\aep,u^n_2\in\aep}} \sum_{\substack{(\ut^n_1, y^n):\\(\ut^n_1, u^n_2, y^n)\in\aep}}\P\{Y^n=y^n| U^n_1(0,0)=u^n_1, U^n_2(0,0)=u^n_2, \Mc\}\\
&\quad\times \P\{U^n_1(m_1, l_1) = \ut^n_1, U^n_1(0,0)=u^n_1, U^n_2(0,0)=u^n_2 |\Mc\}\\
&\stackrel{(b)}{\le}\sum_{\substack{(u^n_1,u^n_2):\\u^n_1\in\aep,u^n_2\in\aep}} \sum_{\substack{(\ut^n_1, y^n):\\(\ut^n_1, u^n_2, y^n)\in\aep}}\P\{Y^n=y^n| U^n_1(0,0)=u^n_1, U^n_2(0,0)=u^n_2, \Mc\}\\
&\quad\times 2^{n(\Rh_1+\Rh_2)}\P\{U^n_1(m_1, l_1) = \ut^n_1, U^n_1(0,0)=u^n_1, U^n_2(0,0)=u^n_2\}\\
&\stackrel{(c)}{=}2^{n(\Rh_1+\Rh_2)}\sum_{\substack{(u^n_1,u^n_2):\\u^n_1\in\aep,u^n_2\in\aep}} \sum_{\substack{(\ut^n_1, y^n):\\(\ut^n_1, u^n_2, y^n)\in\aep}} p(y^n| u^n_1,u^n_2) p_\q(\ut_1^n)p_\q(u^n_1)p_\q(u^n_2)\\
&\le2^{n(\Rh_1+\Rh_2)}\sum_{\substack{(u^n_1,u^n_2):\\u^n_1\in\aep,u^n_2\in\aep}} \sum_{\substack{y^n:\\(u^n_2, y^n)\in\aep}} p(y^n| u^n_1,u^n_2)\sum_{\substack{\ut^n_1:\\(\ut^n_1, u^n_2, y^n)\in\aep}} p_\q(\ut_1^n) p_\q(u^n_1)p_\q(u^n_2)\\
&\le2^{n(\Rh_1+\Rh_2)}\sum_{\substack{(u^n_1,u^n_2):\\u^n_1\in\aep,u^n_2\in\aep}} \sum_{\substack{y^n:\\(u^n_2, y^n)\in\aep}} p(y^n| u^n_1,u^n_2) \\
&\quad\times 2^{n(H(U_1|Y, U_2)+\d(\e))} 2^{-n(2H(U_1)+2D(p_{U_1}\|p_\q))}2^{-n(H(U_2)+D(p_{U_2}\|p_\q))}\\
&\le2^{n(\Rh_1+\Rh_2)}\sum_{\substack{(u^n_1,u^n_2):\\u^n_1\in\aep,u^n_2\in\aep}} 2^{n(H(U_1|Y, U_2)+\d(\e))} 2^{-n(2H(U_1)+2D(p_{U_1}\|p_\q))}2^{-n(H(U_2)+D(p_{U_2}\|p_\q))}\\
&\le2^{n(\Rh_1+\Rh_2)} 2^{-n(I(U_1; Y, U_2)+D(p_{U_1}\|p_\q)-\d(\e))} 2^{-n(\tilde{D}_U-\d(\e))}\\
&\stackrel{(d)}{=} 2^{n(\Rh_1+\Rh_2)} 2^{-n(I(U_1; Y| U_2)+D(p_{U_1}\|p_\q)-\d(\e))} 2^{-n(\tilde{D}_U-\d(\e))}
%%%%%%%%%%%%%%%%%%%%%%%%%%%%%%%%%%%%%%%%%%%%%%%%%%%%%
\end{align*}
where step $(a)$ follows from the fact that conditioned on $\Mc$, we have the Markov relation $Y^n\to(U^n_1(0,0), U^n_2(0,0))\to U^n_1(m_1, l_1)$, step $(b)$ follows from Lemma~\ref{lem:message}, step $(c)$ follows from the independent construction of dithers $d^n_k$, $k=1,2,$ and \cite[Theorem~1]{Domb--Zamir--Feder2013}, and step $(d)$ follows from the independence of $U_1$ and $U_2$.

\subsection{Case $(m_1, l_1, m_2, l_2)\in\Ac_2$:}
By symmetry with the case above, 
\begin{align*}
\P\{&(U^n_1(0,0), U^n_2(m_2, l_2), Y^n)\in\aep, U^n_2(0,0)\in\aepvar |\Mc\}\\
&\le 2^{n(\Rh_1+\Rh_2)} 2^{-n(I(U_2; Y | U_1)+D(p_{U_2}\|p_\q)-\d(\e))} 2^{-n(\tilde{D}_U-\d(\e))}.
\end{align*}

\subsection{Case $(m_1, l_1, m_2, l_2)\in\Lc^c$:}
\begin{align*}
\P\{&(U^n_1(m_1, l_1), U^n_2(m_2, l_2), Y^n)\in\aep, U^n_k(0,0)\in\aepvar, k=1,2 |\Mc\}\\
&\le \sum_{\substack{(u^n_1,u^n_2):\\u^n_1\in\aep,u^n_2\in\aep}} \P\{(U^n_1(m_1, l_1), U^n_2(m_2, l_2), Y^n)\in\aep, U^n_k(0,0)=u^n_k, k=1,2 |\Mc\}\\
&\stackrel{(a)}{=}\sum_{\substack{(u^n_1,u^n_2):\\u^n_1\in\aep,u^n_2\in\aep}} \sum_{\substack{(\ut^n_1, \ut^n_2, y^n):\\(\ut^n_1, \ut^n_2, y^n)\in\aep}}\P\{Y^n=y^n| U^n_k(0,0)=u^n_k, k=1,2 , \Mc\}\\
&\quad\times P\{U^n_k(m_k, l_k) = \ut^n_k, U^n_k(0,0)=u^n_k, k=1,2 |\Mc\}\\
&\stackrel{(b)}{\le}\sum_{\substack{(u^n_1,u^n_2):\\u^n_1\in\aep,u^n_2\in\aep}} \sum_{\substack{(\ut^n_1, \ut^n_2, y^n):\\(\ut^n_1, \ut^n_2, y^n)\in\aep}}\P\{Y^n=y^n| U^n_k(0,0)=u^n_k, k=1,2 , \Mc\}\\
&\quad\times 2^{n(\Rh_1+\Rh_2)}\P\{U^n_k(m_k, l_k) = \ut^n_k, U^n_k(0,0)=u^n_k, k=1,2 \}\\
&\stackrel{(c)}{=}2^{n(\Rh_1+\Rh_2)}\sum_{\substack{(u^n_1,u^n_2):\\u^n_1\in\aep,u^n_2\in\aep}} \sum_{\substack{(\ut^n_1, \ut^n_2, y^n):\\(\ut^n_1, \ut^n_2, y^n)\in\aep}}p(y^n| u^n_1,u^n_2) \prod_{k=1}^2 p_\q(\ut_k^n)p_\q(u^n_k)\\
&\le2^{n(\Rh_1+\Rh_2)}\sum_{\substack{(u^n_1,u^n_2):\\u^n_1\in\aep,u^n_2\in\aep}} \sum_{\substack{y^n\in\aep}} p(y^n| u^n_1,u^n_2)\sum_{\substack{(\ut^n_1, \ut^n_2):\\(\ut^n_1, \ut^n_2, y^n)\in\aep}} \prod_{k=1}^2 p_\q(\ut_k^n)p_\q(u^n_k)\\
&\le 2^{n(\Rh_1+\Rh_2)}\sum_{\substack{(u^n_1,u^n_2):\\u^n_1\in\aep,u^n_2\in\aep}}  \sum_{\substack{y^n\in\aep}}  p(y^n| u^n_1,u^n_2) \\
&\quad\times 2^{n(H(U_1, U_2|Y)+\d(\e))} 2^{-n2(H(U_1)+D(p_{U_1}\|p_\q))}2^{-n2(H(U_2)+D(p_{U_2}\|p_\q))}\\
&=2^{n(\Rh_1+\Rh_2)}\sum_{\substack{(u^n_1,u^n_2):\\u^n_1\in\aep,u^n_2\in\aep}}  2^{n(H(U_1, U_2|Y)+\d(\e))} 2^{-n2(H(U_1)+D(p_{U_1}\|p_\q))}2^{-n2(H(U_2)+D(p_{U_2}\|p_\q))}\\
&\le2^{n(\Rh_1+\Rh_2)} 2^{-n(I(U_1, U_1; Y)-\d(\e))} 2^{-n(2 \tilde{D}_U-\d(\e))}
\end{align*}
where step $(a)$ follows from the fact that conditioned on $\Mc$, we have the Markov relation $Y^n\to(U^n_1(0,0), U^n_2(0,0))\to (U^n_1(m_1, l_1), U^n_2(m_2, l_2))$, step $(b)$ follows from Lemma~\ref{lem:message}, and step $(c)$ follows from the independent construction of dithers $d^n_k$, $k=1,2$ and statistical independence of linearly independent codewords \cite[Theorem~1]{Domb--Zamir--Feder2013}.

\subsection{Case $(m_1, l_1, m_2, l_2)\in\Lc_1(\bb)$:}

Let $W_{\bb}=b_{1}U_1\oplus b_{2} U_2$ and let $s_{\bb} \in [2^{n\Rt(\bb)}]$  be the index whose $\q$-ary expansion satisfies 
\begin{align}
[\nub(s_{\bb})~\mathbf{0}] = b_{1}\etab(m_1, l_1)\oplus b_{2}\etab(m_2, l_2) . 
\end{align} We can also uniquely associate each index $s_{\bb}$ with a linear combination of the codewords
\begin{align*}
W_{\bb}^n(s_{\bb}) :=b_{1}U^n_1(m_1, l_1)\oplus b_{2}U^n_2(m_2, l_2) . 
\end{align*}
Then,
\begin{align*}
\P\{&(U^n_1(m_1, l_1), U^n_2(m_2, l_2), Y^n)\in\aep, U^n_k(0,0)\in\aepvar, k=1,2  |\Mc\}\\
&\le \P\{(U^n_1(m_1, l_1), U^n_2(m_2, l_2), Y^n)\in\aep, U^n_k(0,0)\in\aep, k=1,2  |\Mc\}\\
&\stackrel{(a)}{=}\P\{(W^n_{\bb}(s_{\bb}), U^n_1(m_1, l_1), U^n_2(m_2, l_2), Y^n)\in\aep, U^n_k(0,0)\in\aep, k=1,2 |\Mc\}\\
&\stackrel{(b)}{\le}\P\{(W^n_{\bb}(s_{\bb}), Y^n)\in\aep, U^n_k(0,0)\in\aep, k=1,2 |\Mc\}\\
&\stackrel{(c)}{\le}2^{n(\Rh_1+\Rh_2)}2^{-n(I(W_{\bb}; Y)+D(p_{W_{\bb}}\|p_\q)-\d(\e))}\prod_{k=1}^22^{-n(D(p_{U_k}\|p_\q)-\d(\e))},
\end{align*} where step $(a)$ follows from the fact that $W^n_{\bb}(s_{\bb})$ is a function of $(U^n_1(m_1, l_1), U^n_2(m_2, l_2))$, step $(b)$ follows from the fact that the event $(W_{\bb}^n(s_{\bb}), U^n_1(m_1, l_1), U^n_2(m_2, l_2), Y^n)\in\aep$ implies $(W_{\bb}^n(s_{\bb}), Y^n)\in\aep$, and step $(c)$ follows from Lemma~\ref{lem:cfbound}.

\subsection{Case $(m_1, l_1, m_2, l_2)\in\Lc_2(\bb)$:}
Consider some non-zero vector $\cb=[c_{1},\,c_{2}]\in\Fq^2$ that is linearly independent of $\bb$. Define $s_{\cb} \in [2^{n\Rt(\cb)}]$ as the index whose $\q$-ary expansion satisfies 
\begin{align}
[\nub(s_{\cb})~\mathbf{0}] = c_{1}\etab(m_1, l_1)\oplus c_{2}\etab(m_2, l_2) \ . 
\end{align} and let \begin{align*}
W_{\cb}^n(s_{\cb}) &:=c_{1}U^n_1(m_1, l_1)\oplus c_{2}U^n_2(m_2, l_2) \\
W_{\cb}& :=c_{1}U_1\oplus c_{2} U_2 . 
\end{align*}
Note that by definition, for $(m_1, l_1, m_2, l_2)\in\Lc_2(\bb)$, $W_{\bb}^n(0)=b_{1}U^n_1(m_1, l_1)\oplus b_{2}U^n_2(m_2, l_2)$. Then,
\begin{align*}
\P\{&(U^n_1(m_1, l_1), U^n_2(m_2, l_2), Y^n)\in\aep, U^n_k(0,0)\in\aepvar, k=1,2  |\Mc\}\\
&\le \P\{(U^n_1(m_1, l_1), U^n_2(m_2, l_2), Y^n)\in\aep, U^n_k(0,0)\in\aep, k=1,2  |\Mc\}\\
&\stackrel{(a)}{=} \P\{(W^n_{\bb}(0), W^n_{\cb}(s_{\cb}), U^n_1(m_1, l_1), U^n_2(m_2, l_2), Y^n)\in\aep, U^n_k(0,0)\in\aep, k=1,2  |\Mc\}\\
&\stackrel{(b)}{\le} \P\{(W^n_{\bb}(0), W^n_{\cb}(s_{\cb}), Y^n)\in\aep, U^n_k(0,0)\in\aep, k=1,2  |\Mc\}\\
&\stackrel{(c)}{=} 2^{n(\Rh_1+\Rh_2)}2^{-n(I(W_{\cb}; Y, W_{\bb})+D(p_{W_{\cb}}\|p_\q)-\d(\e))}\prod_{k=1}^22^{-n(D(p_{U_k}\|p_\q)-\d(\e))},
\end{align*}
where step $(a)$ follows from the fact that $W^n_{\bb}(0), W^n_{\cb}(s_{\cb})$ are deterministic functions of the linear codewords $U^n_1(m_1, l_1), U^n_2(m_2, l_2)$ for $(m_1, l_1, m_2, l_2)\in\Lc_2(\bb)$, step $(b)$ follows from the fact that
the event $(W^n_{\bb}(0), W^n_{\cb}(s_{\cb}), U^n_1(m_1, l_1), U^n_2(m_2, l_2), Y^n)\in\aep$ implies $(W^n_{\bb}(0), W^n_{\cb}(s_{\cb}), Y^n)\in\aep$, and step $(c)$ follows from Lemma~\ref{lem:cfbound} with $Y^n$ replaced by $(Y^n, W^n_{\bb}(0))$.
%%%%%%%%%%%%%%%%%%%%%%%%%%%%%%%%%%%%%%

\section{Proof of the equivalence of $\Rc_{\mathsf{LMAC}}$ and rate region \eqref{eq:equiv-rr}}\label{app:equiv-rr}

Define the rate regions
\begin{align}
\Rc_{0} = \{&(R_1, R_2): \nn\\
&R_1 < I(X_1; Y| X_2),\label{eq:sb1}\\ 
&R_2 < I(X_2; Y| X_1),\label{eq:sb2}\\ 
&R_1+R_2 < I(X_1, X_2; Y) \}.
\end{align}
\begin{align*}
\hat{\Rc} = \{&(R_1, R_2):\\
&\min(R_1-H(U_1), R_2-H(U_2)) < I(W_{\bb}; Y)-H(W_{\bb}),\\
&\min(R_1-H(U_1), R_2-H(U_2)) < I(W_{\cb}; Y, W_{\bb})-H(W_{\cb})\},
\end{align*}
and
\begin{align*}
\hat{\Rc}_1&=\{  (R_1, R_2): R_1 < \min\{I_{\mathsf{CF},1}(\bb), I(X_1, X_2; Y)-I_{\mathsf{CF},2}(\bb)\}\},\\
\hat{\Rc}_2&=\{  (R_1, R_2): R_2 < \min\{I_{\mathsf{CF},2}(\bb), I(X_1, X_2; Y)-I_{\mathsf{CF},1}(\bb)\}\},
\end{align*}
where $I_{\mathsf{CF},k}(\bb)$, $k=1,2,$ is defined in~\eqref{eq:I1} and~\eqref{eq:I2}.

First, note that due to the following inequality between \eqref{eq:sb1} and $\hat{\Rc}_1$ (and also between \eqref{eq:sb2} and $\hat{\Rc}_2$),
\begin{align*}
 H(U_1)-H(U_1|Y, U_2) &=  H(U_1)-H(W_{\bb}|Y, U_2) \\
 &\ge  H(U_1)-H(W_{\bb}|Y), 
\end{align*}
we have $\Rc_{\mathsf{LMAC}}=\Rc_{0}\cap(\hat{\Rc}_1\cup\hat{\Rc}_2)$.

Next, note that
\begin{align*}
I(U_1; Y|U_2) &= I(U_1, X_1; Y|U_2, X_2)\\
&= H(Y|U_2, X_2)-H(Y|U_1, X_1, U_2, X_2)\\
&= H(Y|X_2)-H(Y|X_1, X_2)\\
&= I(X_1; Y|X_2),
\end{align*}
where we have used the Markov relations $U_2\to X_2 \to Y$ and $(U_1, U_2)\to (X_1, X_2)\to Y$. Similarly, we have
\begin{align*}
I(U_2; Y|U_1) &= I(X_2; Y|X_1),\\
I(U_1, U_2; Y) &= I(X_1, X_2; Y),
\end{align*}
and thus, the rate region in \eqref{eq:equiv-rr} is $\Rc_{0}\cap\hat{\Rc}$.
Thus, it is sufficient to show that $\hat{\Rc}=(\hat{\Rc}_1\cup\hat{\Rc}_2)$.

To this end, first consider $(R_1, R_2)\in\hat{\Rc}$ such that $R_1-H(U_1)\le R_2-H(U_2)$. Then, we have $(R_1, R_2) \in \hat{\Rc}_1$ since
\begin{align*}
R_1 &< H(U_1)+I(W_{\bb}; Y)-H(W_{\bb})\\
&= H(U_1)-H(W_{\bb}| Y),
\end{align*}
and 
\begin{align}
R_1 &< H(U_1)+I(W_{\cb}; Y, W_{\bb})-H(W_{\cb})\nonumber \\
&= H(U_1)+H(U_2)-H(U_2)-H(W_{\cb}| Y, W_{\bb})\nonumber \\
&= H(U_1, U_2)-H(U_2)-H(W_{\bb}, W_{\cb}| Y)+H(W_{\bb}| Y)\nonumber\\
&= H(U_1, U_2)-H(U_1, U_2| Y)-H(U_2)+H(W_{\bb}| Y)\nonumber\\
&= I(U_1, U_2; Y)-H(U_2)+H(W_{\bb}| Y)\nonumber \\
&= I(X_1, X_2; Y)-I_{\mathsf{CF},2}(\bb).  \label{eqn:sumcapminuscf}
\end{align}
Similarly, for $(R_1, R_2)\in\hat{\Rc}$ such that $R_2-H(U_2)\le R_1-H(U_1)$, we have $(R_1, R_2) \in \hat{\Rc}_2$. Clearly, $\hat{\Rc} \subseteq (\hat{\Rc}_1\cup \hat{\Rc}_2)$. 

To show the inclusion in the other direction, it is sufficient to show following:
\begin{enumerate}
\item For the rate tuples $(R_1, R_2)\in\hat{\Rc}_2$ such that $R_1-H(U_1)\le R_2-H(U_2)$,
\begin{align}
(R_1, R_2)\in \hat{\Rc}, \label{eq:include1}
\end{align}
\item and for the rate tuples $(R_1, R_2)\in\hat{\Rc}_1$ such that $R_1-H(U_1)\ge R_2-H(U_2)$,
\begin{align}
(R_1, R_2) \in \hat{\Rc}. \label{eq:include2}
\end{align}
\end{enumerate}
We begin by considering the first case and assume that a rate pair $(R_1, R_2)$ satisfies $R_1-H(U_1)\le R_2-H(U_2)$
and that $(R_1, R_2)\in\hat{\Rc}_2$. Since 
\begin{align*}
R_1&\le R_2-H(U_2)+H(U_1)\\
&\stackrel{(a)}{<} \min\{I(W_{\bb}; Y)-H(W_{\bb}), I(W_{\cb}; Y, W_{\bb})-H(W_{\cb})\}+H(U_1),\\
&\stackrel{(b)}{=}  \min\{I_{\mathsf{CF},1}(\bb), I(X_1, X_2; Y)-I_{\mathsf{CF},2}(\bb)\},
\end{align*}
$(R_1, R_2)$ is also included in $\hat{\Rc}$, where step $(a)$ follows from the fact that $(R_1, R_2)\in\hat{\Rc}_2$ and step $(b)$ uses~\eqref{eqn:sumcapminuscf}.
The second case \eqref{eq:include2} can also be shown in the same manner.

%-----------------------------------------------------------------------------

\section{Markov Lemma for Nested Linear Codes}\label{app:ML-LC}

Without loss of generality, we assume that the message indices are set to zero and focus on the effect of the auxiliary indices. With a slight abuse of notation, we let $\etab(l_k) = [\nub(l_k)~\mathbf{0}]$ denote the $\q$-ary expansion of the index $l_k$ followed by zero padding to length $\kappa = \max_k n \Rh_k$. 

Consider a nested linear code 
\begin{align*}
\Cc_k=\{u_k^n(l_k): \nub(l_k)\mathbf{G}\oplus D^n_k, l_k\in[2^{n\Rh_k}]\}, \quad k\in[1:K], 
\end{align*}
where $\mathbf{G} \in \Fq^{\kappa \times n}$ is the random generator matrix and $D^n_k \in \Fq^n$ are the random dithers. Each entry of $\mathbf{G}$ and $D^n_k$ is drawn uniformly and independently from $\Fq$. We denote the realization of $\mathbf{G}$ and $D^n_k$ by $\Gb$ and $d^n_k$, respectively.

Let $(X, U_1, \ldots,U_K)\sim p(x)\prod_{k=1}^Kp(u_k|x)$ and consider the following encoding procedure. 

\noindent{\bf Encoding:} For each $x^n\in\aepvar$, find an index $l_k\in[2^{n\Rh_k}]$ such that
\begin{align*}
(x^n, U^n_k(l_k))\in\aepvar.
\end{align*}
If there is more than one index, choose one at random from the available options. If there is none, choose one at random from $[2^{n\Rh_k}]$.
Define the random variable $L_k$ as the chosen index.

\begin{lemma}[Markov Lemma for Nested Linear Codes]\label{lem:ML-LC}
For sufficiently small $\e'<\e$ and any $x^n \in \aepvar(X)$,
\begin{align*}
\lim_{n\to\infty}\P\{(x^n, U_1^n(L_1), \ldots, U_K^n(L_k))\in\aep(X, U_1,\ldots,U_K ) \}= 1,
\end{align*}
if 
\begin{align*}
\Rh_k &> I(U_k; X)+D(p_{U_k}\|p_\q)+\d(\e'), \quad k\in[1:K].
\end{align*}
\end{lemma} As noted earlier, the codebooks share a generator matrix, which means that the auxiliary indices $L_1,\ldots,L_K$ are not conditionally independent given $X^n$, even though the target distribution for $U_1,\ldots,U_K$ is conditionally independent given $X$. This precludes a standard application of the Markov lemma~\cite[Lemma 12.1]{El-Gamal--Kim2011}. Below, we develop a proof from first principles, beginning with some linear algebra definitions.

To simplify our notation, we define $n_k := n \Rh_k \log_2(\q)$, which allows us to write $l_k \in [\q^{n_k}]$ rather than $l_k \in [2^{n \Rh_k}]$.  
Furthermore, let 
\begin{align*}
\tilde{\mathbf{G}}=\left[
\begin{array}{c}
D^n_1 \\
\vdots \\
D^n_K \\
\mathbf{G}
\end{array}\right],
\end{align*}
and for some $(l_1,\ldots,l_K,\tilde{l}_{j_1},\ldots,\tilde{l}_{j_t})\in[\q^{n_1}]\times\cdots\times[\q^{n_K}]\times[\q^{n_{j_1}}]\times\cdots\times[\q^{n_{j_t}}]$, and $1 \leq j_1 < \cdots < j_t \leq K$, define
\begin{align} \label{eq:rank-matrix2} 
\mathsf{H}(l_1,\ldots,l_K,\tilde{l}_{j_1},\ldots,\tilde{l}_{j_t})=\begin{bmatrix}
\mathbf{e}_1 & \etab(l_1) \\
\vdots & \vdots \\
\mathbf{e}_K & \etab(l_K) \\
\mathbf{e}_{j_1} & \etab(\tilde{l}_{j_1}) \\
\vdots & \vdots \\
\mathbf{e}_{j_t} & \etab(\tilde{l}_{j_t}) \\
\end{bmatrix},
\end{align}
where $\mathbf{e}_k$ is the $k^{\text{th}}$ standard basis vector in $\mathbb{F}_q^K$, i.e., its $k^{\text{th}}$ entry is $1$ while the rest are $0$.
We will use the notation $\rank(l_1,\ldots,l_K,\tilde{l}_{j_1},\ldots,\tilde{l}_{j_t})$ to denote the rank of $\mathsf{H}(l_1,\ldots,l_K,\tilde{l}_{j_1},\ldots,\tilde{l}_{j_t})$. Note that, with this notation at hand, the codeword tuple
\begin{align*}
(U^n_1(l_1),\ldots, U^n_K, U^n_{j_1}(\lt_{j_1}),\ldots, U^n_{j_t}(\lt_{j_t}))
\end{align*} 
can be represented by $\mathsf{H}(l_1,\ldots,l_K,\tilde{l}_{j_1},\ldots,\tilde{l}_{j_t}) \cdot \tilde{\mathbf{G}}$.

We can now state two basic statistical properties of nested linear codes. 

\begin{lemma}[Uniformity] \label{lem:uniformity}
For any choice of indices $(l_1,\ldots, l_K)\in[\q^{n_1}]\times\cdots\times[\q^{n_K}]$ and $(u^n_1,\ldots,u^n_K)\in \Fq^{n}\times\cdots\times \Fq^{n}$,
\begin{align*}
\P\{U^n_1(l_1)=u^n_1,\ldots,U^n_K(l_K)=u^n_K\} %&=\prod_{j=1}^K p(u^n_j)\\
&=\frac{1}{\q^{nK}}.
\end{align*}
\end{lemma}
Lemma~\ref{lem:uniformity} is a direct consequence of the independent random dithers.

\begin{lemma}[Linear Independence $\implies$ Statistical Independence]\label{lem:implication}
For  indices satisfying\\ {$\rank(l_1,\ldots,l_K,\tilde{l}_{j_1},\ldots,\tilde{l}_{j_t}) = K + t$}, the random linear codewords 
\[
	({U^n_1(l_1),\ldots,U^n_K(l_K),U^n_{j_1}(\tilde{l}_{j_1}),\ldots,U^n_{j_t}(\tilde{l}_{j_t})})
\] are statistically independent. 
\end{lemma}

\begin{IEEEproof}
For $(u^n_1,\ldots,u^n_K, \ut^n_{j_1}, \ldots, , \ut^n_{j_t})\in \Fq^{n}\times\cdots\times \Fq^{n}$,
\begin{align*}
\P&\{U^n_1(l_1)=u^n_1,\ldots,U^n_K(l_K)=u^n_K,U^n_{j_1}(\tilde{l}_{j_1})=\ut^n_{j_1},\ldots,U^n_{j_t}(\tilde{l}_{j_t})=\ut^n_{j_t}\}\\
&\stackrel{(a)}{=}\frac{1}{\q^{nK}}\P\{U^n_{j_1}(\tilde{l}_{j_1})=\ut^n_{j_1},\ldots,U^n_{j_t}(\tilde{l}_{j_t})=\ut^n_{j_t}|U^n_1(l_1)=u^n_1,\ldots,U^n_K(l_K)=u^n_K\}\\
&=\frac{1}{\q^{nK}}\P\{U^n_{j_1}(\tilde{l}_{j_1})=\ut^n_{j_1},\ldots,U^n_{j_t}(\tilde{l}_{j_t})=\ut^n_{j_t}| D^n_{1}=u_1^n \ominus \etab(l_1)\mathbf{G},\ldots, D^n_{K}=u_K^n \ominus \etab(l_K)\mathbf{G}\}\\
&=\frac{1}{\q^{nK}}\P\{(\etab(\tilde{l}_{j_1})\ominus\etab(l_{j_1}))\mathbf{G}=\ut^n_{j_1}\ominus u^n_{j_1},\ldots,(\etab(\tilde{l}_{j_t})\ominus\etab(l_{j_t}))\mathbf{G}=\ut^n_{j_t}\ominus u^n_{j_t} \\[-0.5em]
&\qquad\qquad\qquad | D^n_{1}=u_1^n \ominus \etab(l_1)\mathbf{G},\ldots, D^n_{K}=u_K^n \ominus \etab(l_K)\mathbf{G}\}\\
&\stackrel{(b)}{=}\frac{1}{\q^{nK}}\P\{(\etab(\tilde{l}_{j_1})\ominus\etab(l_{j_1}))\mathbf{G}=\ut^n_{j_1}\ominus u^n_{j_1},\ldots,(\etab(\tilde{l}_{j_t})\ominus\etab(l_{j_t}))\mathbf{G}=\ut^n_{j_t}\ominus u^n_{j_t} \}\\
&\stackrel{(c)}{=}\frac{1}{\q^{n(K+t)}},
\end{align*}
where step $(a)$ follows from Lemma~\ref{lem:uniformity}, step $(b)$ follows from the fact that $\mathbf{G}$ and the dithers are independent, and step $(c)$ follows from the fact that $(\etab(\lt_{j_1})\ominus \etab(l_{j_1})), \ldots, (\etab(\lt_{j_t})\ominus\etab(l_{j_t}))$ are linearly independent due to the assumption that $\rank(l_1,\ldots,l_K,\tilde{l}_{j_1},\ldots,\tilde{l}_{j_t}) =K+t$ and~\cite[Theorem 1]{Domb--Zamir--Feder2013}.
\end{IEEEproof}

It will be useful to classify codewords according the rank of their auxiliary indices. Define the index set of rank $r$ as 
\begin{align*}
\mathcal{I}_r := \big\{ (l_1,\ldots,l_K,\tilde{l}_{1},\ldots,\tilde{l}_{K})  : \rank(l_1,\ldots,l_K,\tilde{l}_{1},\ldots,\tilde{l}_{K}) = r \big\}.
\end{align*} Note that, by definition, $|\mathcal{I}_0| = \cdots = |\mathcal{I}_{K-1}| = 0$ and $|\mathcal{I}_K| = \q^{n_1 + \cdots + n_K}$.

\begin{lemma}\label{l:indexsetbounds} The size of an index set $\mathcal{I}_r$ of rank $K < r \le 2K$ is upper bounded as follows
\begin{enumerate}
\item $|\mathcal{I}_{K+t}| \le \q^{n_1 + \cdots + n_K}\, \q^{K^2} \displaystyle\sum_{1 \le j_1 < \cdots < j_t \le K} \q^{(n_{j_1} + \cdots + n_{j_t})}$ for $t = 1, \ldots, K-1$,
\item $|\mathcal{I}_{2K}| \le \q^{2 (n_1 + \cdots +  n_K)}$. 
\end{enumerate}
\end{lemma}
\begin{IEEEproof} The latter bound on $| \mathcal{I}_{2K} |$ is trivial since there are only $\q^{2 (n_1 + \cdots +  n_K)}$ possible index tuples. To establish the former bound, we begin by defining
\begin{align} \label{e:indexsubset} \mathcal{I}_{K+t}^{\{j_1, \ldots, j_t\}} := \{ (l_1, \ldots, l_K, \tilde{l}_1, \ldots, \tilde{l}_K) \in \mathcal{I}_{K+t}
 :  \rank(l_1, \ldots, l_K, \tilde{l}_{j_1}, \ldots, \tilde{l}_{j_t}) = K+t\},
 \end{align} which is a subset of $\mathcal{I}_{K+t}$. Therefore, by the union bound, 
\begin{align} \label{e:indexunionbound}
|\mathcal{I}_{K + t}| \le \sum_{1 \le j_1 < \cdots < j_t \le K} \big|\mathcal{I}_{K+t}^{\{j_1, \ldots, j_t\}} \big|.
\end{align}

The following construction can be used to generate all possible index tuples in $\mathcal{I}_{K+t}^{\{j_1, \ldots, j_t\}}$: 
\begin{enumerate}
\item Choose $K$ arbitrary indices $(l_1, \ldots, l_K) \in [\q^{n_1} ] \times \cdots \times [ \q^{n_K} ]$,
\item Choose $t$ indices $( \tilde{l}_{j_1}, \ldots, \tilde{l}_{j_t}  ) \in [\q^{n_{j_1}}] \times \cdots \times [\q^{n_{j_t}}]$  such that $\rank(l_1, \ldots, l_K, \tilde{l}_{j_1}, \ldots, \tilde{l}_{j_t}) = K+t$, and
\item For each $\ell \in \{1, \ldots, K\} \setminus \{ j_1, \ldots, j_t \}$, choose an index $\tilde{l}_\ell \in [\q^{n_{\ell}}]$ such that the row vector $[ \mathbf{e}_\ell ~ \etab(\tilde{l}_\ell)]$ is a linear combination of the $K + t$ row vectors in \eqref{eq:rank-matrix2}.
\end{enumerate} 

We now upper bound the number of choices in each step of the construction above. First, the number of choices in Step 1) is $\q^{n_1 + \cdots + n_K}$. Second, the number of choices in Step 2) is upper bounded by $\q^{n_{j_1} + \cdots + n_{j_t}}$. Third, for any $\ell \in \{1, \ldots, K \} \setminus \{ j_1, \ldots, j_t \}$, the number of choices for $\tilde{l}_\ell$ is upper bounded by $\q^{K+t}$, because $[\mathbf{e}_\ell ~ \etab(\tilde{l}_\ell)]$ is linearly dependent with respect to $K + t$ row vectors. As such, the total number of choices in Step 3) is at most 
$\q^{(K + t)(K - t)}$, which is in turn bounded by $\q^{K^2}$. The total number of choices leads to the following upper bound, 
$$|\mathcal{I}_{K+t}^{\{j_1, \ldots, j_t\}} | \le \q^{n_1 + \cdots + n_K} \q^{n_{j_1} + \cdots + n_{j_t}} \q^{K^2}.$$ Plugging this into~\eqref{e:indexunionbound} gives us the desired upper bound.
\end{IEEEproof}

We now bound the probability that the random linear codewords land in certain subsets. It will be useful to define 
\begin{align}\label{e:subsetcount}
Z_{\mathcal{S}} := \sum_{(l_1, \ldots, l_K)} \ind((U_1^n(l_1), \ldots, U_K^n(l_K) ) \in \Sc)
\end{align}
to represent the number of codeword tuples that fall in $\mathcal{S}$. Since the codewords are uniformly distributed, the mean of $Z_{\mathcal{S}}$ is  
$$\mu_{\mathcal{S}} = \frac{ | \mathcal{S} |}{\q^{Kn - (n_1 + \cdots + n_K)}}.$$ 

\begin{lemma} \label{l:meandeviatebound}
For $k \in [1:K]$, let $\mathcal{S}_k$ be a subset of $\Fq^n$ and let $\mathcal{S}$ be a subset of  $\mathcal{S}_1 \times \cdots \times \mathcal{S}_K$. For any $\gamma > 0$, the probability that $Z_{\mathcal{S}}$ deviates from its mean is bounded as follows 
\begin{align}
&\P\bigg\{ |Z_{\mathcal{S}} - \mu_{\mathcal{S}} | \ge \frac{\gamma |\Sc_1| \cdots |\Sc_K| }{\q^{Kn - (n_1 + \cdots  +n_K)}} \bigg\} \\ &\le \frac{1}{\gamma^2} \left( \frac{\q^{Kn - (n_1 + \cdots + n_K)}}{|\Sc_1| \cdots |\Sc_K| } + \q^{K^2} \sum_{t =1}^{K-1}  \sum_{1 \le j_1 < \cdots < j_t \le K} 
\frac{\q^{n-n_{j_1}}} {|\mathcal{S}_{j_1}|}\cdots \frac{\q^{n-n_{j_t}}}{|\mathcal{S}_{j_t}|}\right). \label{e:probdeviate}
\end{align}
\end{lemma} 

\begin{IEEEproof} We begin by calculating the variance of $Z_{\mathcal{S}}$,
\begin{align}
\sigma_{\mathcal{S}}^2 &:= \var(Z_{\mathcal{S}}) \nonumber\\
&= \E(Z_{\mathcal{S}}^2) - \mu^2_{\mathcal{S}} \nonumber\\
&= \sum_{l_1,\ldots,l_K,\tilde{l}_1,\ldots,\tilde{l}_K} \P\big\{ (U_1^n(l_1), \ldots, U_K^n(l_K)) \in \mathcal{S}, (U_1^n(\tilde{l}_1), \ldots, U_K^n(\tilde{l}_K)) \in \mathcal{S} \big\}- \mu^2_{\mathcal{S}} \nonumber\\
&= \sum_{r = K}^{2K} \varphi(\mathcal{I}_r) - \mu^2_{\mathcal{S}} \label{e:zsvar}
\end{align} where
$$\varphi(\mathcal{I}) := \sum_{(l_1,\ldots,l_K,\tilde{l}_1,\ldots,\tilde{l}_K) \in \mathcal{I}} \P\big\{ (U_1^n(l_1), \ldots, U_K^n(l_K)) \in \mathcal{S}, (U_1^n(\tilde{l}_1), \ldots, U_K^n(\tilde{l}_K)) \in \mathcal{S} \big\}.$$

Note that $(l_1,\ldots,l_K,\tilde{l}_1,\ldots,\tilde{l}_{K}) \in \mathcal{I}_K$ if and only if $l_k = \tilde{l}_k$ for all $k \in [1:K]$. Therefore, 
\begin{align}
\varphi(\mathcal{I}_K) &= \sum_{l_1,\ldots,l_K} \P\{ (U_1^n(l_1), \ldots, U_K^n(l_K)) \in \mathcal{S}\}\nonumber \\
&= \frac{ | \mathcal{S} |}{\q^{Kn - (n_1 + \cdots + n_K)}}\nonumber\\
&= \mu_{\mathcal{S}}. \label{e:indexsetprob1}
\end{align} Next, by Lemma~\ref{lem:implication}, we observe that, for $(l_1,\ldots,l_K,\tilde{l}_1,\ldots, \tilde{l}_K) \in \mathcal{I}_{2K}$, the resulting random codewords are independent. Therefore, 
\begin{align}
\varphi(\mathcal{I}_{2K}) &= \sum_{(l_1,\ldots,l_K,\tilde{l}_1,\ldots,\tilde{l}_K) \in \mathcal{I}_{2K}} \P\{ (U_1^n(l_1), \ldots, U_K^n(l_K)) \in \mathcal{S}\} \P\{ (U_1^n(\tilde{l}_1), \ldots, U_K^n(\tilde{l}_K)) \in \mathcal{S}\} \nonumber\\
&= \sum_{(l_1,\ldots,l_K,\tilde{l}_1,\ldots,\tilde{l}_K) \in \mathcal{I}_{2K}} \frac{|\mathcal{S}|^2}{\q^{2Kn}}\nonumber \\
&= | \mathcal{I}_{2K} |  \frac{|\mathcal{S}|^2}{\q^{2Kn}}\nonumber \\
&\leq \q^{2(n_1 + \cdots + n_K)} \frac{|\mathcal{S}|^2}{\q^{2Kn}} = \mu_{\mathcal{S}}^2, \label{e:indexsetprob2}
\end{align} where the inequality follows from Lemma~\ref{l:indexsetbounds}.

For the remaining terms, we use the subsets defined in~\eqref{e:indexsubset} to obtain a union bound,
\begin{align} \label{e:indexsetunionbound}
\varphi(\mathcal{I}_{K + t}) \le \sum_{1 \le j_1 < \cdots < j_t \le K} \varphi\big(\mathcal{I}_{K+t}^{\{ j_1, \ldots, j_t \}}\big) ,
\end{align}
and then upper bound each term in the sum,
\begin{align*}
& \varphi\big(\mathcal{I}_{K+t}^{\{ j_1, \ldots, j_t \}}\big) \\
&= \sum_{(l_1, \ldots,  l_K, \tilde{l}_1,\ldots,  \tilde{l}_K) \in\mathcal{I}_{K+t}^{\{ j_1, \ldots, j_t \}}}  \P\big\{ (U_1^n(l_1), \ldots, U_K^n(l_K)) \in \mathcal{S} , (U_1^n(\tilde{l}_1), \ldots, U_K^n(\tilde{l}_K)) \in \mathcal{S} \big\}\\
&\le \sum_{(l_1, \ldots,  l_K, \tilde{l}_1,\ldots,  \tilde{l}_K) \in\mathcal{I}_{K+t}^{\{ j_1, \ldots, j_t \}}}  \P\big\{ (U_1^n(l_1), \ldots, U_K^n(l_K)) \in \mathcal{S} , U_{j_1}^n(\tilde{l}_{j_1}) \in \mathcal{S}_{j_1}, \cdots, U_{j_t}^n(\tilde{l}_{j_t}) \in \mathcal{S}_{j_t} \big\} \\
&\overset{(a)}{=} \sum_{(l_1, \ldots,  l_K, \tilde{l}_1,\ldots,  \tilde{l}_K) \in\mathcal{I}_{K+t}^{\{ j_1, \ldots, j_t \}}}  \P\{ (U_1^n(l_1), \ldots, U_K^n(l_K)) \in \mathcal{S} \} \P\{ U_{j_1}^n(\tilde{l}_{j_1}) \in \mathcal{S}_{j_1} \}  \cdots  \P\{ U_{j_t}^n(\tilde{l}_{j_t}) \in \mathcal{S}_{j_t} \} \\
&= \sum_{(l_1, \ldots,  l_K, \tilde{l}_1,\ldots,  \tilde{l}_K) \in\mathcal{I}_{K+t}^{\{ j_1, \ldots, j_t \}}} \frac{|\Sc|}{\q^{Kn}} \frac{|\Sc_{j_1}|}{\q^n} \cdots \frac{|\Sc_{j_t}|}{\q^n} \\
&\overset{(b)}{\le} \q^{n_1 + \cdots + n_k} \q^{n_{j_1} + \cdots + n_{j_t}} \q^{K^2} \frac{|\Sc|}{\q^{Kn}} \frac{|\Sc_{j_1}|}{\q^n} \cdots \frac{|\Sc_{j_t}|}{\q^n} \\
&= \q^{K^2} \mu_{\Sc} \frac{|\Sc_{j_1}|}{q^{n-n_{j_1}}} \cdots \frac{|\Sc_{j_t}|}{\q^{n-n_{j_t}}},
\end{align*} 
where step $(a)$ follows from Lemma~\ref{lem:implication} and step $(b)$ follows from the upper bound in Lemma~\ref{l:indexsetbounds}. Plugging back into~\eqref{e:indexsetunionbound}, we obtain
\begin{align} \label{e:indexsetprob3}
\varphi(\mathcal{I}_{K + t}) \le \q^{K^2} \mu_{\Sc} \sum_{1 \le j_1 < \cdots < j_t \le K} \frac{|\Sc_{j_1}|}{\q^{n-n_{j_1}}} \cdots \frac{|\Sc_{j_t}|}{\q^{n-n_{j_t}}}.
\end{align}

Now, plugging~\eqref{e:indexsetprob1},~\eqref{e:indexsetprob2},~and \eqref{e:indexsetprob3} back into~\eqref{e:zsvar}, we obtain an upper bound on the variance 
\begin{align}
\sigma^2_{\Sc} &\le \mu_{\Sc}  +  \mu_{\Sc} \, \q^{K^2} \sum_{t =1}^{K-1}   \left( \sum_{1 \le j_1 < \cdots < j_t \le K} \frac{|\mathcal{S}_{j_1}|}{\q^{n-n_{j_1}}} \cdots \frac{|\mathcal{S}_{j_t}|}{\q^{n-n_{j_t}}} \right)  + \mu_{\Sc}^2 - \mu_{\Sc}^2 \nonumber \\
&= \mu_{\Sc} \left( 1 +  \q^{K^2} \sum_{t =1}^{K-1}  \left( \sum_{1 \le j_1 < \cdots < j_t \le K} \frac{|\mathcal{S}_{j_1}|}{\q^{n-n_{j_1}}} \cdots \frac{|\mathcal{S}_{j_t}|}{\q^{n-n_{j_t}}}  \right) \right) \nonumber \\
&\leq \frac{ | \Sc_1 | \cdots | \Sc_K| }{q^{Kn - (n_1 + \cdots + n_K)}}  \left( 1 +  \q^{K^2} \sum_{t =1}^{K-1}  \left( \sum_{1 \le j_1 < \cdots < j_t \le K} \frac{|\mathcal{S}_{j_1}|}{\q^{n-n_{j_1}}} \cdots \frac{|\mathcal{S}_{j_t}|}{\q^{n-n_{j_t}}}  \right) \right), \label{e:varupperbound}
\end{align} 
where the last step uses the fact that $| \Sc | \leq |\Sc_1| \cdots |\Sc_K|$. 

Finally, we obtain the desired upper bound via Chebyshev's inequality,
\begin{align*}
&\P\bigg\{ |Z_{\Sc} - \mu_{\Sc} | \ge \frac{\gamma |\Sc_1| \cdots | \Sc_K| }{\q^{Kn - (n_1 + \cdots + n_K)}} \bigg\} \\ 
&\leq \frac{1}{\gamma^2} \bigg(\frac{\q^{Kn - (n_1 + \cdots + n_K)}}{|\Sc_1| \cdots |\Sc_K|} \bigg)^2 \sigma_{\Sc}^2 \\
&\overset{(a)}{\leq} \frac{1}{\gamma^2} \frac{\q^{Kn - (n_1 + \cdots + n_K)}}{|\Sc_1| \cdots |\Sc_K|}  \left( 1 +  \q^{K^2} \sum_{t =1}^{K-1}  \left( \sum_{1 \le j_1 < \cdots < j_t \le K} \frac{|\mathcal{S}_{j_1}|}{\q^{n-n_{j_1}}} \cdots \frac{|\mathcal{S}_{j_t}|}{\q^{n-n_{j_t}}}  \right) \right) \\
&= \frac{1}{\gamma^2}  \left( \frac{\q^{Kn - (n_1 + \cdots + n_K)}}{|\Sc_1| \cdots |\Sc_K|}  +  \q^{K^2} \sum_{t =1}^{K-1}  \left( \sum_{1 \le i_1 < \cdots < i_{K-t} \le K} \frac{\q^{n-n_{i_1}}}{|\mathcal{S}_{i_1}|} \cdots \frac{\q^{n-n_{i_{K-t}}}}{|\mathcal{S}_{i_{K-t}}|}  \right) \right) \\
&=  \frac{1}{\gamma^2} \left( \frac{\q^{Kn - (n_1 + \cdots + n_K)}}{|\Sc_1| \cdots |\Sc_K| } + \q^{K^2} \sum_{t =1}^{K-1}  \sum_{1 \le j_1 < \cdots < j_t \le K} 
\frac{\q^{n-n_{j_1}}} {|\mathcal{S}_{j_1}|}\cdots \frac{\q^{n-n_{j_t}}}{|\mathcal{S}_{j_t}|}\right)
\end{align*} 
where $(a)$ follows from~\eqref{e:varupperbound}.
\end{IEEEproof}

The next lemma, which is a $K$-user generalization of Problem 2.9 in~\cite{Csiszar--Korner2011}, argues that most sequences in the Cartesian product of marginally typical sets belongs to a certain jointly typical set (for conditionally independent random variables). 

\begin{lemma} \label{l:typicalsetoverlap}
Let $V_1,\ldots,V_K$ be random variables that are conditionally independent given the random variable $X$. Then, for sufficiently small $\e' < \e$ and $x^n \in \aepvar(X)$,
\begin{align*}
\lim_{n \rightarrow \infty} \frac{ \big| \aepvar(V_1 | x^n) \times \cdots \times \aepvar(V_K | x^n) \cap \big(\aep(V_1,\ldots,V_K | x^n)\big)^{\mathsf{c}} \big| }{\big|  \aepvar(V_1 | x^n) \times \cdots \times \aepvar(V_K | x^n) \big|} = 0 .
\end{align*} 
\end{lemma}

\begin{IEEEproof}
Lemma~\ref{l:typicalsetoverlap} is a simple consequence of Lemma 12.1 in~\cite{El-Gamal--Kim2011} once we have the following relation.
For some $x^n\in\aepvar(X)$, let $V^n_1,\ldots,V^n_K$ be independent\footnote{This independence assumption does not hold for nested linear codes, which precludes a direct application of the Markov Lemma in our achievability proof.} random sequences uniformly distributed in $\aepvar(V_1|x^n),\ldots, \aepvar(V_K|x^n)$, respectively. Then,
\begin{align*}
&\P\{(x^n, V_1^n,\ldots, V_K^n)\in\aep \}\\ 
&= \sum_{v_1^n\in\aepvar,\ldots, v_K^n\in\aepvar}\P\{(x^n, v_1^n, \ldots, v_K^n)\in\aep, V^n_1=v^n_1,\ldots, V^n_K=v^n_K\}\\
&= \sum_{\substack{v_1^n\in\aepvar,\ldots, v_K^n\in\aepvar:\\
(x^n, v_1^n, \ldots, v_K^n)\in\aep}}\P\{V^n_1=v^n_1,\ldots, V^n_K=v^n_K\}\\
&= \frac{\Big|\big(\aepvar(V_1|x^n) \times \cdots \times \aepvar(V_K|x^n)\big) \cap \aep(V_1, \ldots,V_K|x^n)\Big|}{\big|\aepvar(V_1|x^n) \times \cdots \times \aepvar(V_K|x^n)\big|}.
\end{align*}

It remains to show that the left-hand side of the relation above tends to $1$. For some $\e_1<\e_2<\cdots<\e_K$ where $\e_1=\e'$ and $\e_K = \e$, we have that 
\begin{align*}
\P&\{(x^n, V_1^n, \ldots, V_K^n)\in\aep \} \\
&\ge \P\{(x^n,V_1^n)\in\Tc_{\e_1}^{(n)}, (x^n,V_1^n, V_2^n)\in\Tc_{\e_2}^{(n)}, \ldots, (x^n, V_1^n,V_2^n, \ldots, V_{K}^n)\in\Tc_{\e_K}^{(n)} \} \\
&=\prod_{k=1}^{K}\P\{(x^n,V_1^n,\ldots, V_k^n)\in\Tc_{\e_k}^{(n)}|(x^n,V_1^n,\ldots, V_{k-1}^n)\in\Tc_{\e_{k-1}}^{(n)},\ldots,(x^n, V_1^n)\in\Tc_{\e_1}^{(n)}\}\\
&\stackrel{(a)}{\ge} (1-\delta_n)^{K-1}
\end{align*} where step $(a)$ follows from $K-1$ applications of~\cite[Lemma 12.1]{El-Gamal--Kim2011} and $\delta_n\to 0$ as $n\to\infty$.
\end{IEEEproof}

We are now ready to assemble a proof for the Markov Lemma for Nested Linear Codes.

\noindent\textit{Proof of Lemma~\ref{lem:ML-LC}:} 
Select $0 < \e' < \e$. Define 
\begin{align*}
\mathcal{S}_k &= \aepvar(U_k | x^n) \\
\mathcal{S} &= \big(\mathcal{S}_1 \times \cdots \times \mathcal{S}_K \big) \cap  \big(\aep(U_1,\ldots,U_K | x^n)\big)^{\mathsf{c}} .
\end{align*} Also, define the intersection of the codebooks with the marginally typical sets, 
\begin{align*}
\mathcal{A} = \big(\mathcal{C}_1 \times \cdots \times \mathcal{C}_K\big) \cap \big(\mathcal{S}_1 \times \cdots \times \mathcal{S}_K\big),
\end{align*} as well as the subset that is not jointly typical,
\begin{align*}
\mathcal{B} &= \big(\mathcal{C}_1 \times \cdots \times \mathcal{C}_K\big) \cap \mathcal{S} \\
&= \mathcal{A} \cap \big(\aep(U_1,\ldots,U_K | x^n)\big)^{\mathsf{c}} .
\end{align*} We need to show that, with high probability, there are many choices of marginally typically codewords (i.e., $|\mathcal{A}|$ is large), but relatively few of them are not jointly typical (i.e., $| \mathcal{B} | / |\mathcal{A}|$ is small). 

Define $\mathbf{U}^n = (U_1^n(L_1),\ldots,U_K^n(L_K))$. We have that 
\begin{align*}
&\P\big\{ \mathbf{U}^n  \in \aep(U_1,\ldots,U_K | x^n) \big\} \\
&\geq \P\big\{ \mathbf{U}^n  \in (\mathcal{S}_1 \times \cdots \times \mathcal{S}_K) \cap \aep(U_1,\ldots,U_K | x^n) \big\} \\
&=  \P\big\{ \mathbf{U}^n  \in \mathcal{S}_1 \times \cdots \times \mathcal{S}_K \} - \P\{ \mathbf{U}^n \in \mathcal{B} \}.
\end{align*} The first term is lower bounded as follows:
\begin{align*}
\P\big\{ \mathbf{U}^n  \in \mathcal{S}_1 \times \cdots \mathcal{S}_K \} \geq 1 - \sum_{k=1}^K \P\big\{U_k^n(L_k) \notin \aepvar(U_k|x^n)\big\}.
\end{align*} By Lemma~\ref{lem:mm-covering} in Appendix~\ref{app:mm-covering-packing}, each term in the summation tends to zero as $n \rightarrow \infty$ since, by assumption, $\Rh_k > I(U_k; X)+D(p_{U_k}\|p_\q) + \d(\e')$. 

It remains to show that $\P\{\mathbf{U}^n \in \mathcal{B}\}$ tends to zero. To this end, for some $\gamma > 0$ to be specified later, define 
\begin{align*}
a_n &:= (1 - \gamma) \frac{ | \mathcal{S}_1| \cdots |\mathcal{S}_K|}{\q^{Kn - (n_1 + \cdots + n_K)}} \\ 
b_n &:=  \frac{|\mathcal{S}| + \gamma | \mathcal{S}_1| \cdots |\mathcal{S}_K|}{\q^{Kn - (n_1 + \cdots + n_K)}} .
\end{align*} 
We have that

\begin{align*}
&\P\{\mathbf{U}^n \in \mathcal{B}\} \\
&\leq \P\big\{\mathbf{U}^n \in \mathcal{B} \, \big| \,  | \mathcal{A}|  > a_n,~| \mathcal{B}| < b_n \big\} + \P\big\{ \{ | \mathcal{A}|  > a_n,~| \mathcal{B}| < b_n \}^{\mathsf{c}} \big\}\\
&\leq \P\big\{\mathbf{U}^n \in \mathcal{B} \, \big| \,  | \mathcal{A}|  > a_n,~| \mathcal{B}| < b_n \big\} + \P\{ | \mathcal{A}|  \leq a_n \} + \P\{| \mathcal{B}| \geq b_n \} \\
&< \frac{b_n}{a_n} + \P\{ | \mathcal{A}|  \leq a_n \} + \P\{| \mathcal{B}| \geq b_n \}  
\end{align*} 
where the last step is due to the fact that $\mathbf{U}^n$ is uniformly distributed in $\Ac$ conditioned on $|\Ac|\ge1$, combined with the fact that $\mathcal{B} \subset \mathcal{A}$. The first term can be written as
\begin{align*}
\frac{b_n}{a_n} = \frac{1}{1 - \gamma} \bigg(\gamma + \frac{ | \mathcal{S}| }{ | \mathcal{S}_1| ~ \cdots ~ | \mathcal{S}_K|}\bigg)
\end{align*} and we know, from Lemma~\ref{l:typicalsetoverlap}, that $\lim_{n \rightarrow \infty} \frac{ | \mathcal{S}| }{ | \mathcal{S}_1| ~ \cdots ~ | \mathcal{S}_K|} = 0$. 

For the second and third terms, note that 
$$\frac{\q^{n - n_k}}{| \mathcal{S}_k|} \leq 2^{-n(\Rh_k - (I(U_k; X)+D(p_{U_k}\|p_\q)+\d(\e'))},$$ which tends to $0$ as $n \rightarrow \infty$. For the remainder of the proof, we will assume $n$ is large enough such that the upper bound~\eqref{e:probdeviate} from Lemma~\ref{l:meandeviatebound} is at most $\gamma$. Recall that, from~\eqref{e:subsetcount}, $Z_{\mathcal{A}} = | \mathcal{A}|$ and $Z_{\mathcal{B}} = | \mathcal{B}|$. It follows that
\begin{align*}
\P\{Z_{\mathcal{A}} \leq a_n \} &= \P\bigg\{Z_{\mathcal{A}} - \mu_{\mathcal{A}} \leq -\frac{\gamma | \mathcal{S}_1| ~ \cdots ~ | \mathcal{S}_K|}{ \q^{Kn - (n_1 + \cdots + n_K)}} \bigg\} \\
&\leq  \P\bigg\{|Z_{\mathcal{A}} - \mu_{\mathcal{A}}| \geq \frac{\gamma | \mathcal{S}_1| ~ \cdots ~ | \mathcal{S}_K|}{ \q^{Kn - (n_1 + \cdots + n_K)}} \bigg\} \\ 
&\le \gamma 
\end{align*} where the last step follows from Lemma~\ref{l:meandeviatebound}. Similarly, we have that 
\begin{align*}
\P\{Z_{\mathcal{B}} \geq b_n \} &= \P\bigg\{Z_{\mathcal{B}} - \mu_{\mathcal{B}} \geq \frac{\gamma | \mathcal{S}_1| ~ \cdots ~ | \mathcal{S}_K|}{ \q^{Kn - (n_1 + \cdots + n_K)}} \bigg\} \\
&\leq  \P\bigg\{|Z_{\mathcal{B}} - \mu_{\mathcal{B}}| \geq \frac{\gamma | \mathcal{S}_1| ~ \cdots ~ | \mathcal{S}_K|}{ \q^{Kn - (n_1 + \cdots + n_K)}} \bigg\} \\ 
&\le \gamma .
\end{align*} Finally, by letting $\gamma$ tend to zero as $n \rightarrow \infty$, we obtain the desired result.

\bibliographystyle{IEEEtran}
\bibliography{nit.bib}
\end{document}

%% file: defns.tex
\let\oldbrace\{
\def\{{\oldbrace\kern0.5pt}

\def\var{\mathop{\rm Var}\nolimits}%
\def\rank{\mathop{\rm rank}\nolimits}%
\newcommand{\nn}{\nonumber}

\newcommand{\q}{\mathsf{q}}
\newcommand{\Fq}{\Field_\q}

\newcommand{\Ac}{\mathcal{A}}

\newcommand{\Cc}{\mathcal{C}}

\newcommand{\Ec}{\mathcal{E}}

\newcommand{\Kc}{\mathcal{K}}
\newcommand{\Lc}{\mathcal{L}}
\newcommand{\Mc}{\mathcal{M}}
\newcommand{\Nc}{\mathcal{N}}

\newcommand{\Rc}{\mathcal{R}}
\newcommand{\Sc}{\mathcal{S}}
\newcommand{\Tc}{\mathcal{T}}
\newcommand{\Uc}{\mathcal{U}}

\newcommand{\Xc}{\mathcal{X}}
\newcommand{\Yc}{\mathcal{Y}}

\newcommand{\Xv}{\boldsymbol{X}}

\newcommand{\Uv}{\boldsymbol{U}}

\newcommand{\tv}{\boldsymbol{t}}

\newcommand{\uv}{\boldsymbol{u}}

\newcommand{\pen}{{P_e^{(n)}}}

\newcommand{\aep}{{\mathcal{T}_{\epsilon}^{(n)}}}
\newcommand{\aepvar}{{\mathcal{T}_{\epsilon'}^{(n)}}}

\newcommand{\Mh}{{\hat{M}}}
\newcommand{\Rh}{{\hat{R}}}
\newcommand{\Sh}{{\hat{S}}}
\newcommand{\Uh}{{\hat{U}}}

\newcommand{\Wh}{{\hat{W}}}
\newcommand{\Xh}{{\hat{X}}}

\newcommand{\lh}{{\hat{l}}}

\newcommand{\xh}{{\hat{x}}}

\newcommand{\Rt}{{\tilde{R}}}

\newcommand{\Xt}{{\tilde{X}}}
\newcommand{\Yt}{{\tilde{Y}}}

\newcommand{\lt}{{\tilde{l}}}
\newcommand{\mt}{{\tilde{m}}}

\newcommand{\ut}{{\tilde{u}}}

\newcommand{\xt}{{\tilde{x}}}
\newcommand{\yt}{{\tilde{y}}}

\def\d{\delta}
\def\e{\epsilon}

\DeclareMathOperator\E{\sf E}
\let\P\relax
\DeclareMathOperator\P{\sf P}
\DeclareMathOperator\C{C}

\newcommand{\Bern}{\mathrm{Bern}}

\newcommand{\U}{\mathrm{Unif}}

\newcommand{\Real}{\mathbb{R}}
\newcommand{\Integer}{\mathbb{Z}}
\newcommand{\Field}{\mathbb{F}}
\newcommand{\ind}{\boldsymbol{1}}

\DeclarePairedDelimiter{\ceil}{\lceil}{\rceil}

\DeclareMathOperator*{\argmin}{\arg\min}

%% file: probstatementfig.tex
\begin{figure}[h]
\begin{center}
\psset{unit=0.70mm}
\begin{pspicture}(-42,-15)(189,41)
%\psframe[linecolor=blue](-42,-15)(189,41)
%user 1
\rput(-38,25){$M_1$}  \psline{->}(-34,25)(-25,25)
\psframe[linestyle=dashed](-30,15)(57,35) \rput(13.5,39){\small{Encoder $1$}}
\psframe(-25,17)(11,33) \rput(-7,28){{\small{Bijective}}} \rput(-7,22){{\small{Mapping to $\Field^n$}}}
\psline{->}(11,25)(30,25) \rput(20.5,29){$U_1^n$}
\psframe(30,20)(52,30) \rput(41,25){{$x_1^n(u_1^n)$}}
\rput(63,29){$X_1^n$} \psline{->}(52,25)(68,25)(73,17)(78,17)

\rput(-7,11.5){$\vdots$}
\rput(41.25,11.5){$\vdots$}

%user K
\rput(0,-30){
\rput(-38,25){$M_K$}  \psline{->}(-33,25)(-25,25)
\psframe[linestyle=dashed](-30,15)(57,35) \rput(13.5,39){\small{Encoder $K$}}
\psframe(-25,17)(11,33) \rput(-7,28){{\small{Bijective}}} \rput(-7,22){{\small{Mapping to $\Field^n$}}}
\psline{->}(11,25)(30,25) \rput(20.5,29){$U_K^n$}
\psframe(30,20)(52,30) \rput(41.25,25){{$x_K^n(u_K^n)$}}
\rput(63,29){$X_K^n$} \psline{->}(52,25)(68,25)(73,38)(78,38)
}

%channel
\psframe(78,5)(112,20) \rput(95,12.5){$P_{Y|X_1,\ldots,X_K}$}
\psline{->}(112,12.5)(126,12.5) \rput(119,16.5){$Y^n$}

\psframe(126,5)(148,20) \rput(137,12.5){\small{Decoder}}
\psline{->}(148,12.5)(189,12.5)\rput(169,17.5){$\hat{W}_{\ab_1}^n,\ldots,\hat{W}_{\ab_K}^n$}

\end{pspicture}
\end{center}
\caption{Block diagram of the compute--forward problem. Each transmitter has a message $M_k$ drawn independently and uniformly from $[2^{nR_k}]$ that is bijectively mapped to a representative sequence $U^n_k(M_k)$ over a vector space $\Field^n$, and then into a channel input $X^n_k(M_k) \in \mathcal{X}_k^n$. The $K$ channel inputs pass through a memoryless MAC described by conditional probability distribution $P_{Y|X_1,\ldots,X_K}$ resulting in channel output $Y^n$. Finally, the decoder computes $\hat{W}_{\ab_1}^n,\ldots,\hat{W}_{\ab_K}^n$ of the linear combinations $W_{\ab_\ell}^n(M_1,\ldots,M_K) = \sum_{k} a_{\ell, k} U_k^n(M_k)$.} \label{fig:probstatement}
\end{figure}

%% file: cfrateregion.tex
\begin{figure}[t]
\begin{center}
\psset{unit=.98mm}
\begin{pspicture}(-17,8)(155,134)
\small
%\psframe(-17,8)(155,134)

%All rates scaled up by a factor of 20.

\rput(3,80){
\psline[linewidth=1.5pt,linecolor=black]{->}(0,0)(62,0)  \rput(62,3){$R_1$}
\psline[linewidth=1.5pt]{->}(0,0)(0,48)  \rput(3.5,48){$R_2$}

\psline[linestyle=none,fillstyle=solid,fillcolor=lineblue!25!white]{-}(0.25,0.25)(0.25,34)(40,34)(40,0.25)
\psline[linewidth=3pt,linecolor=lineblue]{-}(0.25,34)(40,34)(40,0.25)

%\psline(22,-1.5)(22,1.5) \rput(26,-4.5){\footnotesize{$I(X_1,X_2;Y)$}}
%\rput(23,-9){\footnotesize{$- I_2(\ab)$}}
%\psline(40,-1.5)(40,1.5) \rput(42,-4.5){\footnotesize{$I_1(\ab)$}}
%\psline(51,-1.5)(51,1.5) \rput(58,-4.5){\footnotesize{$I(X_1;Y|X_2)$}}
%\psline(-1.5,16)(1.5,16) \rput(-11,16){\footnotesize{$I(X_1,X_2;Y)$}}
%\rput(-14,12){\footnotesize{$- I_1(\ab)$}}
%\psline(-1.5,42)(1.5,42) \rput(-11,42){\footnotesize{$I(X_2;Y|X_1)$}}
%\psline(-1.5,34)(1.5,34) \rput(-5.75,34){\footnotesize{$I_2(\ab)$}}

\psline(22,-1.5)(22,1.5) \rput(24,-4.5){\footnotesize{$I(X_1,X_2;Y)$}}
\rput(23,-9){\footnotesize{$- I_{\mathsf{CF},2}(\ab)$}}
\psline(40,-1.5)(40,1.5) \rput(41,-4.5){\footnotesize{$I_{\mathsf{CF},1}(\ab)$}}
\psline(51,-1.5)(51,1.5) \rput(58,-4.5){\footnotesize{$I(X_1;Y|X_2)$}}
\psline(-1.5,16)(1.5,16) \rput(-11,16){\footnotesize{$I(X_1,X_2;Y)$}}
\rput(-12.5,12){\footnotesize{$- I_{\mathsf{CF},2}(\ab)$}}
\psline(-1.5,42)(1.5,42) \rput(-11,42){\footnotesize{$I(X_2;Y|X_1)$}}
\psline(-1.5,34)(1.5,34) \rput(-7.5,34){\footnotesize{$I_{\mathsf{CF},2}(\ab)$}}

\rput(12,12){\large{$\Rc_{\mathsf{CF}}(\ab)$}}
}

\rput(3,18){
\psline[linewidth=1.5pt,linecolor=black]{->}(0,0)(62,0)  \rput(62,3){$R_1$}
\psline[linewidth=1.5pt]{->}(0,0)(0,48)  \rput(3.5,48){$R_2$}

\psline[linestyle=none,fillstyle=solid,fillcolor=linered!25!white]{-}(0.25,42)(36,0.25)(0.25,0.25)
\psline[linestyle=none,fillstyle=solid,fillcolor=linered!25!white]{-}(0.25,28)(51,0.25)(0.25,0.25)
\psline[linewidth=3pt,linecolor=linered,fillstyle=solid,fillcolor=linered!25!white]{-}(0.25,42)(14,42)(22,34)(22,16)
\psline[linewidth=3pt,linecolor=linered,fillstyle=solid,fillcolor=linered!25!white]{-}(22,16)(40,16)(51,5)(51,0.25)
\psline[linestyle=dashed,linecolor=black!70!white](10,46)(56,0)
\psline[linestyle=dashed,linecolor=black!70!white](0,34)(51,34)
\psline[linestyle=dashed,linecolor=black!70!white](0,16)(51,16)
\psline[linestyle=dashed,linecolor=black!70!white](40,0)(40,45)
\psline[linestyle=dashed,linecolor=black!70!white](22,0)(22,45)

\psline[linewidth=1.5pt]{->}(42.5,25)(32,25)
\rput(62,25){$R_1 + R_2=I(X_1,X_2;Y)$}

\psline(22,-1.5)(22,1.5) \rput(24,-4.5){\footnotesize{$I(X_1,X_2;Y)$}}
\rput(23,-9){\footnotesize{$- I_{\mathsf{CF},2}(\ab)$}}
\psline(40,-1.5)(40,1.5) \rput(41,-4.5){\footnotesize{$I_{\mathsf{CF},1}(\ab)$}}
\psline(51,-1.5)(51,1.5) \rput(58,-4.5){\footnotesize{$I(X_1;Y|X_2)$}}
\psline(-1.5,16)(1.5,16) \rput(-11,16){\footnotesize{$I(X_1,X_2;Y)$}}
\rput(-12.5,12){\footnotesize{$- I_{\mathsf{CF},2}(\ab)$}}
\psline(-1.5,42)(1.5,42) \rput(-11,42){\footnotesize{$I(X_2;Y|X_1)$}}
\psline(-1.5,34)(1.5,34) \rput(-7.5,34){\footnotesize{$I_{\mathsf{CF},2}(\ab)$}}

\rput(12,12){\large{$\Rc_{\mathsf{LMAC}}$}}
}

\rput(88,50){
\psline[linewidth=1.5pt,linecolor=black]{->}(0,0)(62,0)  \rput(62,3){$R_1$}
\psline[linewidth=1.5pt]{->}(0,0)(0,48)  \rput(3.5,48){$R_2$}

\psline[linestyle=none,fillstyle=solid,fillcolor=linepurple!25!white]{-}(0.25,42)(51,0.25)(0.25,0.25)
%\psline[linestyle=none,fillstyle=solid,fillcolor=linepurple!25!white]{-}(0.25,28)(51,0.25)(0.25,0.25)
\psline[linewidth=3pt,linecolor=linepurple,fillstyle=solid,fillcolor=linepurple!25!white]{-}(0.25,42)(14,42)(22,34)(40,34)(40,16)(51,5)(51,0.25)
%\psline[linewidth=3pt,linecolor=linepurple,fillstyle=solid,fillcolor=linepurple!25!white]{-}(22,16)(40,16)(51,5)(51,0.25)
\psline[linestyle=dashed,linecolor=black!70!white](10,46)(56,0)
\psline[linestyle=dashed,linecolor=black!70!white](0,34)(51,34)
\psline[linestyle=dashed,linecolor=black!70!white](0,16)(51,16)
\psline[linestyle=dashed,linecolor=black!70!white](40,0)(40,45)
\psline[linestyle=dashed,linecolor=black!70!white](22,0)(22,45)

\psline(22,-1.5)(22,1.5) \rput(24,-4.5){\footnotesize{$I(X_1,X_2;Y)$}}
\rput(23,-9){\footnotesize{$- I_{\mathsf{CF},2}(\ab)$}}
\psline(40,-1.5)(40,1.5) \rput(41,-4.5){\footnotesize{$I_{\mathsf{CF},1}(\ab)$}}
\psline(51,-1.5)(51,1.5) \rput(58,-4.5){\footnotesize{$I(X_1;Y|X_2)$}}
\psline(-1.5,16)(1.5,16) \rput(-11,16){\footnotesize{$I(X_1,X_2;Y)$}}
\rput(-12.5,12){\footnotesize{$- I_{\mathsf{CF},2}(\ab)$}}
\psline(-1.5,42)(1.5,42) \rput(-11,42){\footnotesize{$I(X_2;Y|X_1)$}}
\psline(-1.5,34)(1.5,34) \rput(-7.5,34){\footnotesize{$I_{\mathsf{CF},2}(\ab)$}}

\rput(19,10){\large{$\Rc_{\mathsf{CF}}(\ab) \cup \Rc_{\mathsf{LMAC}}$}}
}

%Connection arrows
\psline[linewidth=5pt,linecolor=LineGray]{->}(56,102)(67,95)
\psline[linewidth=5pt,linecolor=LineGray]{->}(56,48)(67,55)
%\rput(81,94){\textcolor{black!80!white}{Intersect}}
%\rput(81,88){\textcolor{black!80!white}{Rate Regions}}

\end{pspicture}
\end{center}
\caption{Illustration of the rate region from Theorems~\ref{thm:comp-Fq},~\ref{thm:comp-real-discrete}, and~\ref{thm:comp-real} for the special case when the coefficient vector $\ab$ is chosen to (simultaneously) maximize $I_{\mathsf{CF},1}(\ab)$ and $I_{\mathsf{CF},2}(\ab)$ and we assume that $I_{\mathsf{CF},1}(\ab) + I_{\mathsf{CF},2}(\ab) \geq I(X_1,X_2;Y)$. In the top left, we have the rate region $\Rc_{\mathsf{CF}}(\ab)$ for directly recovering a linear combination via ``single--user'' decoding. In the bottom left, we have the rate region $\Rc_{\mathsf{LMAC}}$  for multiple--access with a shared linear codebook. The rate region from Theorems~\ref{thm:comp-Fq},~\ref{thm:comp-real-discrete}, and~\ref{thm:comp-real} are the union of these two regions and is shown on the right.}

\label{fig:cfrateregion}
\end{figure}

%% file: gaussianmacplot_sumrate.tex
% This file was created by matlab2tikz v0.4.7 running on MATLAB 8.4.
% Copyright (c) 2008--2014, Nico Schlmer <nico.schloemer@gmail.com>
% All rights reserved.
% Minimal pgfplots version: 1.3
% 
% The latest updates can be retrieved from
%   http://www.mathworks.com/matlabcentral/fileexchange/22022-matlab2tikz
% where you can also make suggestions and rate matlab2tikz.
% 
%
% defining custom colors
\definecolor{mycolor1}{rgb}{1.00000,0.00000,1.00000}%
\definecolor{mycolor2}{rgb}{0.00000,0.49804,0.00000}%
\begin{tikzpicture}

\begin{axis}[%
width=4in,
height=3in,
scale only axis,
separate axis lines,
every outer x axis line/.append style={white!15!black},
every x tick label/.append style={font=\color{white!15!black}},
xmin=-5,
xmax=15,
xlabel={SNR in dB},
every outer y axis line/.append style={white!15!black},
every y tick label/.append style={font=\color{white!15!black}},
ymin=0,
ymax=5,
ylabel={Sum Rate},
legend style={at={(0.03,0.97)},anchor=north west,draw=white!15!black,fill=white,legend cell align=left}
]
\addplot [color=LinePurple,dash pattern=on 10pt off 11pt,line width=2.0pt]
  table[row sep=crcr]{%
-5	0.396409161163114\\
-4.48717948717949	0.439210728243855\\
-3.97435897435897	0.485905488775673\\
-3.46153846153846	0.536706475877323\\
-2.94871794871795	0.591813815530336\\
-2.43589743589744	0.651410532898869\\
-1.92307692307692	0.715658429038967\\
-1.41025641025641	0.784694206851237\\
-0.897435897435898	0.858626030677391\\
-0.384615384615385	0.937530695617295\\
0.128205128205129	1.02145155962347\\
0.641025641025641	1.11039735465245\\
1.15384615384615	1.20434194537848\\
1.66666666666667	1.3032250495182\\
2.17948717948718	1.40695387797045\\
2.69230769230769	1.51540560124267\\
3.2051282051282	1.62843050590752\\
3.71794871794872	1.74585567470087\\
4.23076923076923	1.86748900819878\\
4.74358974358974	1.99312340482519\\
5.25641025641026	2.12254092764942\\
5.76923076923077	2.25551680825415\\
6.28205128205128	2.39182316647796\\
6.7948717948718	2.5312323565664\\
7.30769230769231	2.67351988206178\\
7.82051282051282	2.81846685114209\\
8.33333333333333	2.96586196938177\\
8.84615384615385	3.11550308713151\\
9.35897435897436	3.26719833364976\\
9.87179487179487	3.42076688003002\\
10.3846153846154	3.57603937846033\\
10.8974358974359	3.73285812721064\\
11.4102564102564	3.89107700981225\\
11.9230769230769	4.05056125397178\\
12.4358974358974	4.21118705155926\\
12.9487179487179	4.37284107610706\\
13.4615384615385	4.53541992910734\\
13.974358974359	4.69882954132999\\
14.4871794871795	4.8629845506241\\
15	5.02780767335052\\
};
\addlegendentry{Upper Bound};

\addplot [color=LineBlue,solid,line width=2.0pt]
  table[row sep=crcr]{%
-5	0.353521846701088\\
-4.48717948717949	0.387725350635672\\
-3.97435897435897	0.424370711334082\\
-3.46153846153846	0.463498431240649\\
-2.94871794871795	0.505132404730003\\
-2.43589743589744	0.549279264093283\\
-1.92307692307692	0.59592819634831\\
-1.41025641025641	0.645051241533864\\
-0.897435897435898	0.696604055158557\\
-0.384615384615385	0.750527091192929\\
0.128205128205129	0.806747139926282\\
0.641025641025641	0.865179139053184\\
1.15384615384615	0.92572816755865\\
1.66666666666667	0.988291530466549\\
2.17948717948718	1.10552169520122\\
2.69230769230769	1.23804813970712\\
3.2051282051282	1.37393472077375\\
3.71794871794872	1.51295358786901\\
4.23076923076923	1.65487968671179\\
4.74358974358974	1.79949327497459\\
5.25641025641026	1.94658197454597\\
5.76923076923077	2.09594237522193\\
6.28205128205128	2.24738122031226\\
6.7948717948718	2.40071621518342\\
7.30769230769231	2.55577650578374\\
7.82051282051282	2.71240287648103\\
8.33333333333333	2.870447715926\\
8.84615384615385	3.02977479695045\\
9.35897435897436	3.19025891243605\\
9.87179487179487	3.35178540424992\\
10.3846153846154	3.51424961720464\\
10.8974358974359	3.67755630490931\\
11.4102564102564	3.8416190095701\\
11.9230769230769	4.00635943342037\\
12.4358974358974	4.17170681558825\\
12.9487179487179	4.33759732486896\\
13.4615384615385	4.50397347604891\\
13.974358974359	4.670783575095\\
14.4871794871795	4.83798119662904\\
15	5.00552469559907\\
};
\addlegendentry{Corollary 1};

\addplot [color=LineGreen,dotted,line width=2.0pt]
  table[row sep=crcr]{%
-5	0.353173724929279\\
-4.48717948717949	0.387256692439538\\
-3.97435897435897	0.423747413428075\\
-3.46153846153846	0.462679629250266\\
-2.94871794871795	0.504070025651902\\
-2.43589743589744	0.547917762905789\\
-1.92307692307692	0.59420449425423\\
-1.41025641025641	0.642894863540789\\
-0.897435897435898	0.693937448536881\\
-0.384615384615385	0.747266098043293\\
0.128205128205129	0.802801699248164\\
0.641025641025641	0.860454068577825\\
1.15384615384615	0.920123770811046\\
1.66666666666667	0.981704909843814\\
2.17948717948718	1.04508703709435\\
2.69230769230769	1.11015731549246\\
3.2051282051282	1.17680242791101\\
3.71794871794872	1.24491015778676\\
4.23076923076923	1.31746299499642\\
4.74358974358974	1.45887507804832\\
5.25641025641026	1.60256987120417\\
5.76923076923077	1.74834101628748\\
6.28205128205128	1.8959754639939\\
6.7948717948718	2.04523487082105\\
7.30769230769231	2.19582285095698\\
7.82051282051282	2.34733588525501\\
8.33333333333333	2.49920133532411\\
8.84615384615385	2.65061380612564\\
9.35897435897436	2.80048755370505\\
9.87179487179487	2.9474436997572\\
10.3846153846154	3.08984486117054\\
10.8974358974359	3.22587852859986\\
11.4102564102564	3.35367911686284\\
11.9230769230769	3.47147165296143\\
12.4358974358974	3.57771883327946\\
12.9487179487179	3.67125543406446\\
13.4615384615385	3.75139657197546\\
13.974358974359	3.8180076568603\\
14.4871794871795	3.87152515165968\\
15	3.91292061809337\\
};
\addlegendentry{Theorem 2};

\addplot [color=black,line width=0.7pt,mark size=3.0pt,only marks,mark=o,mark options={solid}]
  table[row sep=crcr]{%
-5	0.353173724929279\\
-3.97435897435897	0.423747413428075\\
-2.94871794871795	0.504070025651902\\
-1.92307692307692	0.59420449425423\\
-0.897435897435898	0.693937448536881\\
0.128205128205129	0.802801699248164\\
1.15384615384615	0.920123770811046\\
2.17948717948718	1.04508703709435\\
3.2051282051282	1.17680242791101\\
4.23076923076923	1.3143705597276\\
5.25641025641026	1.45692399783225\\
6.28205128205128	1.60362679422553\\
7.30769230769231	1.75355048770894\\
8.33333333333333	1.90523972989343\\
9.35897435897436	2.13345333319579\\
10.3846153846154	2.40910669296939\\
11.4102564102564	2.66724594916603\\
12.4358974358974	2.8894758262338\\
13.4615384615385	3.06274205674088\\
14.4871794871795	3.18280901633627\\
};
\addlegendentry{Theorem 1, $\q=4$};

\addplot [color=LineRed,dash pattern=on 2pt off 4pt on 8pt off 4pt,line width=2.0pt]
  table[row sep=crcr]{%
-5	0.353521846701088\\
-4.48717948717949	0.387725350635672\\
-3.97435897435897	0.424370711334082\\
-3.46153846153846	0.463498431240649\\
-2.94871794871795	0.505132404730003\\
-2.43589743589744	0.549279264093283\\
-1.92307692307692	0.59592819634831\\
-1.41025641025641	0.645051241533864\\
-0.897435897435898	0.696604055158557\\
-0.384615384615385	0.750527091192929\\
0.128205128205129	0.806747139926282\\
0.641025641025641	0.865179139053184\\
1.15384615384615	0.92572816755865\\
1.66666666666667	0.988291530466549\\
2.17948717948718	1.05276084760061\\
2.69230769230769	1.11902406985356\\
3.2051282051282	1.18696736038687\\
3.71794871794872	1.2564767939345\\
4.23076923076923	1.3274398433559\\
4.74358974358974	1.3997466374873\\
5.25641025641026	1.47329098727298\\
5.76923076923077	1.54797118761096\\
6.28205128205128	1.62369061015613\\
6.7948717948718	1.70035810759171\\
7.30769230769231	1.77788825289187\\
7.82051282051282	1.85620143824052\\
8.33333333333333	1.935223857963\\
8.84615384615385	2.01488739847522\\
9.35897435897436	2.09512945621803\\
9.87179487179487	2.17589270212496\\
10.3846153846154	2.25712480860232\\
10.8974358974359	2.33877815245465\\
11.4102564102564	2.42080950478505\\
11.9230769230769	2.50317971671019\\
12.4358974358974	2.58585340779412\\
12.9487179487179	2.66879866243448\\
13.4615384615385	2.75198673802445\\
13.974358974359	2.8353917875475\\
14.4871794871795	2.91899059831452\\
15	3.00276234779954\\
};
\addlegendentry{i.i.d. Gaussian};

\addplot [color=black,line width=0.7pt,mark size=3.0pt,only marks,mark=x,mark options={solid}]
  table[row sep=crcr]{%
-5	0.35231419713526\\
-3.97435897435897	0.422234457446546\\
-2.94871794871795	0.501402794364757\\
-1.92307692307692	0.589476102416893\\
-0.897435897435898	0.685418664126781\\
0.128205128205129	0.788618415617545\\
1.15384615384615	0.9044784080114\\
2.17948717948718	1.06058634928863\\
3.2051282051282	1.27046084670783\\
4.23076923076923	1.47109872688906\\
5.25641025641026	1.64723710028148\\
6.28205128205128	1.78751095107018\\
7.30769230769231	1.88714409329188\\
8.33333333333333	1.94876362398659\\
9.35897435897436	1.98089018500305\\
10.3846153846154	1.99443270130717\\
11.4102564102564	1.99881160640991\\
12.4358974358974	1.99982860768965\\
13.4615384615385	1.99998493036408\\
14.4871794871795	1.99999929088745\\
};
\addlegendentry{Theorem 1, $\q=2$};

\end{axis}
\end{tikzpicture}%

%% file: magnified.tex
% This file was created by matlab2tikz v0.4.7 running on MATLAB 8.4.
% Copyright (c) 2008--2014, Nico Schlmer <nico.schloemer@gmail.com>
% All rights reserved.
% Minimal pgfplots version: 1.3
% 
% The latest updates can be retrieved from
%   http://www.mathworks.com/matlabcentral/fileexchange/22022-matlab2tikz
% where you can also make suggestions and rate matlab2tikz.
% 
%
% defining custom colors
\definecolor{mycolor1}{rgb}{0.00000,0.49804,0.00000}%
\begin{tikzpicture}

\begin{axis}[%
width=4in,
height=3in,
scale only axis,
separate axis lines,
every outer x axis line/.append style={white!15!black},
every x tick label/.append style={font=\color{white!15!black}},
xmin=1,
xmax=3,
xlabel={SNR in dB},
every outer y axis line/.append style={white!15!black},
every y tick label/.append style={font=\color{white!15!black}},
ymin=0.9,
ymax=1.4,
ylabel={Sum Rate},
legend style={at={(0.03,0.97)},anchor=north west,draw=white!15!black,fill=white,legend cell align=left}
]
\addplot [color=LineBlue,solid,line width=2.0pt]
  table[row sep=crcr]{%
1	0.907347152912549\\
1.05128205128205	0.913453833883914\\
1.1025641025641	0.919580873825244\\
1.15384615384615	0.92572816755865\\
1.20512820512821	0.931895609318166\\
1.25641025641026	0.938083092770541\\
1.30769230769231	0.944290511035962\\
1.35897435897436	0.950517756708687\\
1.41025641025641	0.956764721877593\\
1.46153846153846	0.963031298146625\\
1.51282051282051	0.969317376655141\\
1.56410256410256	0.97562284809815\\
1.61538461538462	0.981947602746435\\
1.66666666666667	0.988291530466549\\
1.71794871794872	0.99465452074069\\
1.76923076923077	1.00207292537287\\
1.82051282051282	1.0148744901527\\
1.87179487179487	1.02771351271484\\
1.92307692307692	1.04058976934078\\
1.97435897435897	1.05350303573777\\
2.02564102564103	1.06645308707682\\
2.07692307692308	1.07943969803036\\
2.12820512820513	1.09246264280954\\
2.17948717948718	1.10552169520122\\
2.23076923076923	1.11861662860455\\
2.28205128205128	1.13174721606722\\
2.33333333333333	1.14491323032137\\
2.38461538461538	1.15811444381903\\
2.43589743589744	1.17135062876729\\
2.48717948717949	1.18462155716294\\
2.53846153846154	1.19792700082688\\
2.58974358974359	1.21126673143793\\
2.64102564102564	1.22464052056639\\
2.69230769230769	1.23804813970712\\
2.74358974358974	1.25148936031213\\
2.79487179487179	1.2649639538229\\
2.84615384615385	1.2784716917021\\
2.8974358974359	1.29201234546497\\
2.94871794871795	1.30558568671024\\
3	1.31919148715061\\
};
\addlegendentry{Corollary 1};

\addplot [color=LineGreen,dotted,line width=2.0pt]
  table[row sep=crcr]{%
1	0.906534870213063\\
1.05128205128205	0.912611669059414\\
1.1025641025641	0.918707327781272\\
1.15384615384615	0.924821654988213\\
1.20512820512821	0.930954454118451\\
1.25641025641026	0.937105523270441\\
1.30769230769231	0.94327465502934\\
1.35897435897436	0.94946163629122\\
1.41025641025641	0.955666248082737\\
1.46153846153846	0.962821872829234\\
1.51282051282051	0.972932048263579\\
1.56410256410256	0.983017811436045\\
1.61538461538462	0.993077701358174\\
1.66666666666667	1.00311024998194\\
1.71794871794872	1.01311398291029\\
1.76923076923077	1.02386796857008\\
1.82051282051282	1.03466860220898\\
1.87179487179487	1.04545021016573\\
1.92307692307692	1.05621131490679\\
1.97435897435897	1.06695042797541\\
2.02564102564103	1.07766605063744\\
2.07692307692308	1.08878760730508\\
2.12820512820513	1.10015288883566\\
2.17948717948718	1.11150314354002\\
2.23076923076923	1.1228368924906\\
2.28205128205128	1.13415264294819\\
2.33333333333333	1.14555166479719\\
2.38461538461538	1.15739505698218\\
2.43589743589744	1.1692276724063\\
2.48717948717949	1.18104805334379\\
2.53846153846154	1.19285472557658\\
2.58974358974359	1.20470215248906\\
2.64102564102564	1.21694134246195\\
2.69230769230769	1.22917294266703\\
2.74358974358974	1.24139552119904\\
2.79487179487179	1.25360762762928\\
2.84615384615385	1.26580779344852\\
2.8974358974359	1.278322076092\\
2.94871794871795	1.29088640681261\\
3	1.30344402301015\\
};
\addlegendentry{Theorem 2};

\addplot [color=LineRed,dash pattern=on 2pt off 4pt on 8pt off 4pt,line width=2.0pt]
  table[row sep=crcr]{%
1	0.907347152912549\\
1.05128205128205	0.913453833883914\\
1.1025641025641	0.919580873825244\\
1.15384615384615	0.92572816755865\\
1.20512820512821	0.931895609318166\\
1.25641025641026	0.938083092770541\\
1.30769230769231	0.944290511035962\\
1.35897435897436	0.950517756708687\\
1.41025641025641	0.956764721877593\\
1.46153846153846	0.963031298146625\\
1.51282051282051	0.969317376655141\\
1.56410256410256	0.97562284809815\\
1.61538461538462	0.981947602746435\\
1.66666666666667	0.988291530466549\\
1.71794871794872	0.99465452074069\\
1.76923076923077	1.00103646268644\\
1.82051282051282	1.00743724507635\\
1.87179487179487	1.01385675635742\\
1.92307692307692	1.02029488467039\\
1.97435897435897	1.02675151786888\\
2.02564102564103	1.03322654353841\\
2.07692307692308	1.03971984901518\\
2.12820512820513	1.04623132140477\\
2.17948717948718	1.05276084760061\\
2.23076923076923	1.05930831430228\\
2.28205128205128	1.06587360803361\\
2.33333333333333	1.07245661516069\\
2.38461538461538	1.07905722190952\\
2.43589743589744	1.08567531438364\\
2.48717948717949	1.09231077858147\\
2.53846153846154	1.09896350041344\\
2.58974358974359	1.10563336571896\\
2.64102564102564	1.1123202602832\\
2.69230769230769	1.11902406985356\\
2.74358974358974	1.12574468015607\\
2.79487179487179	1.13248197691145\\
2.84615384615385	1.13923584585105\\
2.8974358974359	1.14600617273248\\
2.94871794871795	1.15279284335512\\
3	1.1595957435753\\
};
\addlegendentry{i.i.d. Gaussian};

\end{axis}
\end{tikzpicture}%

%% file: gaussianregion.tex
\begin{figure*}[h]
\begin{center}
\psset{unit=.98mm}
\begin{pspicture}(-2,-2)(160,130)
\small
%\psframe(-2,0)(160,130)

%All rates scaled up by a factor of 20.

\rput(3,70){
%\pspolygon[fillstyle=solid,fillcolor=BoxBlue,linestyle=none](0,37.5)(20,37.5)(37.5,20)(37.5,0)(0,0)
\psline[linewidth=1.5pt,linecolor=black]{->}(0,0)(60,0)  \rput(60,3){$R_1$}
\psline[linewidth=1.5pt]{->}(0,0)(0,60)  \rput(3.5,60){$R_2$}

%Rate regions for MAC with H = [1 sqrt(2)] and P = diag(25,18).
%MAC region defined by corner points (0.3724,2.6047) and (2.3502, 0.6269).
%Multiplied by 20 (7.448, 52.094) and (47.004, 12.538)
\psline[linewidth=6pt,linecolor=black!65!white]{-}(0,52.094)(7.448,52.094)(47.004,12.538)(47.004,0)
%LMAC region defined by points (0.3724, 2.6047) (1.3697, 1.6074) (1.8443,1.1328) and (2.3502, 0.6269).
%Multiplied by 20 (7.448, 52.094) (27.394, 32.148) (36.886, 22.656) and (47.004, 12.538)
\psline[linewidth=3.5pt,linecolor=lineblue]{-}(0,52.094)(7.448,52.094)(27.394, 32.148)(27.394, 22.656)(36.886, 22.656) (47.004,12.538)(47.004,0)
%Successive comp MAC region defined by points (0.3724, 2.6047) (1.3697, 1.6074) (1.8443,1.1328) and (2.3502, 0.6269).
%Multiplied by 20 (7.448, 52.094) (27.394, 32.148) (36.886, 22.656) and (47.004, 12.538)
\psline[linewidth=1.5pt,linecolor=linered]{-}(0,52.094)(7.448,52.094)(7.448,32.148)(27.394, 32.148)(27.394, 22.656)(36.886, 22.656)(36.886,12.538) (47.004,12.538)(47.004,0)

\psline(20,-1.5)(20,1.5) \rput(20,-4.5){\small{$1$}}
\psline(40,-1.5)(40,1.5) \rput(40,-4.5){\small{$2$}}
\psline(-1.5,20)(1.5,20) \rput(-3.5,20){\small{$1$}}
\psline(-1.5,40)(1.5,40) \rput(-3.5,40){\small{$2$}}

\rput(34,56){\textbf{Receiver 1 Rate Constraints}}

}

\rput(3,0){
\psline[linewidth=1.5pt,linecolor=black]{->}(0,0)(60,0)  \rput(60,3){$R_1$}
\psline[linewidth=1.5pt]{->}(0,0)(0,55)  \rput(3.5,55){$R_2$}

%Rate regions for recovering a = [1 1] over H = [1 1] with P = diag(25,18).
%MAC region defined by corner points (0.6058,2.1240) and (2.3502, 0.3795).
%Multiplied by 20 (12.116, 42.48) and (47.004, 7.59)
\psline[linewidth=6pt,linecolor=black!65!white]{-}(0,42.48)(12.116,42.48)(47.004,7.59)(47.004,0)
%Joint CF region defined by points (0.6058,2.1240) (2.3385, 2.1015) (2.3502, 0.3795)
%Multiplied by 20 (7.448, 52.094) (46.77,42.03) (47.004, 12.538)
\psline[linewidth=3.5pt,linecolor=lineblue]{-}(0,42.48)(12.116,42.48)(12.566,42.03)(46.77,42.03)(46.77,7.824)(47.004,7.59)(47.004,0)

%Successive CF region defined by points (0.6058,2.1240) (2.3385, 2.1015) (2.3502, 0.3795)
%Multiplied by 20 (7.448, 52.094) (46.77,42.03) (47.004, 12.538)

\psline[linewidth=1.5pt,linecolor=linered]{-}(0,42.48)(12.116,42.48)(12.116,42.03)(12.566,42.03)(46.77,42.03)(46.77,7.824)(46.77,7.59)(47.004,7.59)(47.004,0)

\psline(20,-1.5)(20,1.5) \rput(20,-4.5){\small{$1$}}
\psline(40,-1.5)(40,1.5) \rput(40,-4.5){\small{$2$}}
\psline(-1.5,20)(1.5,20) \rput(-3.5,20){\small{$1$}}
\psline(-1.5,40)(1.5,40) \rput(-3.5,40){\small{$2$}}

\rput(34,48){\textbf{Receiver 2 Rate Constraints}}

}

\rput(95,70){
\psline[linewidth=1.5pt,linecolor=black]{->}(0,0)(60,0)  \rput(60,3){$R_1$}
\psline[linewidth=1.5pt]{->}(0,0)(0,60)  \rput(3.5,60){$R_2$}

%Intersection of rate regions
%Compound MAC region defined by corner points (0.6058,2.1240) and (2.3502, 0.3795).
%Multiplied by 20 (12.116, 42.48) and (47.004, 7.59)
\psline[linewidth=6pt,linecolor=black!65!white]{-}(0,42.48)(12.116,42.48)(47.004,7.59)(47.004,0)

%Joint decoding of linear codes
\psline[linewidth=3.5pt,linecolor=lineblue]{-}(0,42.48)(12.116,42.48)(12.566,42.03)(17.512,42.03)(27.394, 32.148)(27.394, 22.656)(36.886, 22.656) (46.77,12.538)(46.77,7.824)(47.004,7.59)(47.004,0)

%Successive  linear codes
\psline[linewidth=1.5pt,linecolor=linered]{-}(0,42.48)(7.448,42.48)(7.448,32.148)(27.394, 32.148)(27.394, 22.656)(36.886, 22.656)(36.886,12.538) (46.77,12.538)(46.77,7.59)(47.004,7.59)(47.004,0)

\psline(20,-1.5)(20,1.5) \rput(20,-4.5){\small{$1$}}
\psline(40,-1.5)(40,1.5) \rput(40,-4.5){\small{$2$}}
\psline(-1.5,20)(1.5,20) \rput(-3.5,20){\small{$1$}}
\psline(-1.5,40)(1.5,40) \rput(-3.5,40){\small{$2$}}

\rput(34,56){\textbf{Intersection of Receiver 1 and}}
\rput(34,52){\textbf{Receiver 2 Rate Constraints}}

}

\rput(95,0){
\psline[linewidth=1.5pt,linecolor=black]{->}(0,0)(60,0)  \rput(60,3){$R_1$}
\psline[linewidth=1.5pt]{->}(0,0)(0,55)  \rput(3.5,55){$R_2$}
%Compound MAC region defined by corner points (0.6058,2.1240) and (2.3502, 0.3795).
%Multiplied by 20 (12.116, 42.48) and (47.004, 7.59)
\psline[linewidth=6pt,linecolor=black!65!white]{-}(0,42.48)(12.116,42.48)(47.004,7.59)(47.004,0)

%Joint decoding of linear codes
\psline[linewidth=3.5pt,linecolor=lineblue]{-}(0,42.48)(12.116,42.48)(17.512,42.03)(27.394, 32.148) (46.77,12.538)(47.004,7.59)(47.004,0)

%Successive  linear codes
\psline[linewidth=1.5pt,linecolor=linered]{-}(0,42.48)(7.448,42.48)(27.394, 32.148)(36.886, 22.656) (46.77,12.538)(47.004,7.59)(47.004,0)

\psline(20,-1.5)(20,1.5) \rput(20,-4.5){\small{$1$}}
\psline(40,-1.5)(40,1.5) \rput(40,-4.5){\small{$2$}}
\psline(-1.5,20)(1.5,20) \rput(-3.5,20){\small{$1$}}
\psline(-1.5,40)(1.5,40) \rput(-3.5,40){\small{$2$}}

\rput(34,48){\textbf{Achievable Rate Regions}}

}

%Connection arrows
\psline[linewidth=5pt,linecolor=LineGray]{->}(65,100)(85,100)
\psline[linewidth=5pt,linecolor=LineGray]{->}(65,52)(85,83)
\rput(76,94){\textcolor{black!80!white}{Intersect}}
\rput(76,88){\textcolor{black!80!white}{Rate Regions}}

\psline[linewidth=5pt,linecolor=LineGray]{->}(121,66)(121,51)
\rput(134,60){\textcolor{black!80!white}{Time-Sharing}}
\rput(134,56){\textcolor{black!80!white}{(Convexify)}}

\end{pspicture}
\end{center}
\caption{Step-by-step illustration for determining the achievable rate regions for Example~\ref{ex:binary2mac2}. On the left, we have the rate constraints imposed by the receivers $1$ and $2$, respectively. On the top right, we have the intersection of these rate constraints. Time sharing yields the achievable rate regions on the bottom right. The thick black line represents the rate region available to i.i.d.~Gaussian codebooks combined with joint typicality decoding. The blue line represents the rate region available to nested linear codebooks combined with joint typicality decoding (along with a discretization argument to the Gaussian case). The thin red line represents the rate region available to nested lattice codebooks combined with successive cancellation decoding.}

\label{fig:gaussianregion}
\end{figure*}

%% file: linearcombination.tex
\begin{figure}[t]
\psset{unit=1mm}
\begin{center}
\begin{pspicture}(-7,-8)(80,33)
%\psframe[linewidth=0.5pt](-7,-8)(80,33)
\rput(10,20){$a_{\ell 1} \etab(m_1,l_1)$}
\rput(7.5,13){$\bigoplus a_{\ell 2} \etab(m_2,l_2)$}
\rput(9,7){$\nub(s_{\ab})$}
\psline[linewidth=1pt](-6,10)(21,10)

\rput(36,24.5){$a_{\ell 1}$}
\psline[linewidth=1pt]{<->}(41.25,29)(59.75,29)
\rput(51,31){\scriptsize{$\nub(m_1)$}}
\psline[linewidth=1pt]{<->}(61.25,29)(74.75,29)
\rput(68,31){\scriptsize{$\nub(l_1)$}}
\psframe[linewidth=0.5pt](40,22)(76,28)
\pscircle[fillstyle=solid,fillcolor=lineblue,linewidth=0.5pt](43,25){2}
\pscircle[fillstyle=solid,fillcolor=lineblue,linewidth=0.5pt](48,25){2}
\pscircle[fillstyle=solid,fillcolor=lineblue,linewidth=0.5pt](53,25){2}
\pscircle[fillstyle=solid,fillcolor=lineblue,linewidth=0.5pt](58,25){2}
\pscircle[fillstyle=hlines,linewidth=0.5pt,hatchwidth=1pt,hatchcolor=lineblue,hatchsep=1pt](63,25){2}
\pscircle[fillstyle=hlines,linewidth=0.5pt,hatchwidth=1pt,hatchcolor=lineblue,hatchsep=1pt](68,25){2}
\pscircle[fillstyle=hlines,linewidth=0.5pt,hatchwidth=1pt,hatchcolor=lineblue,hatchsep=1pt](73,25){2}

\rput(0,-12){
\rput(33.5,24.75){$\bigoplus a_{\ell 2}$}
\psline[linewidth=1pt]{<->}(41.25,29)(54.75,29)
\rput(48,31){\scriptsize{$\nub(m_2)$}}
\psline[linewidth=1pt]{<->}(56.25,29)(64.75,29)
\rput(60.5,31){\scriptsize{$\nub(l_2)$}}
\psline[linewidth=1pt]{<->}(66.25,29)(74.75,29)
\rput(70.5,31){\scriptsize{$\mathbf{0}$}}
\psframe[linewidth=0.5pt](40,22)(76,28)
\pscircle[fillstyle=solid,fillcolor=linered,linewidth=0.5pt](43,25){2}
\pscircle[fillstyle=solid,fillcolor=linered,linewidth=0.5pt](48,25){2}
\pscircle[fillstyle=solid,fillcolor=linered,linewidth=0.5pt](53,25){2}
\pscircle[fillstyle=hlines,hatchwidth=1pt,hatchcolor=linered,hatchsep=1pt,linewidth=0.5pt](58,25){2}
\pscircle[fillstyle=hlines,hatchwidth=1pt,hatchcolor=linered,hatchsep=1pt,linewidth=0.5pt](63,25){2}
\pscircle[linewidth=0.5pt](68,25){2}
\pscircle[linewidth=0.5pt](73,25){2}
}

\psline[linewidth=1.5pt](28,5)(78,5)

\rput(0,-30){
\psline[linewidth=1pt]{<->}(41.25,29)(74.75,29)
\rput(58,31){\scriptsize{$\nub(s_{\ab_\ell})$}}
\psframe[linewidth=0.5pt](40,22)(76,28)
\pscircle[fillstyle=solid,fillcolor=linepurple,linewidth=0.5pt](43,25){2}
\pscircle[fillstyle=solid,fillcolor=linepurple,linewidth=0.5pt](48,25){2}
\pscircle[fillstyle=solid,fillcolor=linepurple,linewidth=0.5pt](53,25){2}
\pscircle[fillstyle=solid,fillcolor=lineblue,linewidth=0.0pt](58,25){2}
\pscircle[fillstyle=hlines,hatchwidth=1pt,hatchcolor=linepurple,hatchsep=1pt,linewidth=0.5pt](58,25){2}
\pscircle[fillstyle=hlines,hatchwidth=1pt,hatchcolor=linepurple,hatchsep=1pt,linewidth=0.5pt](63,25){2}
\pscircle[fillstyle=hlines,linewidth=0.5pt,hatchwidth=1pt,hatchcolor=lineblue,hatchsep=1pt](68,25){2}
\pscircle[fillstyle=hlines,linewidth=0.5pt,hatchwidth=1pt,hatchcolor=lineblue,hatchsep=1pt](73,25){2}
}

\end{pspicture}
\end{center}
\caption{An illustration of a linear combination of the $\q$-ary expansions of message and auxiliary indices. On the right-hand side, we have used solid colors for message symbols and dashed lines for auxiliary symbols. Transmitter 1's symbols are shown in blue and occupy the entire vector. Transmitter 2's symbols are shown in red and only occupy part of the vector. We have assumed that both $a_{\ell 1}$ and $a_{\ell 2}$ are non-zero so the linear combination occupies the entire vector. (If $a_{\ell 1} = 0$, then the last two entries will be zero.) }
\label{fig:linearcombination}
\end{figure}